%% file: vhmpidLoI_EPJStyle.tex
\pdfoutput=1
\documentclass[epj,referee]{svjour}
\usepackage{graphics}
\usepackage[utf8]{inputenc}
\usepackage{multicol}
\usepackage{xspace}
\usepackage{graphicx}
\usepackage{multicol}
\usepackage{amsmath}
\usepackage{color}
\usepackage{rotating}
\usepackage{cite}
\usepackage[section] {placeins}

\begin{document}
\title{A Very High Momentum Particle Identification Detector}
\author{
	T. V. Acconcia\inst{8} 
	\and A.G. Ag\'ocs\inst{12}
	\and F. Barile\inst{7}
	\and G.G. Barnaf\"oldi\inst{12}
	\and R. Bellwied\inst{9}	
	\and G. Benc\'edi\inst{12}
	\and G. Bencze\inst{12}
	\and D. Ber\'enyi\inst{12}
	\and L. Boldizs\'ar\inst{12}
 	\and S. Chattopadhyay\inst{6}
	\and F. Cindolo\inst{10}\thanks{\emph{Visitor researcher from INFN Bologna, Italy.}}
	\and D. D. Chinellato\inst{9}
	\and S. D'Ambrosio\inst{10}
	\and D. Das\inst{6}
	\and K. Das\inst{6}
	\and L. Das-Bose\inst{6}
	\and A. K. Dash\inst{8}
	\and G. De Cataldo\inst{7}
	\and S. De Pasquale\inst{10}
	\and D. Di Bari\inst{7}
        \and A. Di Mauro\inst{13}
	\and E. Fut\'o\inst{12}
	\and E. Garc\'ia\inst{1}
	\and G. Hamar\inst{12}
	\and A. Harton\inst{1}
	\and G. Iannone\inst{10}
	\and R.T. Jimenez\inst{4}
	\and D.W. Kim\inst{3}
	\and J.S. Kim\inst{3}	
	\and A. Knospe\inst{11}
	\and L. Kov\'acs\inst{12}
	\and P. L\'evai\inst{12}
	\and E. Nappi\inst{7}
	\and C. Markert\inst{11}
        \and P. Martinengo\inst{13}
	\and D. Mayani\inst{4}
	\and L. Moln\'ar\inst{12}
	\and L. Oláh\inst{2}
	\and G. Pai\'c\inst{4}
	\and C. Pastore\inst{7}
	\and G. Patimo\inst{10}
	\and M.E. Patino\inst{4}
	\and V. Peskov\inst{4}
	\and L. Pinsky\inst{9}
        \and F. Piuz\inst{13}
	\and S. Pochybov\'a\inst{12}
	\and I. Sgura\inst{7}
	\and T. Sinha\inst{6}
	\and J.~Song\inst{5}
	\and J. Takahashi\inst{8}
	\and A. Timmins\inst{9}
        \and J.B. Van Beelen\inst{13}
	\and D. Varga\inst{2}
	\and G. Volpe\inst{7}
	\and M. Weber\inst{9}
	\and  L. Xaplanteris\inst{11}
	\and J.~Yi\inst{5}
	\and I.-K.~Yoo\inst{5}
% etc
}                     % Do not remove
\institute{
	Chicago State University, Chicago, IL, USA
	\and E\"otv\"os University, Budapest, Hungary
	\and Gangneung-Wonju National University, Dept. of Physics, Gangneung, South Korea
	\and Instituto de Ciencias Nucleares Universidad Nacional Aut\'onoma de M\'exico, Mexico City, Mexico
	\and Pusan National University, Dept. of Physics, Pusan, South Korea
	\and Saha Institute of Nuclear Physics, Kolkata, India
	\and Universita degli Studi di Bari, Dipartimento Interateneo di Fisica "M. Merlin" \& INFN Sezione di Bari, Bari, Italy
	\and UNICAMP, University of Campinas, Campinas, Brazil
	\and University of Houston, Houston, USA
	\and University of Salerno, Salerno, Italy
	\and University of Texas at Austin, Austin, USA
	\and Wigner RCP of the HAS, Budapest, Hungary
        	\and CERN, CH1211 Geneva 23, Switzerland	
}
\date{Received: date / Revised version: date}
\abstract{
The construction of a new detector is proposed to extend the capabilities of ALICE in the high transverse momentum ($p_\mathrm{T}$) region. This Very High Momentum Particle Identification Detector (VHMPID) performs charged hadron identification on a track-by-track basis in the $5$ GeV/c $ < p < 25$ GeV/c momentum range and provides ALICE with new opportunities to study parton-medium interactions at LHC energies. The VHMPID covers up to $30 \%$ of the ALICE central barrel and presents sufficient acceptance for triggered- and tagged-jet studies, allowing for the first time identified charged hadron measurements in jets. This Letter of Intent summarizes the physics motivations for such a detector as well as its layout and integration into ALICE.
 % end of PACS codes
} %end of abstract
\maketitle
%

%
% For  figures use
%\begin{figure*}
%\vspace*{5cm}       % Give the correct figure height in cm
%\includegraphics{leer.eps}
%\end{figure*}
% or  this
%\begin{figure}
%\centering
%\resizebox{0.75\textwidth}{!}{%
%  \includegraphics{leer.eps}
%\vspace{5cm}       % Give the correct figure height in cm
%\end{figure}
%\begin{table}
%\centering
%\begin{tabular}{lll}
%\hline\noalign{\smallskip}
%\end{tabular}
% Or use
%\vspace*{5cm}  % with the correct table height
%\end{table}

%
% BibTeX users please use
% \bibliographystyle{}
% \bibliography{}
%\begin{thebibliography}{}
%Author, \textit{Book title} (Publisher, place year) page numbers
% etc
%\end{thebibliography}

\input{vhmpidLoI_pub_body.tex}               %%%%%%%%%%% put the body of the article here

\end{document}

%% file: vhmpidLoI_pub_body.tex
./%--------------------------------------------------------------------------
% This is for arxiv submission
% w.r.t. to original document following changes:
% - signed only by ALICE-VHMPID collaboration
% - removed management chapter
%-------------------------------------------------------------------------

%%%%%%%%%%%%%%%%%%%%%%%%%%%%%%%%%%%%%%%%%%%%%%%%%%%%%%%%%%%%%%%%%%%%%%%%%%%%%%%%%%%%%%
\tableofcontents

\newpage
%%%%%%%%%%%%%%%%%%%%%%%%%%%%%%%%%%%%%%%%%%%%%%%%%%%%%%%%%%%%%%%%%%%%%%%%%%%%%%%%%%%%%%
\section{Physics Motivation}

The purpose of the ALICE experiment is to identify and study the quark-gluon plasma (QGP) in heavy ion collisions at LHC energies~\cite{Alessandro:2006yt}. The ALICE detectors were designed in the mid-1990's with the aim to discover the properties of QCD matter at high temperatures in the soft regime. However, after the start of operation of RHIC at BNL in 2000, results from high energy nucleus-nucleus collisions have shown the importance of high-momentum particles as hard probes and the need for particle identification in a very large momentum range. The ALICE detector has a unique capability to identify a wide variety of particles, however its momentum coverage should be extended to meet new physics challenges at LHC.

Recent spectacular measurements using the particle identification capabilities of the TPC in the relativistic rise region of the measured dE/dx enabled ALICE to separate and determine anisotropic flow ($v_2$) and nuclear suppression factors ($R_{AA}$) for protons, kaons and pions out to 25 GeV/c. These measurements are based on a statistical sample of unidentified tracks which is then fit with a global Bethe\,--\,Bloch response based algorithm in order to obtain the particle separated results. To significantly enhance ALICE's particle identification capabilities in the regime of particle-by-particle measurements, we propose to construct a new detector, the {\bf Very High Momentum Particle Identification Detector (VHMPID)}.

The VHMPID aims to identify charged pions, kaons, protons and antiprotons in the momentum range $5$ GeV/$c < p < 25$ GeV/$c$ on a track-by-track basis. Its main purpose is to determine hadron specific effects in the fragmentation of partons and subsequent formation of jets in vacuum and in medium. This gives us a unique view of the hadronization process itself, which is non-perturbative and thus often treated simply through a factorization approach in the perturbative calculations. The underlying physics and the relevant degrees of freedom, in particular in the deconfined medium, are not known and generally approximated schematically on the basis of lattice QCD calculations. Only specific experimental evidence of the modification of the hadronization process in medium will enable us to constrain the schematic models and gain a deeper understanding of the fundamental process of hadron formation during the evolution of the deconfined matter. Since the low momentum region of the hadron spectrum is populated by many competing processes, such as thermal production and recombination in addition to fragmentation, only the high momentum (high-$p_T$) part of the spectrum can be interpreted unambiguously. Historically this region is considered featureless with respect to specific quark configurations, since the fragmentation process and its modification in medium should not be flavor or configuration dependent, at least in the light quark sector.

Many recent theory predictions ~\cite{Sapeta:2007ad,liu-fries,Hwa:2006zq,PL-QM2011,aurenche-2011} have hinted at significant flavor and baryon effects in the medium modified fragmentation of jets in the deconfined matter. This is in part confirmed by recent results from RHIC though~\cite{star1,phenix1}, and early measurements from ALICE in the 5 GeV/c to 10 GeV/c range.

Furthermore even the basic measurements of baryon fragmentation in vacuum have not been explored to a decisive level in past high energy experiments, and none of the competing LHC experiments has the necessary detector capabilities to measure protons at high momentum.

It seems that the measurement of the high momentum hadro-chemistry in jets might enable us to understand baryon-meson formation and flavor specific effects. This has also led to a series of  theoretical predictions for high-$p_T$ particle formation in vacuum and in medium~\cite{Sapeta:2007ad,Hwa:2006zq,PL-QM2011,bellwied-markert}. In order to measure hadron formation in jets we need a detector that can identify hadrons on a track-by-track and event-by-event basis out to $20-30$ GeV/c. A RICH based Cherenkov detector presently is the {\it only} technology capable of such measurements. These measurements will be unique to ALICE, since no other LHC experiment has such specific PID capabilities or is planning to employ a similar device in the future. Since particle identified fragmentation in the vacuum has not been addressed by high energy physics for several decades, there is also a significant relevance to these measurements in proton-proton collisions. Leading (LO) and next-to-leading order (NLO) calculations based on the factorization of the fragmentation process lack the specifics to successfully predict in particular baryon or hadronic resonance formation. Only the input of high momentum identified spectra will improve the theoretical understanding of hadronization to states other than the pion.

%%%%%%%%%%%%%%%%%%%%%%%%%%%%%%%%%%%%%%%%%%%%%%%%%%%%%%%%%
\subsection{Existing ALICE high-$p_T$ PID capabilities}

\begin{figure}[!h]
\centering
\includegraphics[width=0.55\textwidth]{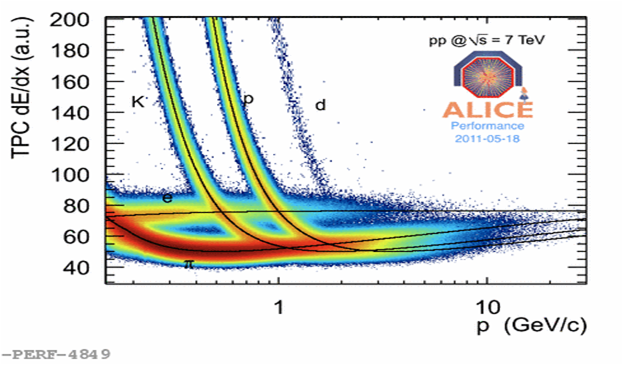}
\includegraphics[width=0.43\textwidth]{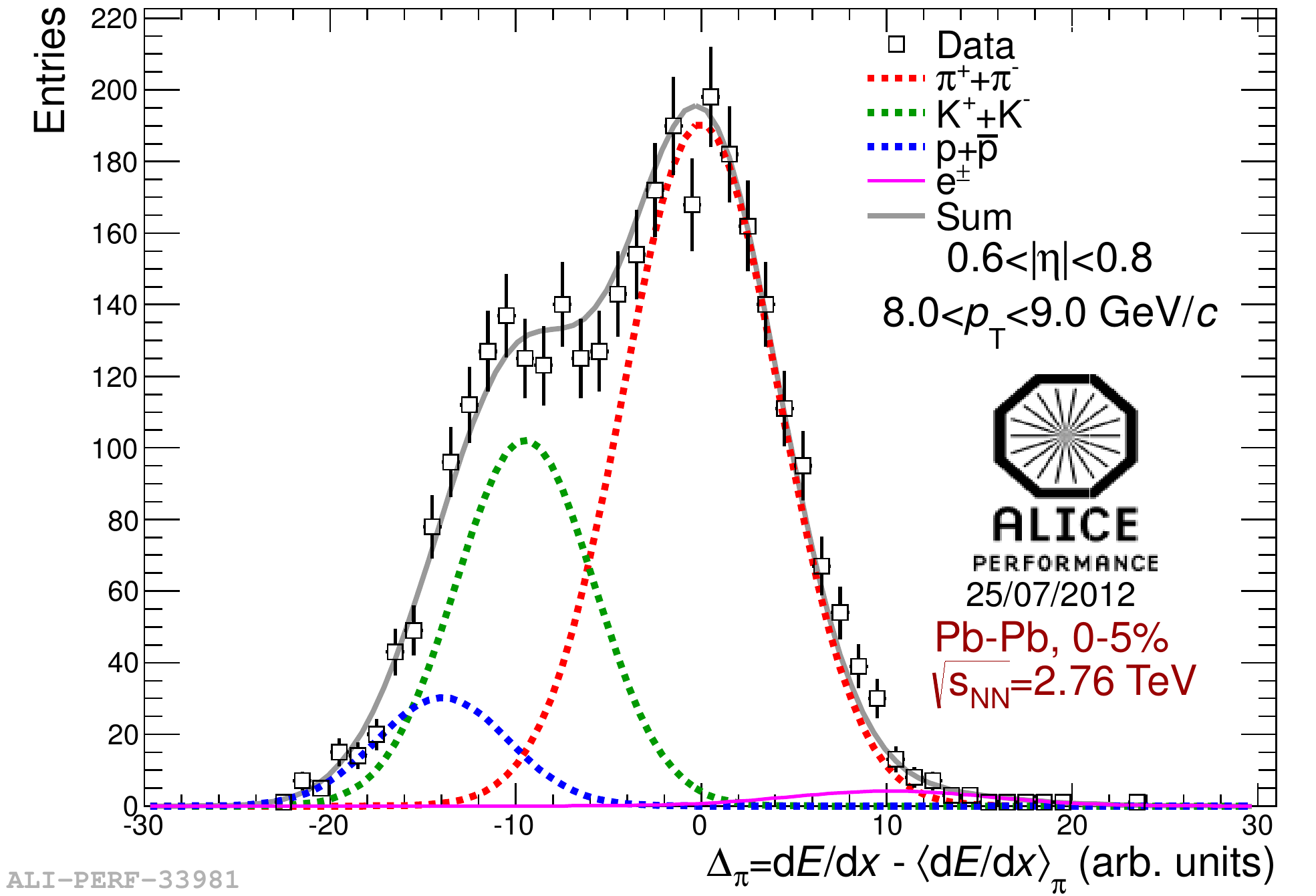}

\caption{Left: Present level of the TPC dE/dx resolution measured in pp collisions in ALICE. Right: Statistical separation fit applied to the TPC dE/dx distribution for 8-9 GeV/c particles measured in $ 0-5\% $ central Pb-Pb collisions in ALICE.}
\label{fig:tpc-perf1}
\end{figure}

During the first two LHC heavy ion runs the ALICE TPC produced significant particle identification results through particle energy loss measurements in the relativistic rise region (dE/dx). This method allows for particle identification on a statistical basis. Fig.~\ref{fig:tpc-perf1} shows the present level of separation obtained for all momenta in proton-proton collisions ({\it left}) and for a specific $p_T$ bin in lead-lead collisions ({\it right}). Clearly the charged pions can be identified in a rather background free way through a specific cut on the dE/dx distribution. 

\begin{figure}[!h]
\centering
\includegraphics[width=0.5\textwidth]{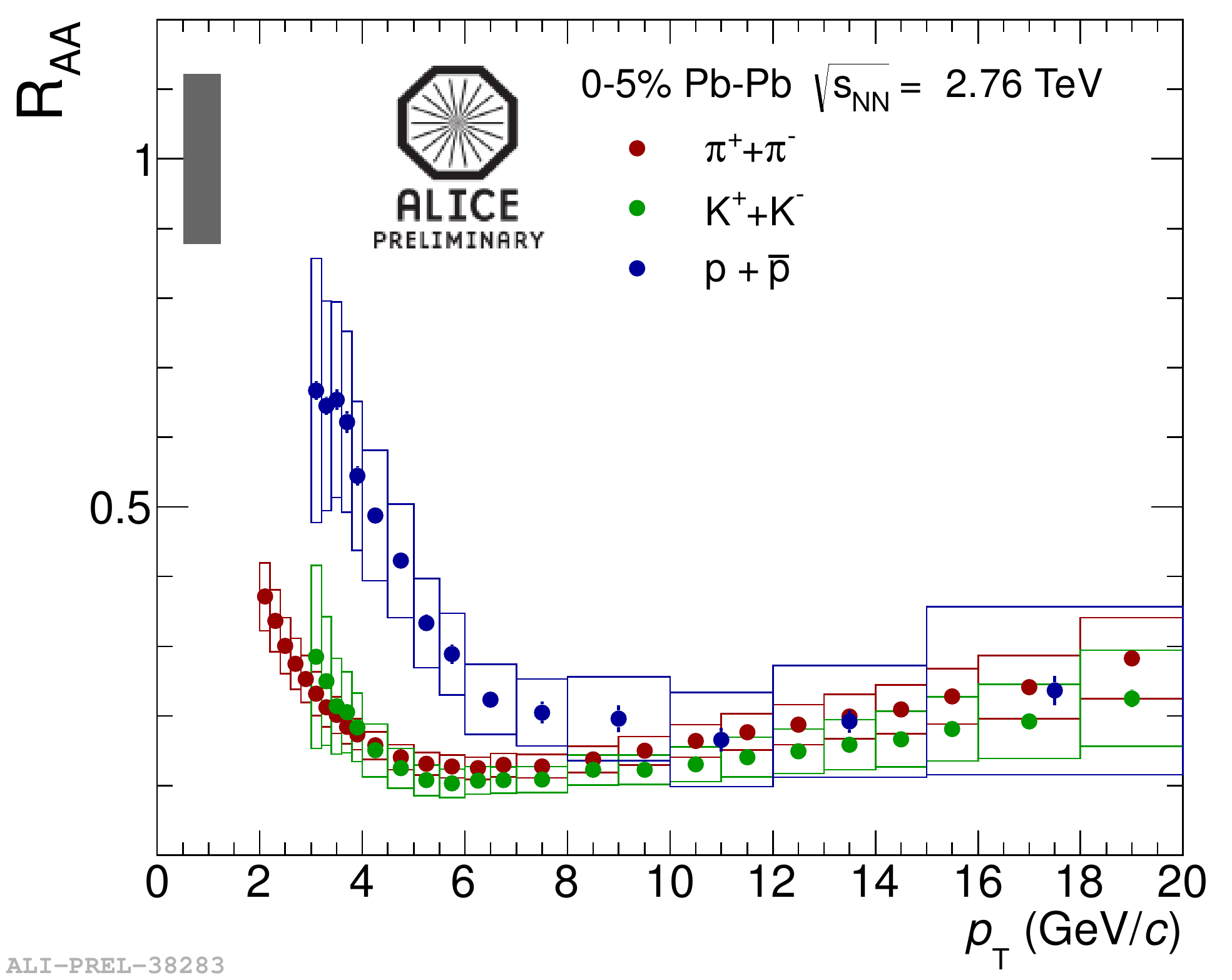}
\caption{Charged pion, kaon and (anti-)proton  $R_{\mathrm{AA}}$ vs. $p_{\mathrm{T}}$ measured in $ 0-5\% $ central Pb-Pb collisions. Statistical (vertical error bars) and systematic (grey and colored boxes) are shown for the charged pion $R_{\mathrm{AA}}$. The gray boxes contain the common systematic error related to pp normalization to INEL and $N_{\mathrm{coll}}$.}
\label{fig:tpc-perf2}
\end{figure}

The TPC PID at high $p_T$ relies on constraining dE/dx vs. $\beta
\gamma$ using a combination of low momentum regions, TOF PID, and
topological PID ($V^0$ daughters). With these constrains statistical PID has
been performed for pions, kaons, and protons to extract $R_{\mathrm{AA}}$ out to $20$~
GeV/c, see Fig. \ref{fig:tpc-perf2} (QM12 preliminary result). The systematic error for
kaons and protons is quite large due to the limited near constant
separation of order 1 sigma.

An alternative method that has been used is to select low contamination
regions in dE/dx to obtain $v_2$ and $v_3$ for pions and protons at high $p_{\mathrm{T}}$~
\cite{highPtFlowALICE}. This latter method is not possible to use for kaons, and for
protons there is a large statistical trade off to have a low kaon
contamination.

\begin{figure}[!h]
  \centering
   \includegraphics[width=0.6\textwidth]{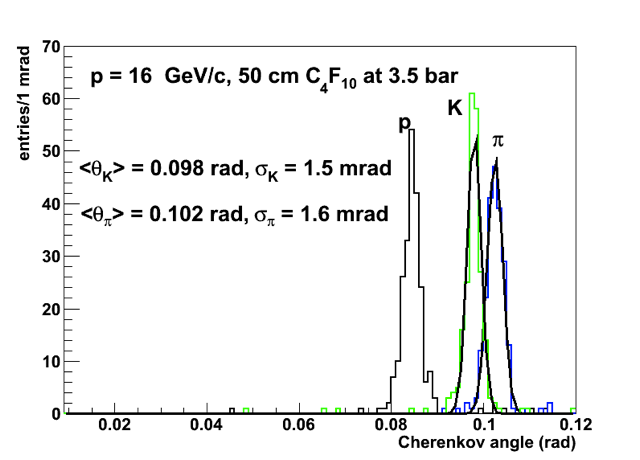}
 \caption{Distribution of ring-averaged Cherenkov angles for pions, kaons, and
protons at $16$ GeV/c at 3.5 atm gas radiator pressure (for details see section~\ref{Subsection:Detector_performance_simulations}).}
      \label{thetaChIntroduction}
\end{figure}

The synergy between two complementary PID measurements in the TPC and the VHMPID is best documented when comparing {\it right panel} of Fig.~\ref{fig:tpc-perf1} with Fig.~\ref{thetaChIntroduction} (for details see section~\ref{Subsection:Detector_performance_simulations}). Clearly, the relativistic dE/dx will provide excellent statistical pion measurements in the $5-25$ GeV/c $p_T$-range. The superior $K/p$ separation in the same momentum range in the VHMPID will add track-by-track measurements not only for the pions but in particular for protons and charged kaons, since the velocity measurements in the RICH are more sensitive to the mass differences between the species. A combined PID from both detectors will yield the best possible particle separation.

%%%%%%%%%%%%%%%%%%%%%%%%%%%%%%%%%%%%%
%%%%%%%%%%%%%%%%%%%%%%%%%%%%%%%%%%%%%

The unique capability of the VHMPID is in the track-by-track particle identification though which lies at the heart of all the jet-jet, jet-hadron and di-hadron correlations shown in the following section. Modified fragmentation or flavor/baryon number dependencies in the hadronization can only be mapped if the identified particles can be unambiguously linked to the generating parton (i.e. the fragmenting jet).

%%%%%%%%%%%%%%%%%%%%%%%%%%%%%%%%%%%%%%%%%%%%%%%%%%%%%%%%%%%%%%%
\subsection{Specific measurements in proton-proton collisions}

Since proton-proton collisions are a primary goal of the LHC, and are presently performed at various center of mass energies, the ALICE collaboration has proposed a wide spectrum of measurements of minimum bias and jet-triggered studies for proton-proton collisions. In such an elementary system like pp the VHMPID measurements at high $p_T$ have a significantly larger signal-to-noise ratio and thus lead to very stringent baseline measurements of the hadron yield for later comparison to the nucleus-nucleus data. The recent RHIC pp analyses, based on measurements of identified particles, addressed two main topics: the aforementioned hadro-chemistry and PID-triggered jet analysis~\cite{Sona,Elezar,AndrasAgocs}. The first set of measurements mainly measures the flavor balance in PID-triggered minimum bias spectra thus testing hadronization/fragmentation models. The latter focuses on the geometrical distribution (e.g. angular distribution) of flavors in event-by-event analysis. The precise measurement of parton fragmentation to protons and anti-protons~\cite{bbar}, which has never been done before out to the relevant momentum transfer scale, will significantly constrain fragmentation functions and enable us to distinguish between quark and gluon fragmentation as well as shed
light on the issue of diquark formation during the fragmentation process \cite{qg1,qg2, qg3, diquark}. Recent theory papers (e.g.~\cite{onium}) have suggested that the onium production mechanism (color singlet vs. color octet) can be determined by mapping the sub-leading jet structure (jet shape, hadron composition, etc.) in triggered heavy flavor jets in proton proton collisions.

Unfortunately, neither the predicted PID differences between quark and gluon jets, nor the different onium production mechanisms have been modeled on an event generator basis, therefore our simulations focus on the statistical and systematic significance of the identification of the rarest ground state light flavor particle, namely the proton. In section~\ref{pp-yields} we show the significance of measuring the proton fragmentation function in elementary collisions (based on PYTHIA). In section~\ref{sec:corr} we show the correlated proton measurements in pp collisions in comparison to PbPb collisions. The proton fragmentation function as well as the correlated production of protons and anti-protons can be measured to great accuracy and is not statistics limited in the 5-25 GeV/c range. These studies can be easily extended to pions and kaons, as shown in Figs.~\ref{fig:PID-triggered_spectra}-\ref{fig:PID-triggered_corr}.

In summary the VHMPID, in conjunction with the central barrel tracking and the calorimeter jet trigger, enables the following new physics measurements in  proton-proton collisions at the LHC:
\begin{itemize}
\item Determination of baryon fragmentation functions via protons and anti-protons in jets
\item Determination of charmonium production process via PID characteristics in subleading heavy quark jet.
\item Determination of quark vs. gluon fragmentation by measuring hadro-chemistry in tagged jets.
\end{itemize}

%%%%%%%%%%%%%%%%%%%%%%%%%%%%%%%%%%%%%%%%%%%%%%%%%%%%%%%%%%
\subsection{Specific measurement in heavy ion collisions}

In heavy ion collisions hadrons emitted from the QGP are predominantly in the transverse momentum region $p_T < 2$ GeV/c, initial studies focused on the lower momentum region to ascertain bulk properties of the medium. These studies have established a rapid build-up of strong collective flow early in the collision and the consistency of the yields of identified hadrons with results of a thermal model~\cite{Sapeta:2007ad} with a temperature $T \approx 165$ MeV, similarly to predictions of lattice QCD calculations at the time of QGP hadronization~\cite{ZFodor:2008,Karsch:2008fe}. The measurement of a strong hadron suppression up to $20$ GeV/c at RHIC, and up to 100 GeV/c at the LHC, has highlighted the importance of the high-$p_T$ regime. This regime might have unexpected particle species dependent features, though. Certain theories, based on recombination for example, have postulated strong particle dependent effects, in particular in the light flavor baryon sector (enhanced formation) out to 10 GeV/c~\cite{Hwa:2006zq, bm}, whereas theories based on enhanced gluon splitting or early formation time~\cite{Sapeta:2007ad,bellwied-markert} predict effects also in the meson sector. Thus PID out to high $p_T$ will unambiguously test certain recombination and fragmentation predictions.

An interesting extension of this program is the reconstruction of leading hadronic resonances. Resonances are likely the relevant degree of freedom near the critical temperature, $T_{c}$, to disentangle the effect of chiral symmetry restoration above $T_{c}$. Hadronic resonance gas models have shown that they agree with lattice QCD data even above $T_{c}$ all the way out to $1.5 \ T_{c}$. The survival probability and medium modification of resonant states above $T_{c}$ can be tested in particular with high-momentum resonances which have been formed early in the fragmentation process. This unique probe is only accessible through reconstruction of the particle identified decay topology of the strongly decaying resonant state, which is uniquely linked to the VHMPID for resonances where both decay daughters can be identified and reconstructed in the detector.

The main advantage of the proposed PID/Calorimeter combination is that all these flavor and baryon number dependent measurements will be based on triggered and identified jets. The analysis of unidentified back-to-back jets in heavy ion collisions via di-hadron correlations have highlighted the role of the non-Abelian jet energy loss (gluon splitting)~\cite{GLV}, however fragmentation properties cannot be extracted from unidentified hadron correlation measurements alone.

Instead, characterization of the underlying event (UE) and full jet reconstruction will provide access to the hard process and the fragmentation~\cite{Korytov:2004iw,Paic:2009,Agocs:2009,Pochybova:2009}. In that context, particle identification is essential in intra- and inter-jet correlations to determine details of the fragmentation (hadronization) process in the medium~\cite{Ellis:1996nv,sickles,Abreu:2000nw,LMolnar:2009} and differentiate it from in-vacuum fragmentation using, for example, the anomalous baryon/meson ratio observed in the intermediate transverse momentum region at RHIC and LHC energies. The relative particle production contribution, from a hadron mass dependent hydrodynamic regime to a flavor dependent recombination regime to a potentially flavor and quark mass dependent fragmentation regime, can be systematically mapped out by employing %particle identification
PID out to the highest $p_T$. %transverse momentum.

High momentum particle correlations have been used extensively at RHIC and the LHC to determine features of the jet energy loss in medium~\cite{jia2009}, the hadronization process in medium~\cite{mbv2008} and the determination of initial conditions of the collisions~\cite{roland2010,gavin2009,nagle2011}. In the past this has been done exclusively on the basis of unidentified charged particles or neutral pions and photons. The VHMPID enables particle identification in the relevant $p_T$-regime from $5-25$ GeV/c. Recent results in di-hadron correlations and jet-hadron correlations, as presented at QM2011 and also at QM2012, indicate that the complex structures in the correlation spectrum might be due to initial state density fluctuations rather than medium modification of a traversing jet, and that  the quenched jet distributes most of its lost energy in the form of low momentum particles outside the jet cone. These two rather surprising, and not yet unambiguously determined, results require a more detailed understanding by studying the particle composition in the harmonics and the jet remnant spectra. Only track-by-track reconstruction of the light quark baryons and mesons will enable us to answer these questions as well as studies of the mechanisms of hadronization in a jet and in the medium~\cite{bm}.

Finally, the unique capability of event-by-event track-by-track particle identification can be
used to determine the canonical nature of each heavy ion event and the cause of space-time fluctuations from event to event. Recent studies, based on Tsallis fits, of the spectral shape and multiplicity distributions in single LHC events have shown that the shape as well as the fluctuations event-by-event can be roughly reproduced across the measured momentum range, but require a more detailed measurement of identified particles at high momentum in order to determine the level of collectivity and the mechanism in particle production event-by-event. 

In summary the VHMPID, in conjunction with the central barrel tracking and the calorimeter jet trigger, enables the following new physics measurements in  heavy ion collisions at the LHC:
\begin{itemize}
\item Determination of cause of baryon enhancement at intermediate to high $p_T$ through measurement of hadro-chemistry in tagged jets.
\item Detailed mapping of gluon splitting process (energy loss in medium) through measurement of hadro-chemistry in tagged jets.
\item Medium modification of onium production and gluon/quark fragmentation through measurement of sub-leading identified hadron distribution in tagged jets.
\item Determination of baryon/anti-baryon imbalance through momentum dependent proton/anti-proton measurement in tagged jets in medium.
\item Determination of hadronic resonance modification in medium at high momenta.
\item Determination of canonical nature of the system and source of space time fluctuations event-by-event.
\end{itemize}

In order to address some of these issues with the available event generators we have simulated and compared identified particle production in PYTHIA and HIJING. Sections~\ref{pbpb-yields} and \ref{Ch-ratios} address the VHMPID capabilities for single particle measurements. Section~\ref{sec:corr} addresses the correlation measurements. As an example for the difference of correlated particle production in the recombination region (5-10 GeV/c) between pp and AA collisions we show the significance for identfied two-particle correlations in PYTHIA and HIJING (Fig.~\ref{fig:PID-triggered_spectra}-\ref{fig:PID-triggered_corr}). Section~\ref{ebyePID} addresses the relevance of the VHMPD for event-by-event track-by-track measurements for pions, kaons and protons. 

%%%%%%%%%%%%%%%%%%%%%%%%%%%%%%%%%%%%%%%%%%%%%%%%%%%%%
\newpage
\section{Detector Layout and Integration}
\label{sec:det-layout}

The detector is a state-of-the-art Ring Imaging CHerenkov detector (RICH), designed to meet the constraints of available space and structure inside ALICE, without compromising the prospect for new physics. We propose a fully modular design which can be integrated in ALICE in a staged approach, leading to a final coverage of about 30\% of the ALICE central barrel. Furthermore we propose to integrate all VHMPID modules with existing calorimeter modules (DCal, PHOS and DCal extension), leading in its final stage to a contiguous area at the bottom of ALICE which is populated with PID modules backed by calorimeter modules. Thus the jet triggering, which is necessary for the track-by-track jet studies, will be performed in the same phase space as the particle identification. The main measurement which will benefit from such an integrated detector system is the determination of particle identified fragmentation functions in heavy ion and proton-proton collisions, since the jet energy can be determined unambiguously and thus leads to a strict analysis of the fractional momentum. In addition, this new integrated detector will reside back to back with the existing EMCal, making
it perfect for di-jet and gamma-jet correlation measurements. Di-jets up to cone radii of $R=0.7$ will thus be fully contained in the central barrel coverage ($\eta = \pm 1.0$) of the upgraded ALICE experiment.

By adding the VHMPID to the ALICE configuration the detector extends the particle identification range beyond the capabilities of the TOF and HMPID detectors which range out at about 6 GeV/c~\cite{Alessandro:2006yt} for kaons, pions and protons. The complementary measurements with the TPC provide identification at larger momenta on a statistical basis by using the energy loss measurements in the relativistic rise region. 

The design of the VHMPID detector is based on earlier experience with the ALICE HMPID detector; however, while the technology is in-hand for the CsI-MWPC photon detector, the Cherenkov radiator system will need engineering studies to meet innovative specifications in terms of operative conditions (specifically pressurization and heating). Data collection and DAQ integration can be performed on the basis of solutions adopted for HMPID. A summary of present HMPID experience in terms of detector performance and PID achievements can be found in \cite{BarileQM11} and in section~\ref{Appendix:A}.

\subsection{Details of the RICH design}

\begin{figure}[!h]
\centering
    \includegraphics[scale=0.6]{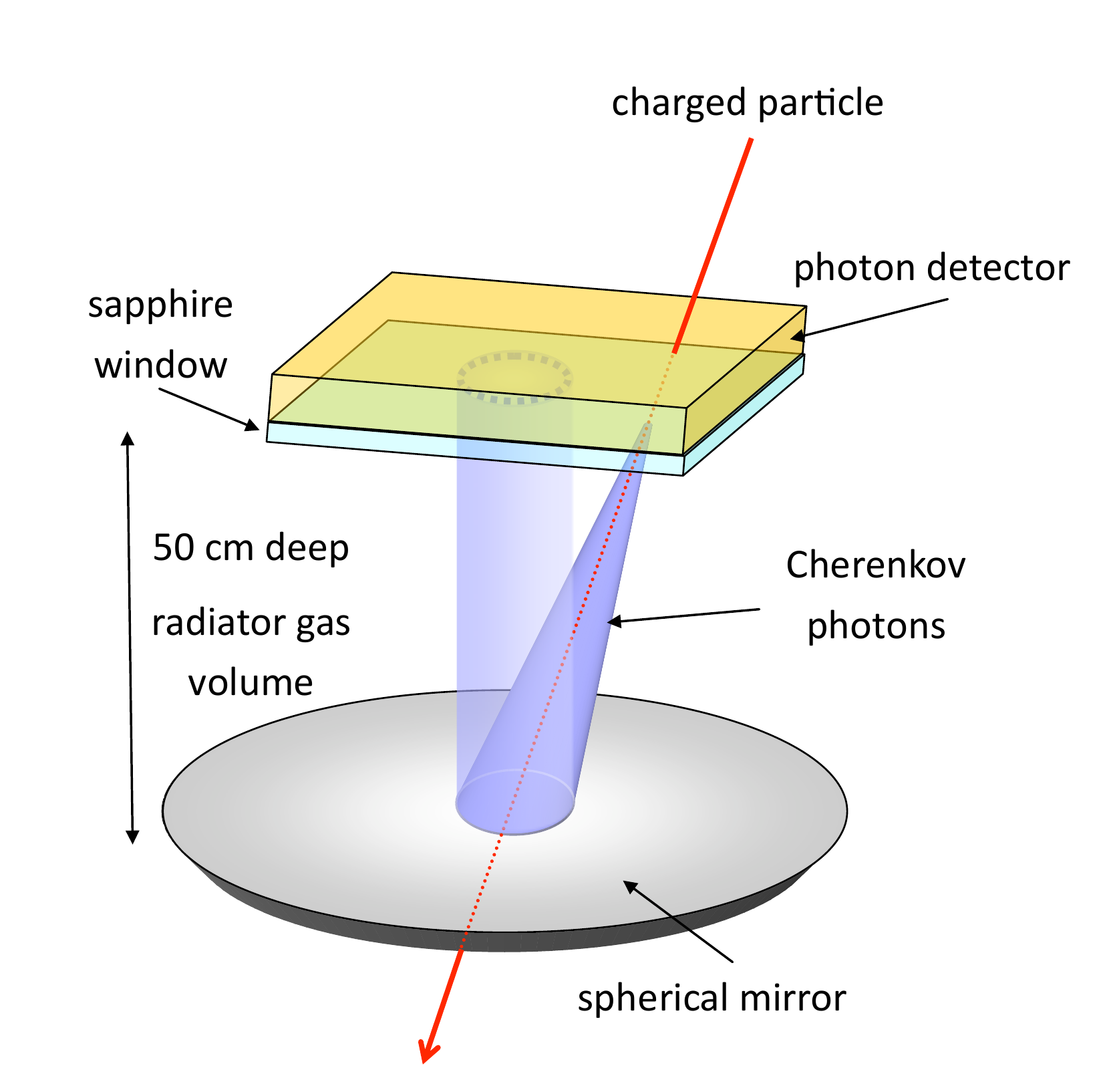}
    %\vspace{-1.0cm}
    \caption{Principle scheme of the focusing RICH configuration of the proposed VHMPID system.}
    \label{focusgeo}%
\end{figure}

The PID momentum range of interest as well as the space constraints of a detector sandwiched between the ALICE TOF and the ALICE calorimeters has driven the choice towards a focusing Ring Imaging Cherenkov (RICH) detector using a pressurized C$_4$F$_8$O gas radiator. Fig.~\ref{focusgeo} shows a schematic view of the detector layout. The Cherenkov photons emitted in the radiator are focused by a spherical mirror (of radius of curvature $R$) on the photodetector plane, located at $R/2$ from the mirror center. The refractive index of the gas can be modified by changing the pressure. By choosing the pressure up to 3.5 atm the PID range ($5-25$ GeV/c for pions, kaons, protons) can be optimized for recombination and fragmentation physics. This will require a special radiator vessel to be constructed and the usage of sapphire windows (see more in section~\ref{gas-radiator}). Furthermore, the operation at pressures larger than 2 atm will require heating of the radiator gas up to $46^{\circ}$C in order to prevent condensation.
%
%%%%%%%%%%%%%%%%%%%%%%%%%%%%%%%%%%%%%
%%%%%%%%%%%%%%%%%%%%%%%%%%%%%%%%%%%%%
%
%%%%%%%%%%%%%%%%%%%%%%%%%%%%%%%%%%%%%
\begin{figure}[!h]
\begin{center}
   \includegraphics[width=0.9\textwidth]{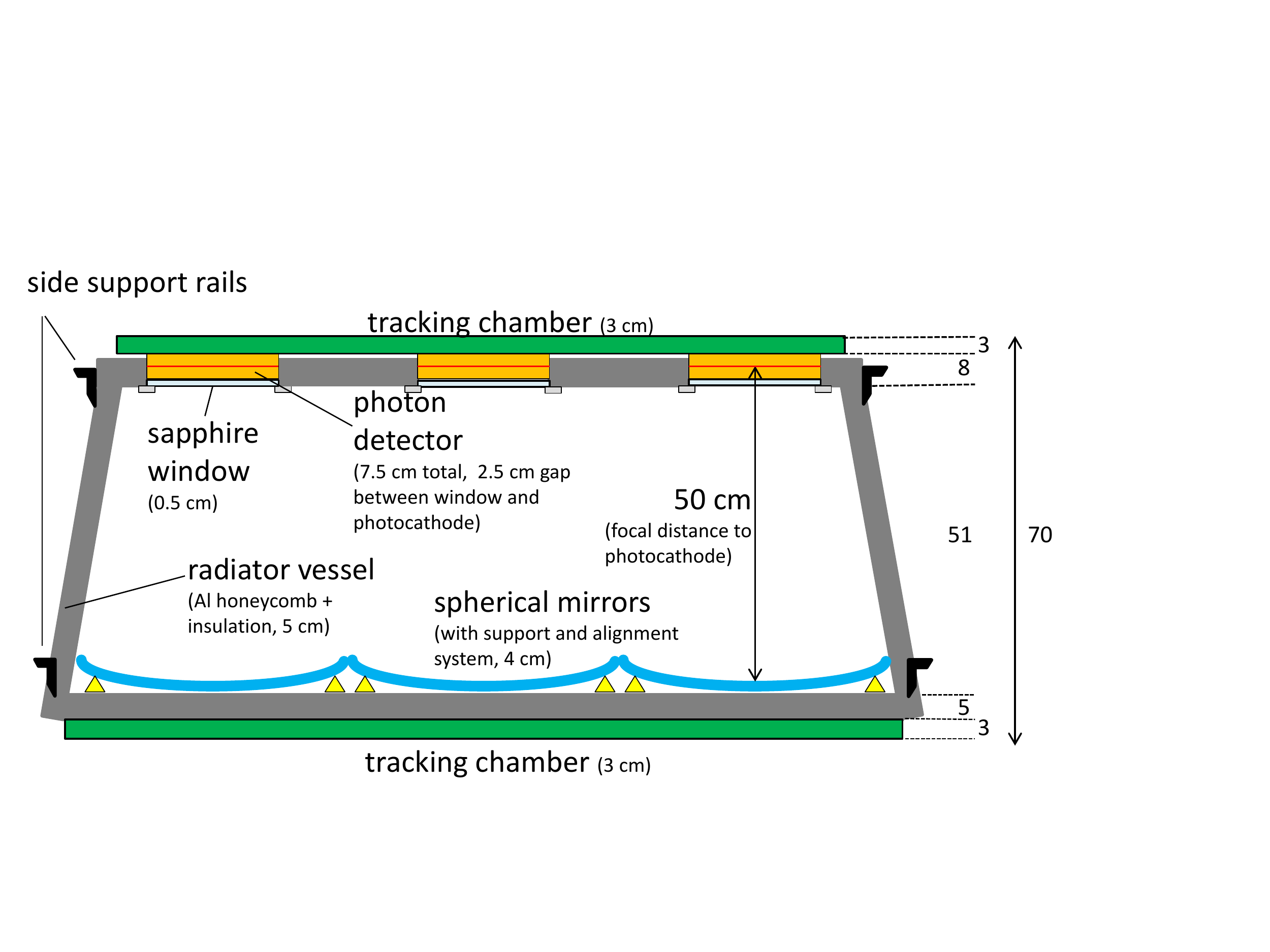}
\end{center}
   \caption{Longitudinal cross section of VHMPID central module, showing the depth or thickness of main components and the sharing among them of the 72 cm available height. }
      \label{intfig1}
\end{figure}
Two tracking layers, upstream and downstream from the RICH detector, will provide additional points to improve the track extrapolation from the TPC to the VHMPID volume (at about 4.5 m from the interaction point). 

The radial space available between TOF and the recessed calorimeter modules is very limited.
A new support structure will be developed which holds the VHMPID and the calorimeter modules in place. Fig.~\ref{intfig1} shows the cross-section in the $\eta$ plane of the central module. Overall 72 cm in radial depth are foreseen for the VHMPID. The pressurized gas vessel will occupy a total of 64 cm: 9 cm at the bottom for the vessel composite panel (including insulation) and mirror system, 47 cm for the actual radiator gas depth from the mirror up to the sapphire window, 8 cm for the vessel top panel and photon detector; plus two tracking layers (one upstream and one downstream) will fill the remaining 6 to 8 cm. The radiation length of such a device can be limited to $22 \%$ which is comparable to the radiation length of the existing TOF and TRD detectors. Studies have shown that effect on lepton measurements in the calorimeter is negligible and that the additional low-momentum background in the photon measurement is manageable.

\begin{figure}[!h]
\centering
   \includegraphics[width=13cm]{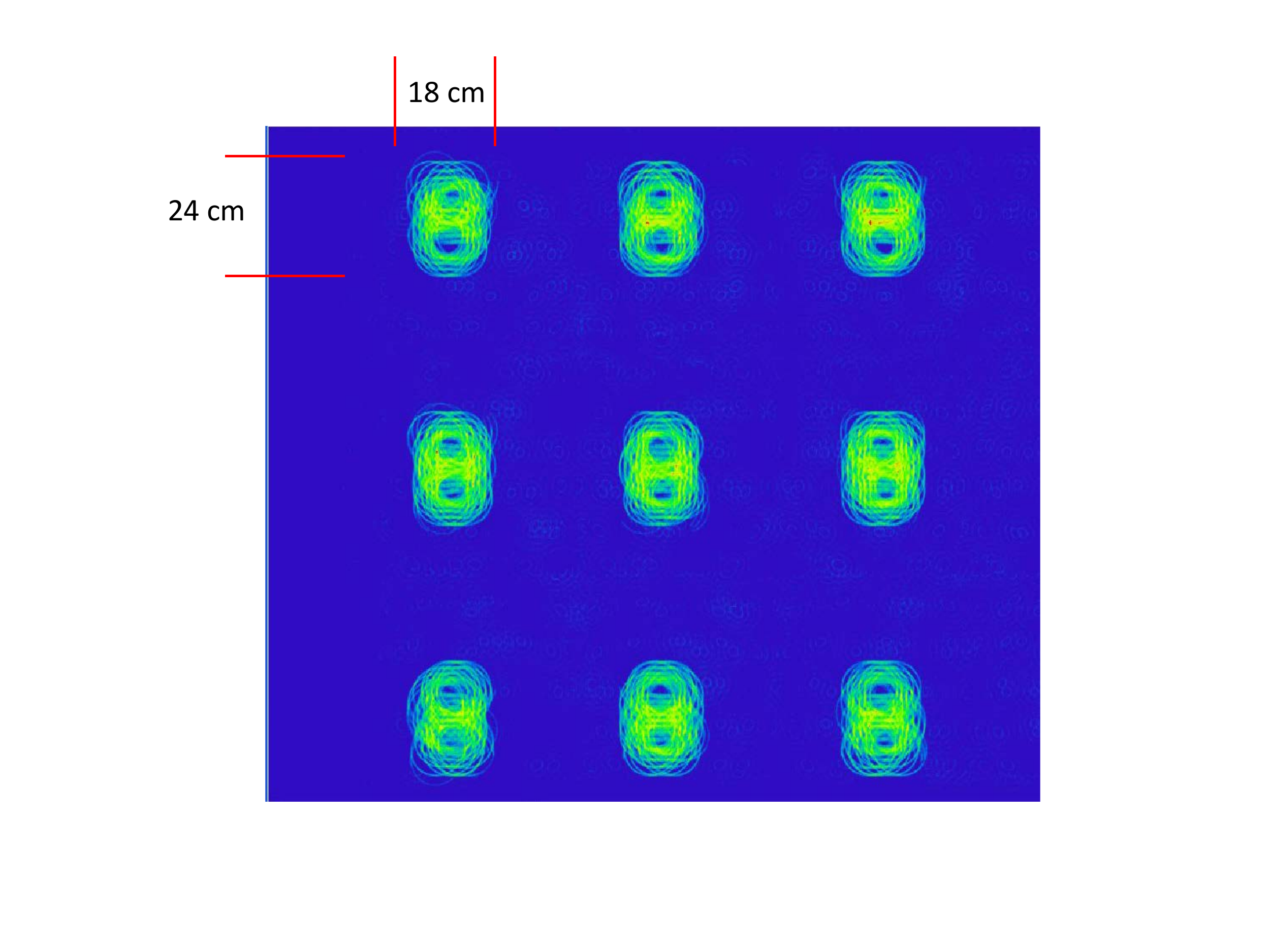}
%\vspace*{-1.0truecm}
   \caption{View of Cherenkov rings fiducial area corresponding to an array of $3\times 3$ mirrors spherical mirrors of $50 \times 50$ cm\textsuperscript{2} and 100 cm curvature radius, obtained using an optical design program (Zemax), for 5 GeV/c charged pions and a magnetic field of 0.5 T.}
      \label{zemax1}
\end{figure}

More details about the module arrangement and integration can be found in Sect.~\ref{VHMPID_integration}.  According to the present layout, each module will be equipped with an array or mirrors of about $50 \times 50$ cm$^2$, each focusing Cherenkov photons on a corresponding photon detector of $18 \times 24$ cm$^2$ area (Fig.~\ref{zemax1}).  
The full coverage of five sectors, each measuring $\sim 7.5 \times 1.5$ m\textsuperscript{2}, will then require 225 mirrors and photon detectors of the above mentioned size, for an overall photosensitive area of $\sim 10 $ m$^2$. An alternative design with larger mirrors is under study to reduce the amount of photon detectors; for example, in case of $0.75 \times 0.75$ m\textsuperscript{2} mirrors, 100 photon detectors of $24 \times 24$ cm$^2$ would be needed (total photosensitive area of $\sim  5.8$ m$^2$).

%%%%%%%%%%%%%%%%%%%%%%%%%%%%%%%%%%%%%
%%%%%%%%%%%%%%%%%%%%%%%%%%%%%%%%%%%%%

Such a large total photosensitive area and its operation inside the magnetic field of $0.5$ T of the ALICE solenoid led to opt for a CsI gaseous photon detector based on the experience gained within the ALICE HMPID RICH project~\cite{HMPID-TDR,Antonello:2010}. This detector will operate in the UV, although a more costly version based on commercial photon detectors which would operate in the visible wavelength range and thus forgo the CsI readout, is still part of the ongoing R\&D.

In the following sections we review the main elements of the detector: the Cherenkov radiator, the photon detector and the front-end and readout electronics.

%%%%%%%%%%%%%%%%%%%%%%%%%%%%%%%%%%
%%%%%%%%%%%%%%%%%%%%%%%%%%%%%%%%%%%%%%%%%%%%%%%%%%%%%%%%%
\subsubsection{Gaseous radiator and mirror system}
\label{gas-radiator}

\begin{figure}[!h]
  \centering
  \includegraphics[width=0.8\textwidth]{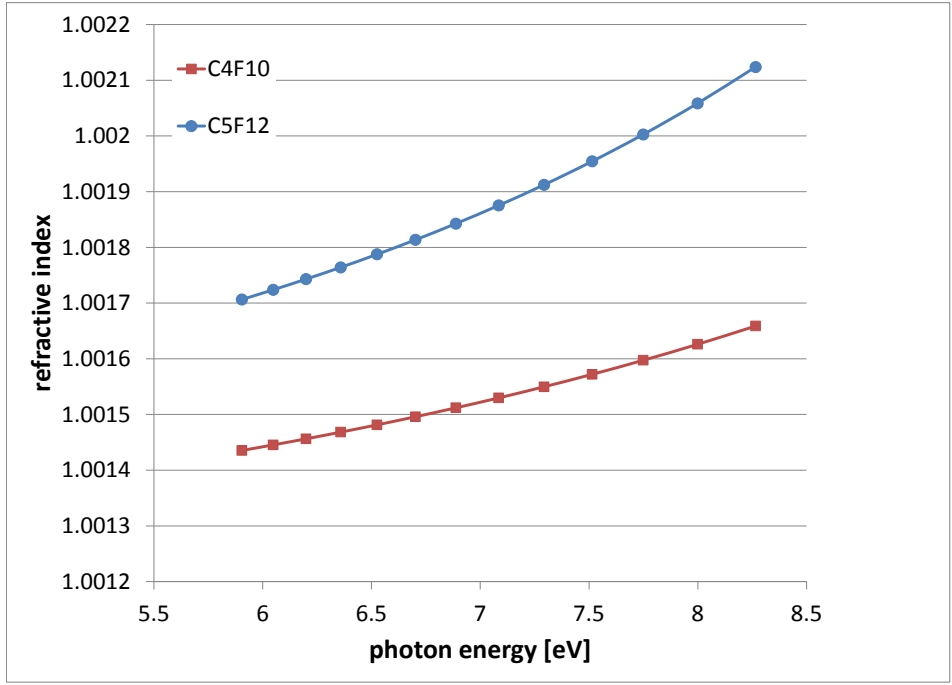}
  \caption{Refractive index of C$_4$F$_{10}$ (C$_4$F$_8$O) and C$_5$F$_{12}$ as a function of photon energy at 1 atm.}
  \label{refindices}
\end{figure}

The radiator refractive index $n$ is the most important parameter since it establishes the threshold for Cherenkov emission $p_{th}$, the angle $\theta_c$ and the amount of produced photons $N_{ph}$ via the well known relations:
$$ \cos \theta_c = \frac{1}{n\beta} \, ; \ \ p_{th} = \frac{m}{\sqrt{n^2-1}} \, ; \ \ N_{ph}= 370 \cdot L \cdot Z^2  \sin^2 \theta_c , $$ 
where $m$ and $Z$ are the particle mass and charge, respectively. Perfluorocarbon gases C$_n$F$_{2n+2}$ are characterized by refractive indices suitable for particle identification above $5$ GeV/c. In particular C$_4$F$_{10}$ ($\langle n \rangle ~\approx 1.0015$ at 175 nm, $\gamma_{th}~\approx18.9$)~\cite{olav} C$_4$F$_8$O and C$_5$F$_{12}$ are the only possible candidates for the momentum range defined by the physics motivations. From the limited data available in literature on C$_4$F$_8$O in the UV range (see for example ~\cite{ando, artuso}), a refractive index similar at per mil level to C$_4$F$_{10}$ can be deduced. Given its larger availability on the market and lower cost it can be considered as a valid alternative to C$_4$F$_{10}$. Furthermore, recent test beam studies have demonstrated good performance with this gas as discussed in section~\ref{testbeam}.
Fig.~\ref{refindices} shows the refractive index of C$_4$F$_{10}$ (C$_4$F$_8$O) and C$_5$F$_{12}$ at atmospheric pressure as a function of the photon energy. 
%%%%%%%%%%%%%%%%%%%%%%%%%%%%%%%%%%%%%

A pressurized radiator vessel, presently under study as baseline option, will allow achieving a "tunable" refractive index and a PID in a large momentum range from 5 up to 40 GeV/c. 
Table ~\ref{tabrefind} shows the C$_4$F$_{10}$ refractive index at 175 nm and Cherenkov emission thresholds for different pressure values: the Cherenkov emission threshold for K at 5 GeV/c corresponds to 3.5 atm radiator gas pressure. 

\begin{table}[h!]
\centering
  \begin{tabular}{clccc}
\hline
{\small Radiator }& \small{Refr. ind.} & \small{ $ \pi $ thresh. }& \small{ $K$ thresh.} & \small{$p$ thresh. } \\
{\small pressure [atm] } & \small{at 175\,nm }&  \small{ [GeV/c] }& \small{ [GeV/c] } & \small{[GeV/c]}  \\ \hline \hline
1.0 & 1.00153 & 2.5 & 9 & 17  \\ 
1.3 & 1.00199 & 2.2 & 7.9 & 15  \\ 
1.5 & 1.002295 & 2.1 & 7.3 & 13.5  \\ 
2.0 & 1.00306 & 1.8 & 6.4 & 12  \\ 
2.3 & 1.00352 & 1.7 & 5.9 & 11.2  \\  
2.5 & 1.00383 & 1.6 & 5.6 & 10.7  \\ 
3.0 & 1.0046 & 1.5 & 5.1 & 9.8  \\ 
3.5 & 1.00535 & 1.3 & 4.8 & 9.1 \\ \hline \hline
\end{tabular}
\caption{Variation with C$_4$F$_{10}$ gas radiator pressure of refractive index and momentum thresholds for Cherenkov emission.}
\label{tabrefind}
\end{table}

\begin{figure}[!h]
  \centering
  \includegraphics[width=0.8\textwidth]{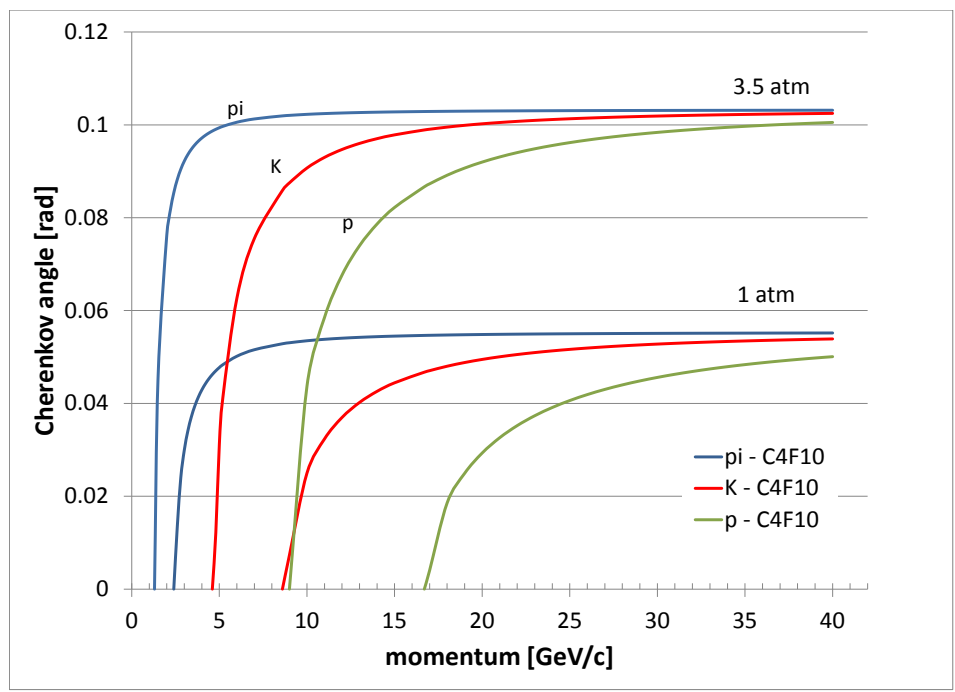}  
  \caption{Cherenkov angle as a function of momentum for $\pi$, $K$, and $p$ in C$_4$F$_{10}$ at 1 and 3.5 atm.}
  \label{thetacer}
\end{figure}

The theoretical Cherenkov angles (at atmospheric pressure and at 3.5 atm) as a function of the momentum, for pions, kaons and protons, are reported in Fig.~\ref{thetacer}.

%%%%%%%%%%%%%%%%%%%%%%%%%%%%%%%%%%%%%

\begin{figure}[!h]
  \centering
  \includegraphics[width=0.8\textwidth]{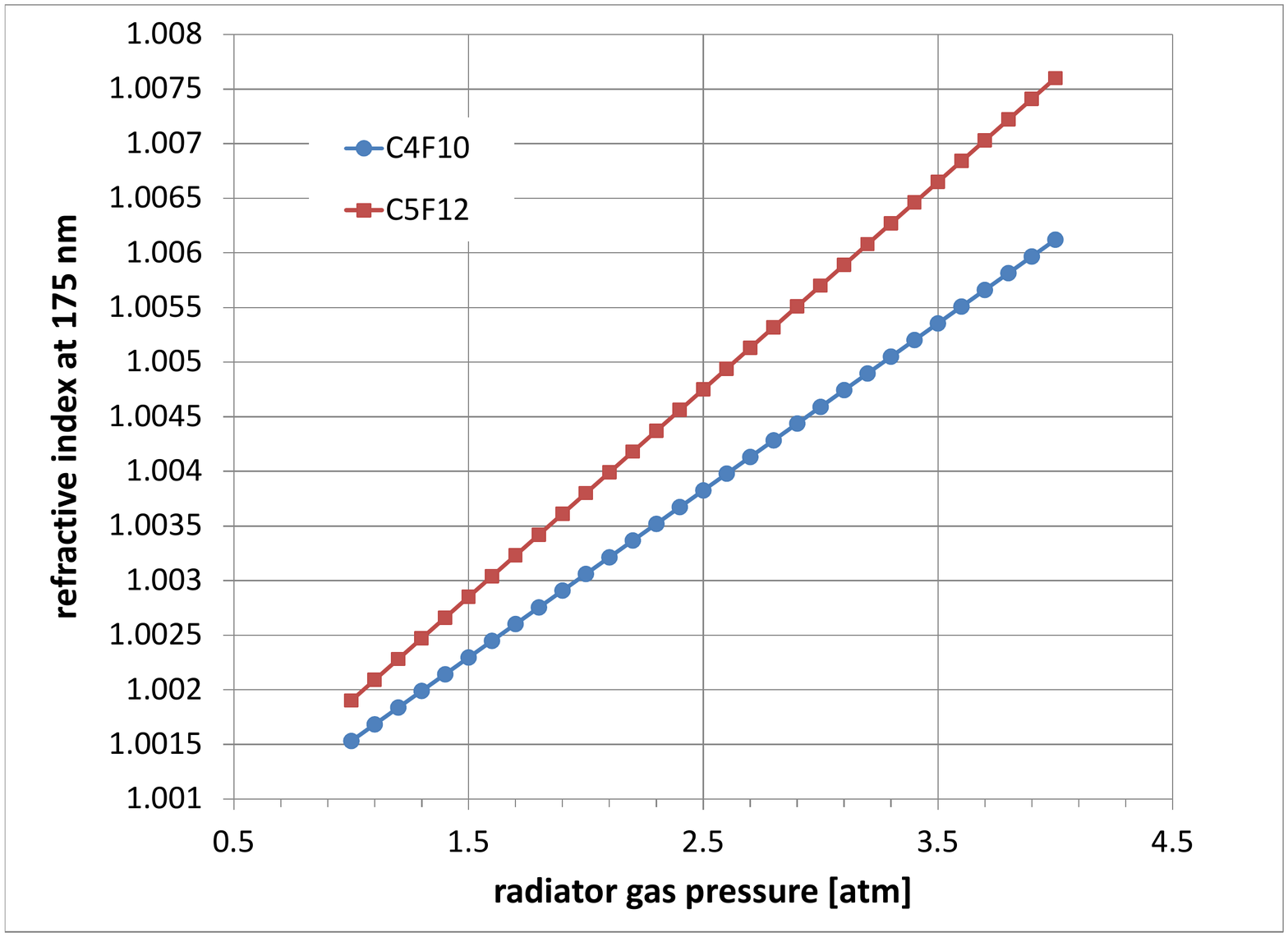}  
  \caption{Refractive index at 175 nm as a function of pressure for C$_4$F$_{10}$ and C$_5$F$_{12}$.}
  \label{rindcomp}
\end{figure}

C$_5$F$_{12}$ is liquid at ambient temperature (29 $^{o}$C boiling point) however its larger refractive index ($\approx 1.0019$  at 175 nm) makes it an attractive alternative as gaseous radiator, since the required refractive index can be obtained at lower pressure ($\approx$ 2.7 atm) compared to C$_4$F$_8$O (Fig.~\ref{rindcomp}).

\begin{figure}[!h]
    \centering
    \includegraphics[width=0.8\textwidth]{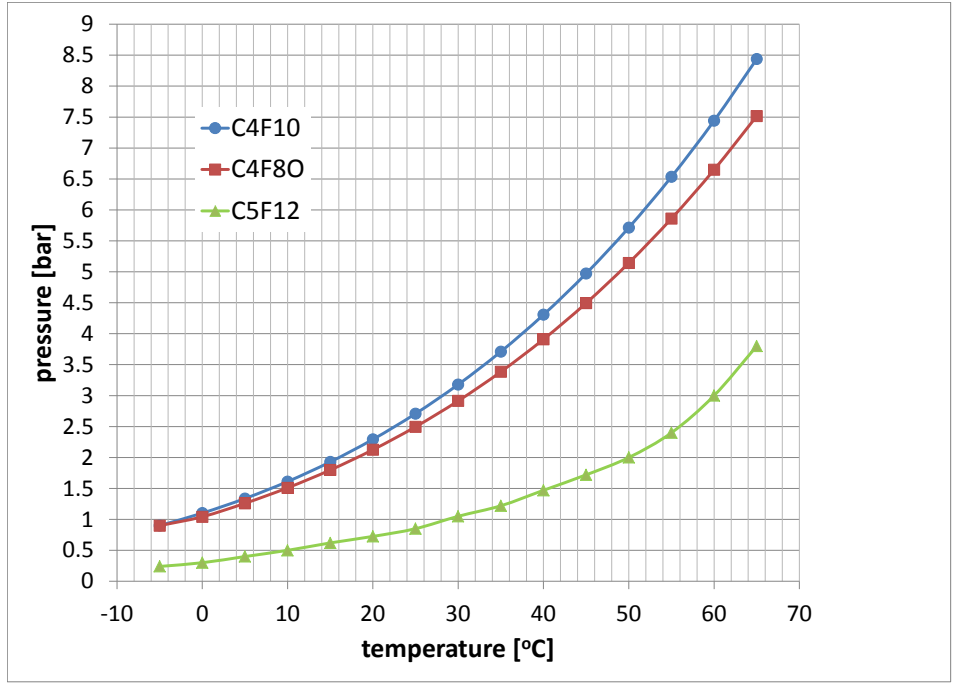}
    \vspace*{-1.0truecm}
    \caption{Phase diagram curves for some fluorocarbon compounds.}
    \label{PT_freon}
\end{figure}

Usage of sapphire window as interface to the photon detector and special reinforcement structure for the vessel will be needed. As already mentioned, this will have a clear impact on the layout (mirror and photo-detector segmentation). 
In addition, the operation of C$_4$F$_8$O at pressures up 3.5 atm (2.7 atm for C$_5$F$_{12}$) will require heating of the radiator gas at 35-40 $^o$C (55-60 $^o$C for C$_5$F$_{12}$) to prevent condensation. 
Fig.~\ref{PT_freon} shows the $P-T$ diagram of the mentioned fluorocarbons.   

%%%%%%%%%%%%%%%%%%%%%%%%%%%%%%%%%%%%%
%%%%%%%%%%%%%%%%%%%%%%%%%%%%%%%%%%%%%

Preliminary engineering studies have confirmed the feasibility of a pressurized RICH vessel and different solutions are available to implement a heating system. Both radiator vessel and gas pipes will need to be heated and insulated. An additional advantage of using composite panels for the radiator vessel is that the insulation function could be implemented in the same structure (for example, Rohacell thermal conductivity is 0.029 W/m/K). 

Concerning the mirror construction, two options are considered: a classical glass substrate and a light\-weight carbon-fiber substrate, a material which minimizes the material for traversing particles and is fluorocarbon-compatible, thus not degraded by C$_4$F$_{8}$O~\cite{Barber}. In both cases, high reflectivity up to VUV (Vacuum Ultra Violet) is achieved by Al/MgF$_2$ coating~\cite{Baillon}. A mirror alignment system will be designed and integrated in the radiator vessel.
The mirror segmentation and orientation are under study with the aim to minimize the photosensitive area while keeping identification efficiency for close tracks; preliminary results can be found in  ~\cite{volpe-rich2010}.

%%%%%%%%%%%%%%%%%%%%%%%%%%%%%%%%%%%%
\subsubsection{The photon detector}
\label{Photon_detector}

The baseline solution for the photon detector is a Multi-Wire Proportional Chamber (MWPC) equipped with a CsI photo-cathode, consisting of pad-segmented cathode coated with a $300$ nm thick CsI layer. The chamber has the same structure and characteristics as in the HMPID~\cite{cozza}. A sapphire window, of thickness from 3 to 5 mm depending on size and segmentation, will be used to separate the photon detector, operated with CH$_4$ at atmospheric pressure from the pressurized radiator gas volume. The pad size is $0.8 \times 0.84$ cm$^2$, the wire pitch is 4.2 mm and the anode-cathode gap is 2 mm. This pad segmentation was optimized for the HMPID detector where the Cherenkov ring radius at saturation is 12 cm. Therefore, a larger photon overlap probability, resulting in a reduction of the number of detected photons,  will correspond to the ring radius of only $5.5$ cm produced by C$_4$F$_{10}$ gaseous radiator. CsI photo-cathodes with smaller pads ($0.4 \times 0.8$ cm$^2$) are presently under test using a MWPC prototype with a smaller anode-cathode gap ($\sim 1$~mm) with the aim of reducing the photon overlap probability and increase the spatial resolution and the PID performance.     

As alternative options, a photon detector based on triple Thick-GEM with reflective CsI photo-converter, a combined CsI-coated TGEM + MWPC detector (called TCPD) or a commercial Micro-Channel Plate photon detector operating in the visible, are under evaluation.
   
%%%%%%%%%%%%%%%%%%%%%%%%%%%%%%%%%%%%%%%%%%%%%%%%%%
\subsubsection{Front-end and readout electronics}

The Front-End Electronics (FEE) is based on the Gassiplex chip in 0.7 $\mu$m technology~\cite{HMPID-TDR,Santiard} used in the HMPID detector,
which has 16 input channels and one output channel, a peaking time 1.1 $\mu$s and achieves $1,000$~$e^-$ noise on detector.
The single electron average pulse height at 2050 V is 35 ADC channels corresponding to about $40,000$~$e^-$,
hence a single electron detection efficiency of 90\% with 4s zero suppression.
The HMPID Gassiplex FEE is capable of reading out an interaction
rate up to 200 KHz a limit set by the analogue multiplexed readout
which requires about 5 $\mu$s for each block of 3 chips (48 channels).
The digital readout chain, presently in use in
the HMPID, is based on the DILOGIC chip, implementing zero-suppression and pedestal substraction,
which achieves a total readout time of 170 $\mu$s in central Pb-Pb events with a parallelized
readout via single DDL of 24 blocks of 480 channels.  We plan to optimize the performance
and readout rate in this section by designing a new FPGA based alternative.
 
A modified version of the GASSIPLEX has been produced for the COMPASS RICH detector
\cite{Abbon2007},  implementing a shaping time reduced to
500 ns and a second output channel; the COMPASS GASSIPLEX is characterized by a 500 KHz rate capability limited by the 
baseline recovery to below 1\% within 2 $\mu$s.
 
Another option is represented by the solution adopted for the COMPASS RICH-1 and the HADES RICH upgrade, namely the APV25 chip which, although widely used for silicon and GEM detectors readout, can be adapted to slower MWPC signals by changing an external bias current \cite{Abbon2008}. The APV-25 performs continuous sampling of the input signal at 40 MHz and the readout time for a typical MWPC signal can be reduced to 400 ns needed to sample the baseline, the rising edge and the peak. This solution would allow to access the MHz range triggering rate.
 
The most obvious solution could be offered by the ALICE common electronics development for the TPC and Muon-arm high luminosity upgrade, since any solution could be easily adapted to the VHMPID photon-detector readout.
 
%The 2 MHz pp collision rate poses a challenge in terms of pile-up in the MWPC; indeed even for a thin anode-cathode
%gap of 0.5 mm the total ion drift time would be 1.2 $\mu$s and pile-up free events would require a trigger rate below 800 KHz. However a reduction of trigger rate could be adopted in the case of modified GASSIPLEX based FEE, while a continuous sampling electronics would allow proper treatment of overlapped events.  
 
In conclusion several solutions are available to meet the trigger rate specifications of the high luminosity upgrade
and the FEE of the VHMPID will be capable of handling both the 50 kHz and 2 MHz interaction rates envisioned in Pb-Pb and pp, respectively.

%%%%%%%%%%%%%%%%%%%%%%%%%%%%%%%%%%%%%%%%%%%%%%%%
\subsection{MIP detector}

The ring pattern recognition critically depends on the known ring center, that is, the known position of the particle from which the Cherenkov light originates. This particle is usually called the MIP signal.

The situation for the ring imaging outline in the focusing geometry of the VHMPID detector is slightly less stringent than in the HMPID proximity focusing case of the HMPID (see section~\ref{sec:ChallengesTrackingAndReconstruction}) . The ring center depends to first order only on the direction of the incident particle, but not so much on its impact point. However, the MIP detection is still an important issue to ensure that the particle observed in the ALICE TPC actually did cross the VHMPID volume. MIP detection does not need to be very precise (order of a few mm is sufficient) but needs to ensure a positive signal on impact. This is of particular interest in the momentum region, where protons are identified by the absence of Cherenkov photons. Two independent pad layers, in front and behind the radiator volume, can be used for MIP detection. 

The proposed devices are based on the newly designed and tested "Close Cathode Chambers"~\cite{CCC_Paper,CCC_Paper2}. These special thin, multi-wire gaseous chambers are similar to classic MWPCs however they have narrow pad response functions. All chambers use the gas mixture of $80 \%$ Ar and $20 \%$ CO$_2$. For cost efficiency, the detectors can have relatively long cathode segments ($4$ mm wide and $10-30$ cm long) and this way use projective geometry: the cathode segments and the chamber wires can be read out independently, which gives a two-dimensional position information for a single layer. For the projective geometry, ambiguities in multiple hits pose a problem, which can be resolved to some extent by projecting the TPC tracks to the MIP detector surface, and in addition to use the direction information from the  measured layer. The actual geometry needs to be optimized ensuring sufficient suppression of the ambiguities.

Due to the minimalization of the space (to leave more for the radiator length), electronics are mounted onto the backside of the pads. The estimated power consumption of the electronics is $3$ mW/channel, meaning $7.5$ W (2,500 pads)/m$^2$. This heat production only requires simple air cooling.

%%%%%%%%%%%%%%%%%%%%%%%%%%%%%%%%%%%%%%%%%%
\subsection{Anticipated occupancy}

\begin{figure}[!h]
\centering
\includegraphics[width=0.8\textwidth]{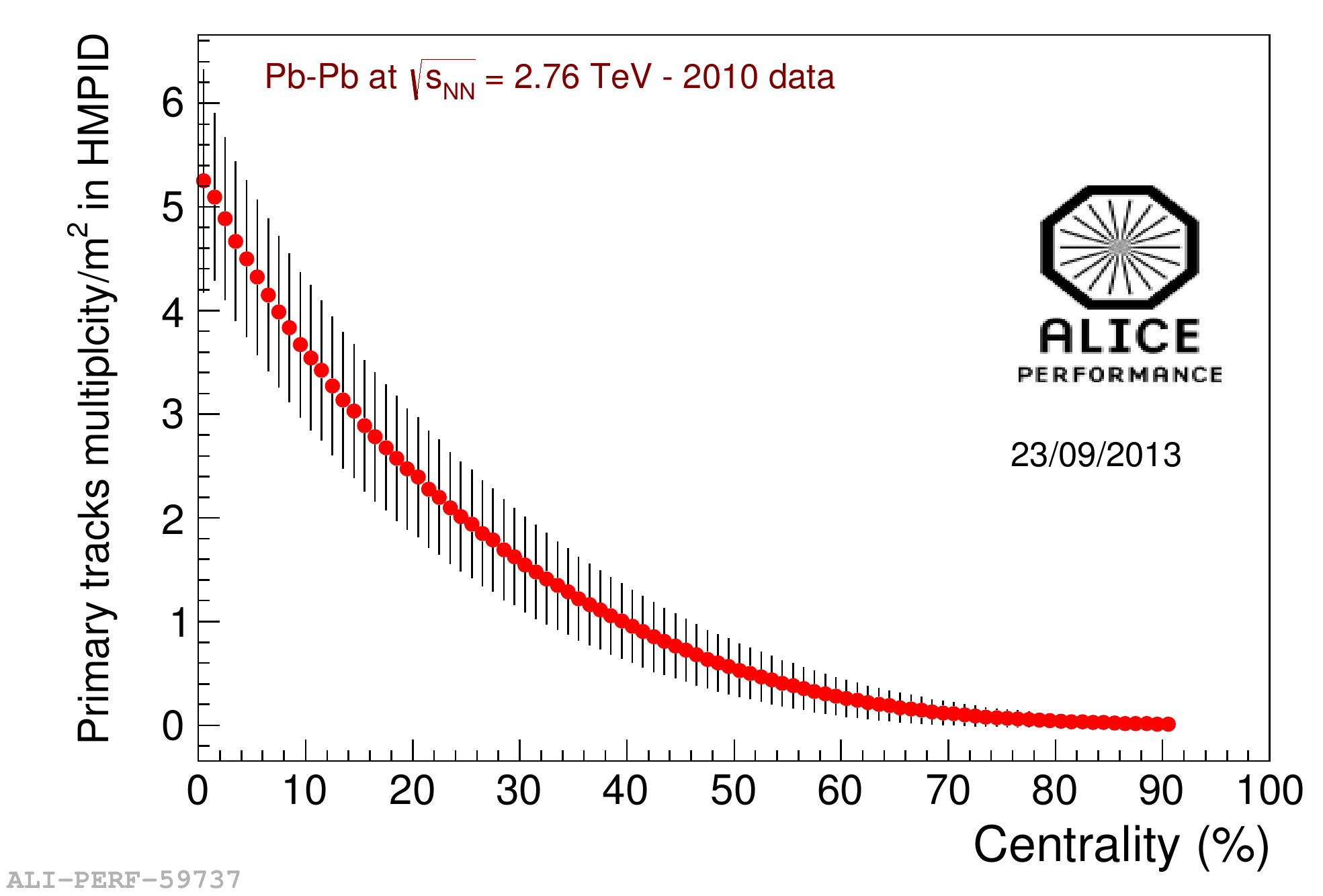}
\caption{Track multiplicity vs. VZERO centrality in the HMPID detector with its 10 m$^2$ acceptance.}
\label{hmpid-multiplicity}
\end{figure}

In the 2010 heavy ion run in ALICE, a primary track surface density of $\sim5/m^{2}$ has been observed in the HMPID acceptance for $0-10\%$ central Pb--Pb collisions (Fig.~\ref{hmpid-multiplicity}). Therefore one can deduce a multiplicity  of around 5 charged particles/m$^2$ for Pb--Pb central collisions at $5.5 \, A$TeV in the proposed VHMPID region. Only $10\%$ of these will be above the threshold for producing Cherenkov light.
In case of $ 50 \times 50$~cm\textsuperscript{2} mirror, the Cherenkov photons produced by all particles crossings will be focused on a photon detector of $24 \times 18$~cm\textsuperscript{2} segmented in $60 \times 22$ pads. Since on average a particle will hit 5 pads and all detected Cherenkov photons in a ring correspond to 25 pads an occupancy of $ < 1\% $ is expected in central Pb-Pb collisions. 
% Did not understand this next sentence, so removed it (MW)!
% Ring overlap could be a more significant issue in a jet cone, but one has to take into account that the detector is about the size of a jet cone, so also in the case of a jet reconstruction the overlap should be minimal.

%%%%%%%%%%%%%%%%%%%%%%%%%%%%%%%%%%%%%%%%%%%%%%%%%%%%
\subsection{Trigger options for the VHMPID}
\label{Trigger_opportunities}

We are anticipating that after the upgrade of all central barrel detectors to continuous
readout the trigger rates discussed in this section have to be revised. Still, in case the
continuous readout cannot be realized for all central subsystem, we are describing here a VHMPID triggering scheme that is based on the present performance of existing ALICE detector components. 
Since the hadron yield drops rapidly in the momentum range of interest, a specialized scheme
should be implemented to fully exploit the VHMPID detector. This trigger needs to be coincident with an EMCal jet trigger for the case of back to back di-jet or hadron-jet correlations to study quenching effects or possible changes in hadro-chemistry due to medium modification of the fragmentation process. Any proposed scheme requires fast triggering which must be at level-0 (L0, for proton-proton collisions within 1.3~$\mu$s) or level-1 (L1, for lead-lead collisions, within 7~$\mu$s). For the VHMPID we can adopt two distinct strategies:
\begin{description}
\item
(i) One is to rely on a high momentum particle trigger, based on the Transition Radiation Detector (TRD)
\item
(ii) the other is to rely on the DCal jet trigger positioned right behind the VHMPID modules
\end{description}

The default option for triggering on high-$p_T$ particles in the VHMPID is based on the trigger capabilities of the Transition Radiation Detector (TRD) which is located directly in front of the VHMPID. The TRD has 6 tracking layers within a depth of $75$ cm. The detector has $400-600$ $\mu$m spatial resolution.

The TRD trigger will be provide a level-1 decision (7~$\mu$s after the interaction) which will be constructed from TRD-only information. The TRD-only trigger is based on the matching of at least 4 TRD track segments (tracklets) in a single TRD stack. These tracklets are combined to a single track via a straight line fit. The L0-trigger rate, thus the input sample to TRD, can in principle be up to 100 kHz. However, then the dead time is almost saturated (7 $\mu$ s between L0 and L1) such that the L0 triggers would sample only a small fraction of the interactions. The specific trigger mix will be determined by the physics program of any particular run year, which will decide whether it is more beneficial to run at high L0 rate and sample more events at the L1 stage, or at smaller dead time and sample more events at L0.

\begin{figure}[!h]
    \centering
    \includegraphics[width=0.8\textwidth]{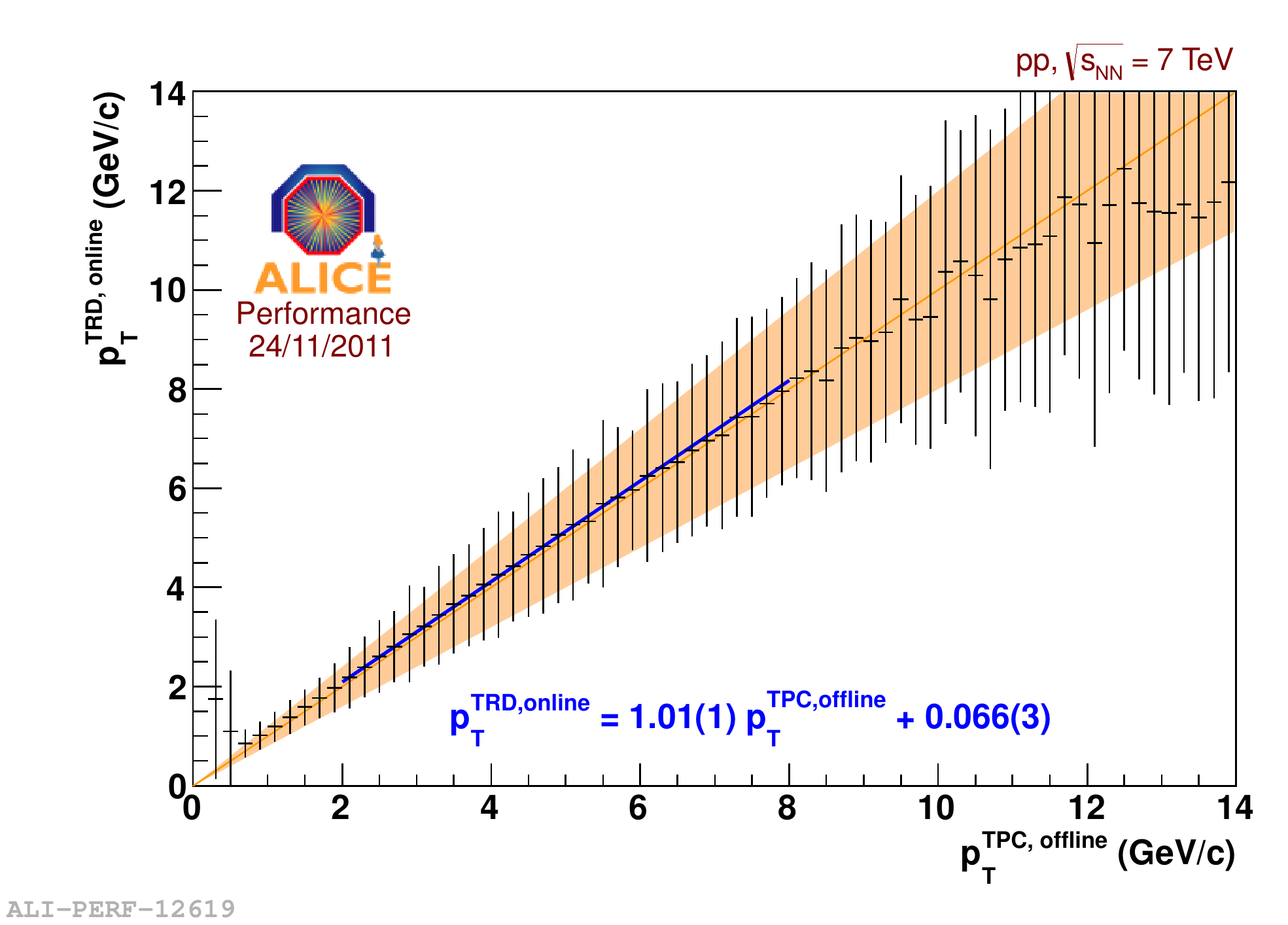}
    \caption{Correlations between online and offline $p_T$ tracks within different momentum ranges.}
    \label{PtCorrelationsVsTracks}%
\end{figure}

The anticipated L1 input rate in central collisions is roughly 1 kHz, which covers all central Pb-Pb collisions at the nominal luminosity. For the near future the plan is to run at about $10 \%$ dead time resulting in a recording rate, i.e. L2a, of about 100 Hz (for the rare trigger time share).
The correlation between the transverse momentum reconstructed online in the TRD and the offline momentum is very good as shown in Fig.~\ref{PtCorrelationsVsTracks}.

%%%%%%%%%%%%%%%%%%%%%%%%%%%%%%%%%%%%%
%%%%%%%%%%%%%%%%%%%%%%%%%%%%%%%%%%%%

Since the VHMPID is only relevant to tracks with momentum above 5 GeV/c, an efficient trigger will require a single track threshold somewhere between 5 GeV/c  and 10 GeV/c and will be chosen during actual data taking, based on the measured high-$p_T$ particle multiplicity.  The tracking efficiency of the TRD levels off at about $p_T \approx 4$  GeV/c at close to $90 \% $, thus any high $p_T$ track in the VHMPID should be reconstructed with maximum efficiency~\cite{Klein:2011}.
For the rates in the physics performance section we assumed a $ 100\% $ efficiency of the TRD trigger for any particle $p_T$ threshold above 5 GeV/c. We anticipate a maximum VHMPID/TRD trigger rate of about 40 Hz with a 5 GeV/c threshold and about 5 Hz with a 9 GeV/c threshold.

The efficiency and rates for the jet trigger using the DCal is described in the DCal Technical Design Report. In general, for jet-jet correlations the data selection will be done through the L0 calorimeter trigger.

%%%%%%%%%%%%%%%%%%%%%%%%%%%%%%%%%%%%%%%%%%%%%%%%%%%%%%%%%%%%%
\subsection{Proposed and ongoing R\&D activities}

Although the detector described in this document is conceived as the baseline VHMPID option, several optimizations are being investigated  as part of the ongoing and proposed R\&D activities, which will be  completed by late 2013:

\begin{itemize}
\item The final fluorocarbon gas for the Cherenkov radiator will depend on photon yield and ease of usage. Presently C$_{4}$F$_{8}$O and C$_{5}$F$_{12}$ are being tested.
\item The pressure range and heating requirement for the radiator vessel 
has to be optimized and tested for different gases, which will define the final vessel design.
\item Photon detection in the visible range rather than the UV range, using commercial vacuum-based devices (specifically large area Micro-Channel-Plate with Bialkali photo-cathode) 
is being investigated as an option to simplify the photon detector system and enhance photon yield.
\item The optical window material will depend on the pressure in the  radiator vessel and the anticipated wavelength range of the photon detector.
\item In the case of photon detection in the UV, GEM detectors (CsI-TGEM or TCPD option) instead of the baseline MWPC option have been proposed for the photon detector.
\item The effect of the photo-cathode pad size (granularity) on spatial resolution is under investigation.
\item The mirror substrate technology (Glass vs composite Carbon-fiber) and the mirror segmentation in the vessel
is under investigation.
\item The FEE/DAQ/TRG implementation of the VHMPID can be modified in accordance with anticipated DAQ/TRG upgrades to the existing ALICE central barrel detectors. A combined development of a new frontend for all gas detectors seems possible and is presently under investigation. 
\end{itemize}

%%%%%%%%%%%%%%%%%%%%%%%%%%%%%%%%%%%%%%%%%%%%%%%%%%%%%%%%%%%%%%%%%%%%%%%%%%%%%%%%%%%%%%

\subsection{The VHMPID integration in ALICE}
\label{VHMPID_integration}

The limited space available inside the ALICE solenoid has a significant impact on the VHMPID detector design and integration. A super-module layout with pressurized radiator vessels, installed in existing space-frame sectors (11-15), has been adopted in order to exploit all the available space and maximize the acceptance. The most recent advances in designing a thin VHMPID enable us to propose a device which will fit in front of the existing ALICE calorimeter modules (EMCal, DCal and PHOS) after some modification to their support structure. For practical purposes we propose here a combined and staged PID/calorimeter project which will cover up to 30\% of the TPC acceptance. We suggest to implement this detector in three stages, two of them in time for the long shutdown at the end of this decade (LS2) plus a final stage in time for LS3. Stage 1 will simply add VHMPID modules in front of the existing DCal structure (around 15\% of the TPC coverage). Stage 2 will populate the presently vacant area adjacent to the DCal and PHOS with new VHMPID/Cal modules (leading to around 25\% of the TPC coverage). The calorimeter part of stage 2 has been proposed by the existing calorimeter groups (the so-called FullCalLite option) and is part of an ALICE upgrade proposal to the U.S. Department of Energy. Finally, stage 3 will populate the area in front of the existing PHOS detector with VHMPID modules. Each stage enables ALICE to perform a specific physics program which is detailed in the upcoming sections. 
Stage 1 and 2 combined will, for the first time in relativistic heavy ion collisions, allow full jet reconstruction with particle
identification back to back with the existing EMCal. 

The integration in front of an electromagnetic calorimeter will require minimization of the material budget in order to prevent a negative impact to the calorimeter performance. According to preliminary engineering studies, the radiator vessel should consist of composite panels, obtained by sandwiching 1.5 mm thick Al foils with Rohacell foam or Al honeycomb panels of $\sim 3$ cm thickness providing the required stiffness to stand the absolute working pressure of 3.5 atm, adding a reduced amount of material in front of the calorimeter. A very preliminary estimate of the radiation length of the main VHMPID detector components is reported in Table~\ref{tabX0}.

\begin{table}[h!]
\centering
\begin{tabular}{lc}
\hline
{\em Detector component} & $X/X_0$ \\
\hline \hline
Bottom and top radiator vessel panel & 7 \% \\ 
Photon detector and tracking layers & 3 \% \\ 
Mirror & 2 \% \\ 
Radiator gas & 4 \% \\
Sapphire window (4 mm) & 6 \% \\ \hline
Total & 22 \% \\ \hline \hline
\end{tabular}
\caption{Estimated radiation length of different components of the VHMPID detector.}
\label{tabX0}
\end{table}

This radiation length is comparable to the thickness of the existing TRD and TOF subsystems in ALICE. Its impact on the
calorimeter performance was studied with detailed GEANT simulations and was found to be negligible for the lepton performance
(no impact on E/p measurement) and have only a less than 10\% effect on the photon detection through generation of additional low momentum photon background.
The integrated detector system will consist of separate VHMPID and calorimeter modules plus a combined support structure which can hold both types of modules. This layout will be slightly different for stages 1,2 and 3. Here we only describe the ongoing work for stages 1 and 2. According to the DCal design drawings the existing modules are sufficiently small in radial direction that only the detector support structure has to be modified.Specifically, the detector frame on one side adjacent to the PHOS detector has to be lowered in order to enhance the available radial distance from 24 cm between TOF and the ALICE space frame to 72 cm.
As already shown in fig. \ref{intfig1}, these 72 cm would be divided as follows:

\begin{table}[h!]
\centering
\begin{tabular}{lc}
\hline
{\em Detector component} & radial distance \\
\hline \hline 
Top tracking layer & 3 cm \\
Photon Detector (incl. top radiator panel) & 8 cm \\
Radiator vessel (incl. mirror structure) & 57 cm \\ 
Bottom tracking layer& 3 cm \\ 
Safety stand-off & 2 cm \\
\hline 
\end{tabular}
\caption{Estimated radial depth of different components of the VHMPID detector.}
\label{tab-bkdown}
\end{table}

The present support structure of the DCal distinguishes between the A- and the C-side surrounding the PHOS. In order to assure easy access to the PHOS, the C-side is elevated. By lowering the C-side down to the A-side levels all DCal modules can be instrumented with VHMPID modules. The access to the PHOS will be restricted, but the front-end electronics can still be accessed from below. The ensuing gaps in the DCal azimuth coverage could be filled with two standard sized DCal modules per gap, if necessary.

\begin{figure}[!h]
\centering
   \includegraphics[width=0.8\textwidth]{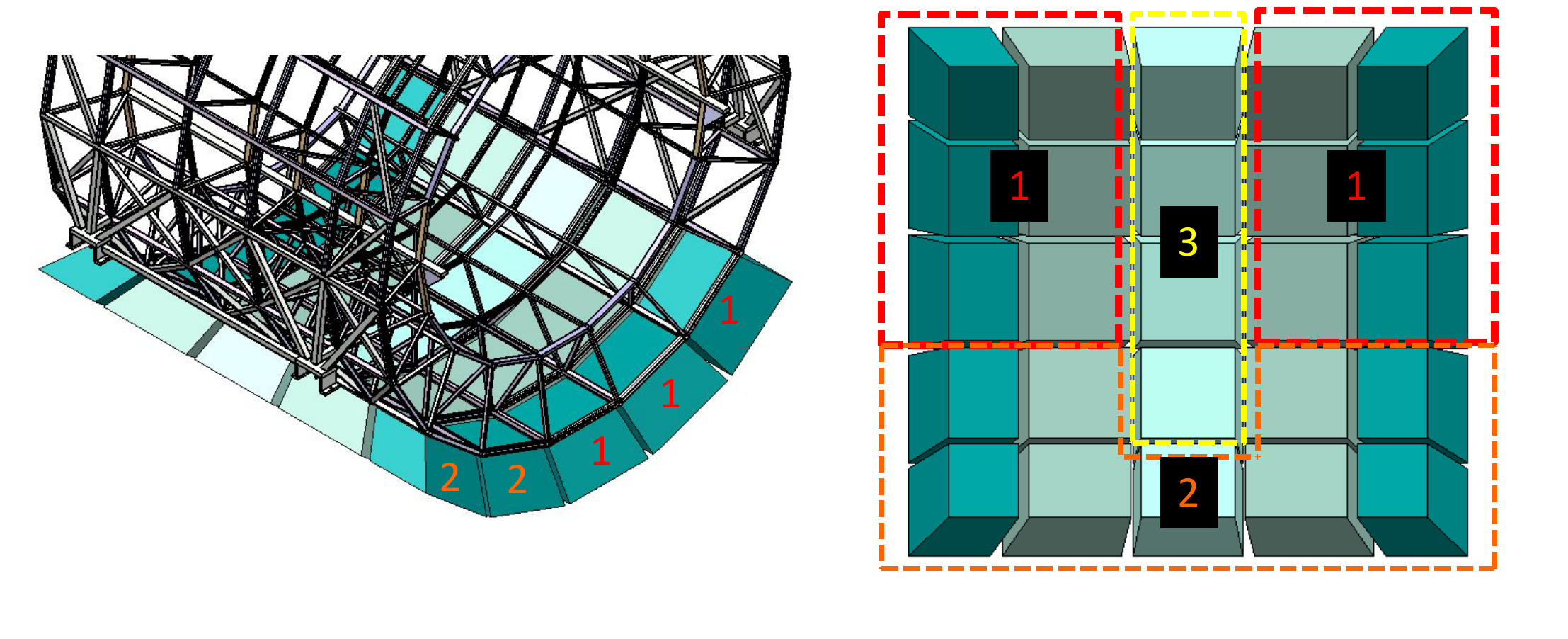}
   \caption{The proposed VHMPID layout, with 25 modules installed in sectors 11 to 15. The picture on the right shows the installation sequence in three stages.}
      \label{intfig2}
\end{figure}

\begin{figure}[!h]
\centering
   \includegraphics[width=0.7\textwidth]{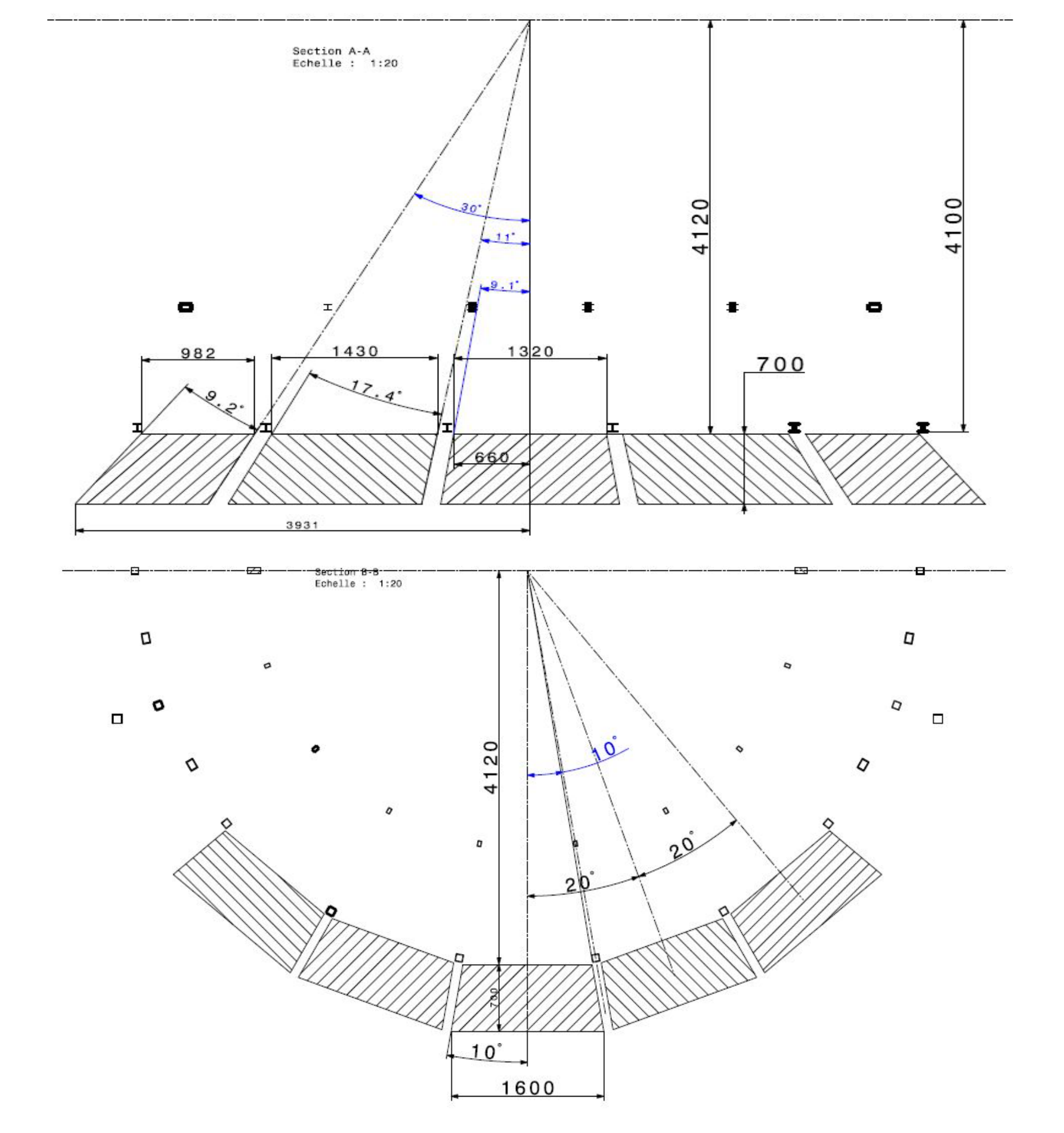}
   \caption{Cross sections of $\eta$ (top) and $\phi$ (bottom) views of the proposed layout.}
      \label{intfig3}
\end{figure}

\begin{figure}[!h]
\centering
   \includegraphics[width=0.8\textwidth]{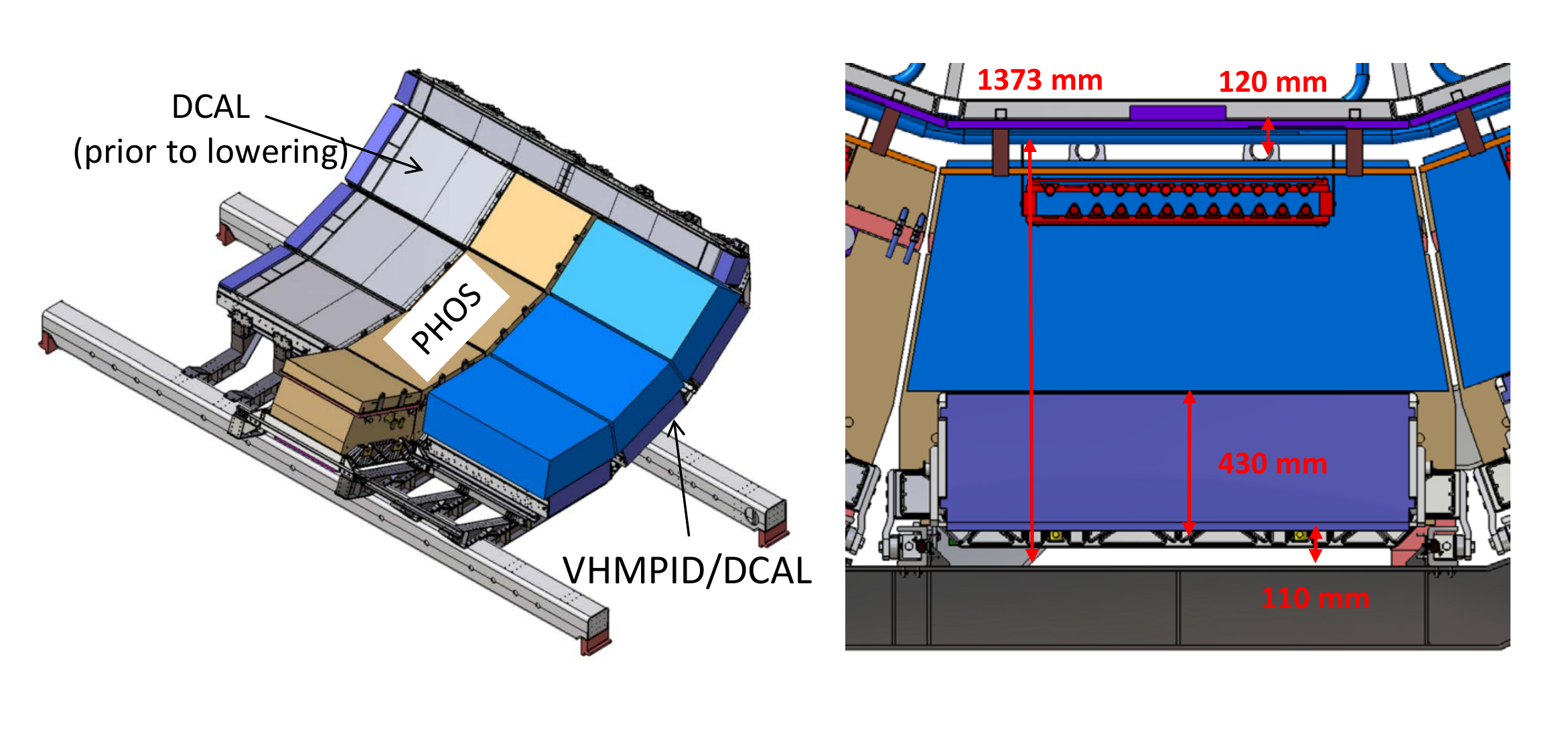}
   \caption{Views of combined DCAL and VHMPID modules during stage 1 installation.}
      \label{intfig4}
\end{figure}

Fig.~\ref{intfig2} shows the layout of the final full VHMPID detector with modules segmentation and grouping corresponding to the three proposed installation stages. Each of the five space-frame sectors (from 11 to 15) will host five modules of different shape and size to minimize losses with respect to the space-frame structure. The present dimensions are shown in Fig.~\ref{intfig3}; however the final segmentation is still under study to take into account constraints on vessel geometry and structure related both to operation at 3.5 atm and mirror system layout. Fig.~\ref{intfig4} shows a view of a possible combined DCAL and VHMPID layout.

%%%%%%%%%%%%%%%%%%%%%%%%%%%%%%%%%%%%%
%%%%%%%%%%%%%%%%%%%%%%%%%%%%%%%%%%%%%

%%%%%%%%%%%%%%%%%%%%%%%%%%%%%%%%%%%%%
%%%%%%%%%%%%%%%%%%%%%%%%%%%%%%%%%%%%%

%%%%%%%%%%%%%%%%%%%%%%%%%%%%%%%%%%%%%
%%%%%%%%%%%%%%%%%%%%%%%%%%%%%%%%%%%%%

As presented in Fig.~\ref{intfig2}, we propose to install 12 modules for stage 1 (covering the existing DCal modules in sectors 13, 14 and 15) and 9 additional modules in sectors 11 and 12 for stage 2. Stage 3 would then add to the design four more central modules in front of the present PHOS detector. With the proposed layout of twenty-five modules covering five space-frame sectors a total acceptance of about $ 30 \% $ of the ALICE central barrel can be achieved in the pseudorapidity range $\eta = \pm 0.75$. 

The handling and installation of any new detector in ALICE is a delicate operation. This is due to the difficult access to crowded areas such as the front part of the solenoid and the volume inside it. The presence of the mini-space-frame limits the use of the crane and consequently the weight of the objects to be integrated. The need of having light elements is a fundamental parameter in the design of both the detector and the support structures since the largest part of the installation work has to be made "by hand". The only tools that can be used inside the solenoid to help the installation are temporary supports and rails and relatively small pulleys. The design of the new DCAL and PHOS cradle has already taken into account the integration of VHMPID modules in the space-frame sectors 11-15. The support structure of the VHMPID could be realized by an extension of the new DCAL and PHOS cradle with bolted modular aluminum elements. The support structure is taking into account the required gas pressure in the radiator vessel and heating elements on the support panels in order to keep the gas from liquefying at high pressure. Finite element analyses of the vessel itself and the support structure are in progress. The integrated support of VHMPID and DCal is a joint program of the VHMPID group and the calorimeter groups proposing an extension of the DCal.

The installation of new services (cables, pipes, etc.) inside the solenoid will also require a very accurate study, given the limited amount of free slots available in the cable-trays between the doors and the magnet itself. Table~\ref{services} shows a preliminary estimation of services which will be needed for the integration in ALICE.
%Two options are considered for the mirror layout (0.5x0.5 m and 0.75x0.75 m, respectively) corresponding to two different photo-detector segmentations.
%
%\begin{sidewaystable}
\begin{table}[h!]
\centering
%{\tiny
\begin{tabular}{llll}
\hline 
& RICH with & tracking \\ 
& 50x50 cm$^2$ mirrors & layers \\ 
\hline \hline
Photo-detectors: & &\\
& 195 pcs. & 120 modules \\ 
& $24\times18$ cm$^2$ & 1 m$^2$\\
& $60\times22$ pads & $\sim 5000 $ \\ 
& of $0.4\times0.8$ cm$^2$ & channel/each\\ 
No. of channels: & 257,400 & 600,000\\ 
Power: & 9 kW & 4 kW \\ 
Radiator pipes in: & $25\times 16/18$ &  \\
Radiator pipes out: & $25\times 20/22$ & \\
Gaseous detectors & &  \\
-pipes in: & $195\times4/6$ & $60\times 6/8$ \\ 
-pipes out: & $195\times 6/8$ & $60 \times 8/10$\\
Cooling pipes: & Water cool. & Air vent.\\
in+out [m] & $25\times20/22$ & \\
Cables: & 195 for HV & 120 for HV \\
& 390 for LV & 240 for LV \\ 
& 250 for pressure & 120 for pressure \\
& 200 for temperature & 120 for temperature \\
& 100 for signals/control & 50 for signals/control\\ 
Data volume: & 20 kB/evt & 300 kB/evt\\
Readout time: & 100 $\mu$s & 70 $\mu$s \\
No. of DDLs: & 50 & 60 \\ \hline \hline
\end{tabular}
%}
\caption{\em Summary of main services specifications for the integration in ALICE.}
\label{services}
\end{table}
%\end{sidewaystable}

%%%%%%%%%%%%%%%%%%%%%%%%%%%%%%%%%%%%%%%%%%%%%%%%%%%%
\clearpage
\section{Detector Performance Studies}
\label{Section:Detector_performance}

The performance of Cherenkov detectors can be summarized by the performance of photon production and detection. In this section the resulting PID capabilities will be discussed by means of Monte Carlo simulations in AliROOT~\cite{volpe1,volpe2}, beam tests as well as the results of the existing HMPID detector.

\subsection{Simulations}
\label{Subsection:Detector_performance_simulations}

The particle identification efficiency in a RICH detector depends on the number of produced Cherenkov photons, their absorption
in the traversed media, the single photon detection efficiency and the corresponding
angular resolution. In the VUV range there is no measurement of the
refractive index of C$_4$F$_8$O, which is needed as an input parameter
for the Monte-Carlo simulations. Since the refraction index of
C$_4$F$_{10}$ is known and the same at a per mil level in the visible (see section~\ref{Subsection:Detector_performance_testbeam}), this radiator gas has been used in the following.

The different contributions to the Cherenkov angle
resolution (radiator chromaticity, photon emission point uncertainty,
photon detector spatial resolution and tracking error) have been
estimated  via theoretical calculation and are shown in
Table~\ref{angerr} for a radiator pressure of 3.5 atm. The resulting single photon angular resolution is 3.6 mrad.

\begin{table}[h!]
\centering
  \begin{tabular}{lc}
\hline
{\em Cherenkov angle error} & $\sigma_{\theta}$ (mrad)\\
\hline \hline
chromatic & 2.7\\ 
emission &  0.5 \\ 
granularity & 1.7  \\ 
tracking & 1.6  \\ \hline
total & 3.6  \\ \hline \hline
\end{tabular}
\caption{\em Theoretical values of different contributions to the Cherenkov angle resolution.}
\label{angerr}
\end{table}

\begin{figure}[htb]
    \centering
    \includegraphics[width=0.6\textwidth]{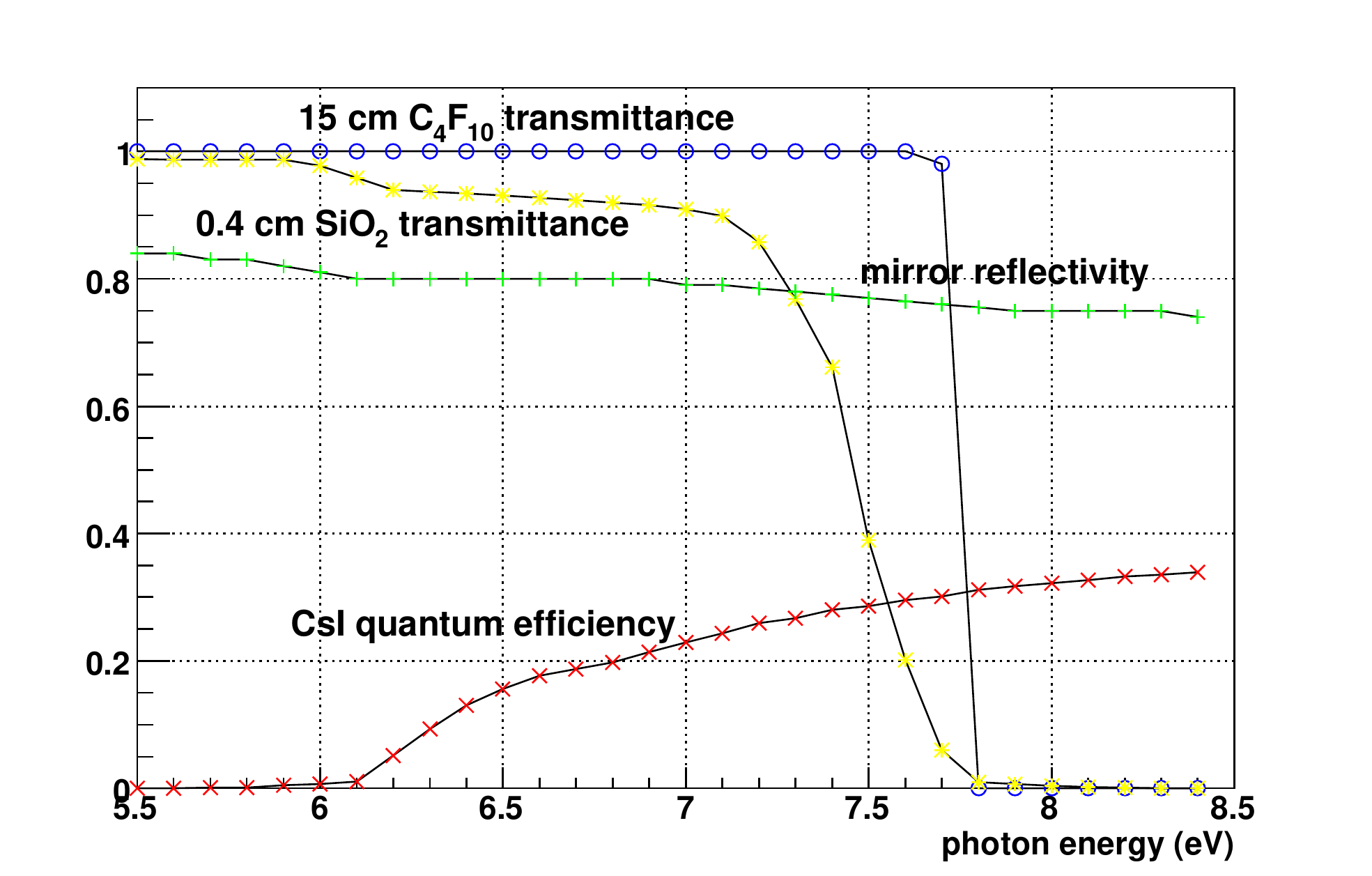}
    \caption{Detector material optical properties and CsI quantum efficiency.}
    \label{convolutionC4F10}%
\end{figure}

\begin{figure}[htb]
    \centering
    \includegraphics[width=0.45\textwidth]{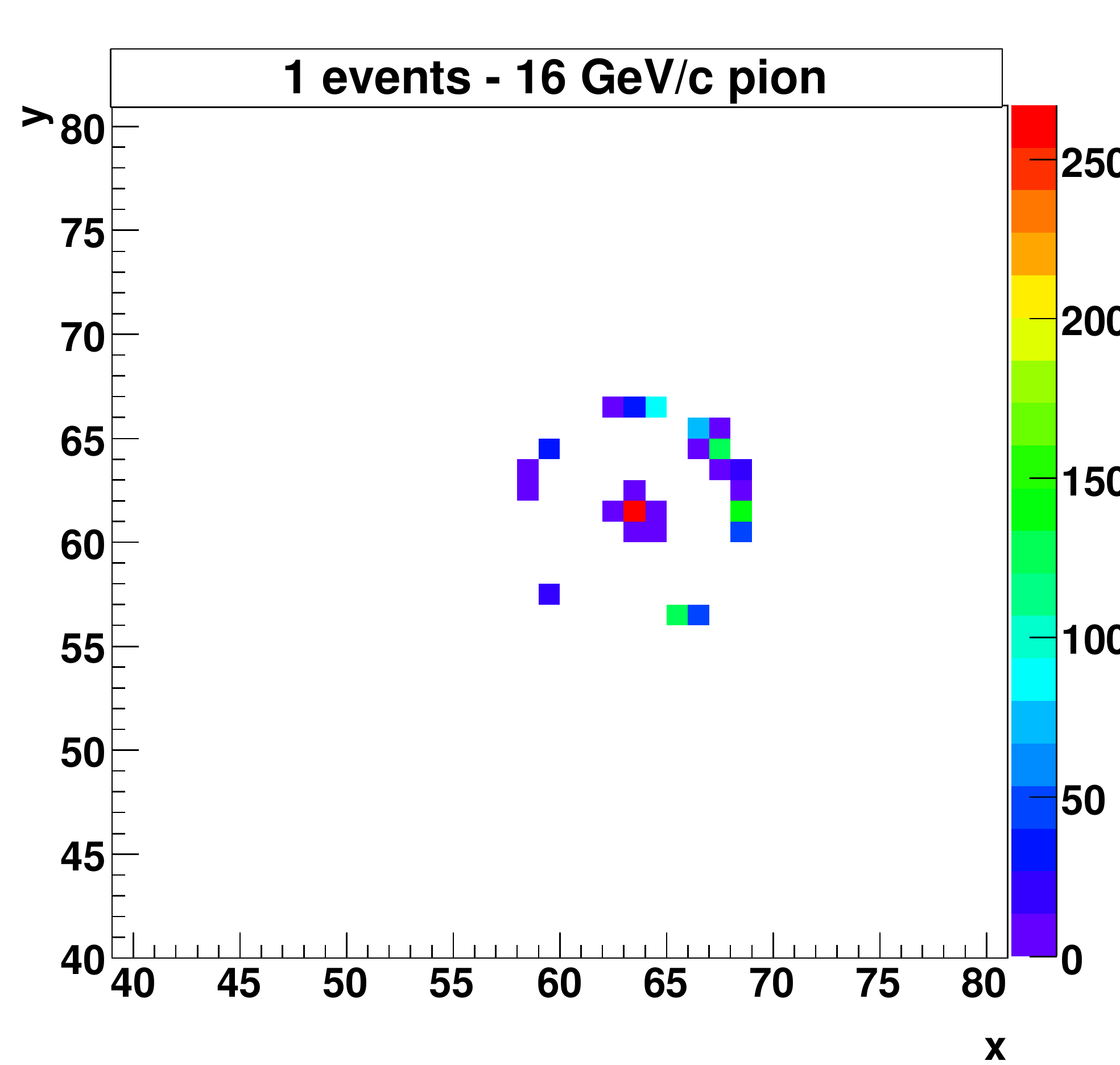}
    \includegraphics[width=0.45\textwidth]{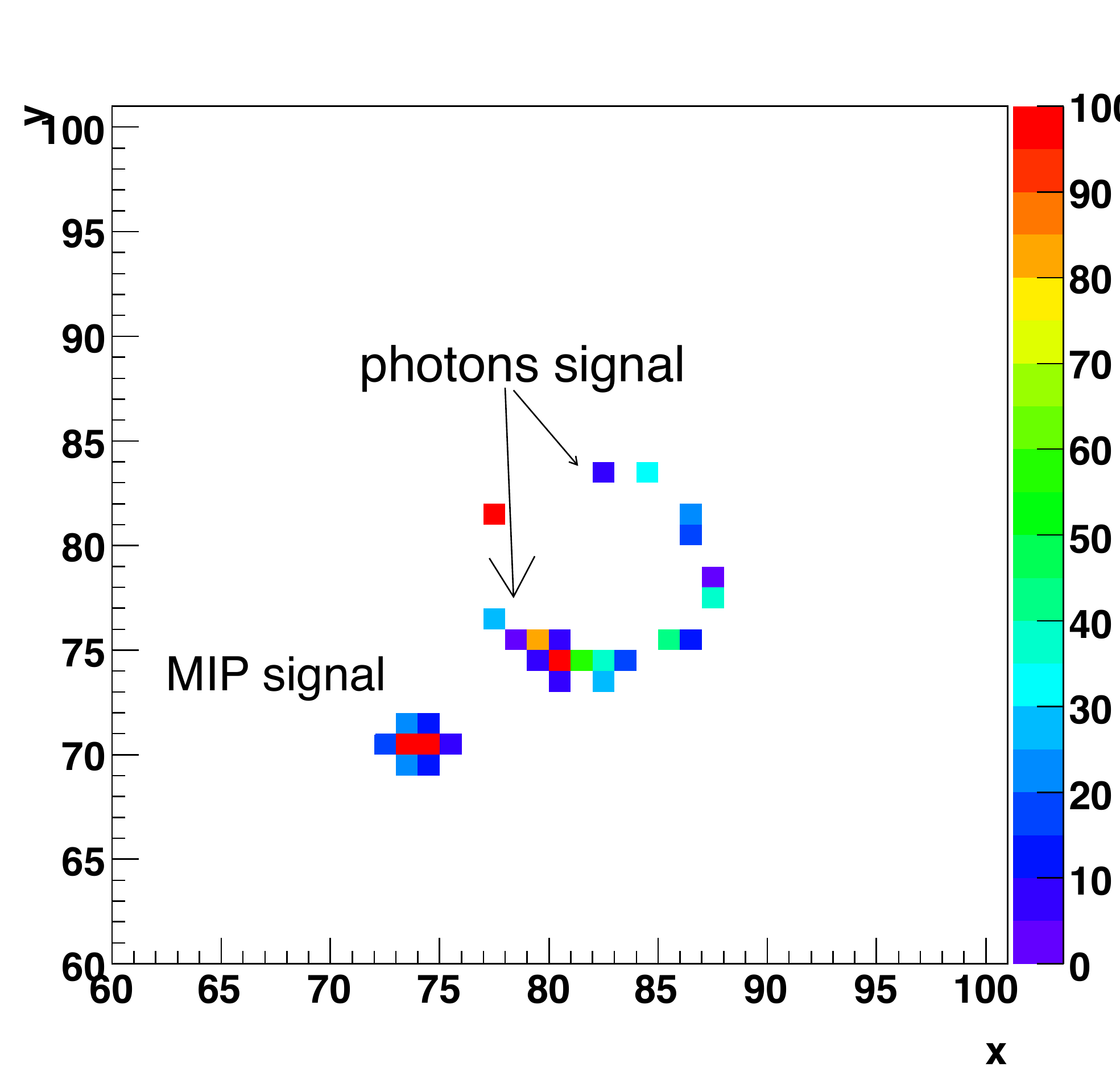}
    \caption{Cherenkov event of single pion at saturation (left) and with incident angle of $15^o$ (right). Axes $x$ and $y$ are given in pad units, the color coding represents the integrated charge in ADC units. }
\label{16gevpion-15degrees}
%    \label{16gevpion}%
\end{figure}

\begin{figure}[!h]
\centering
   \includegraphics[width=0.47\textwidth]{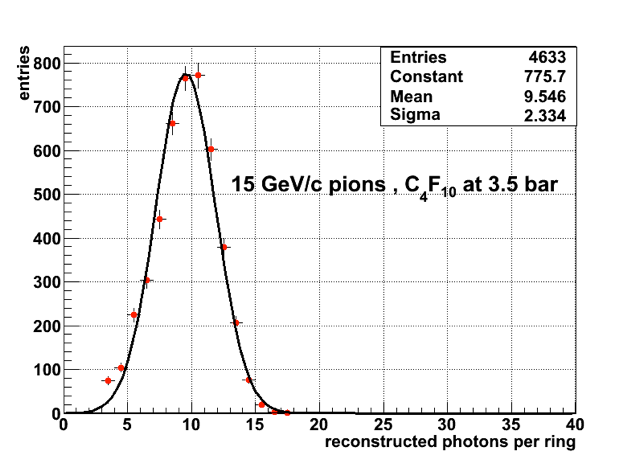}
   \includegraphics[width=0.47\textwidth]{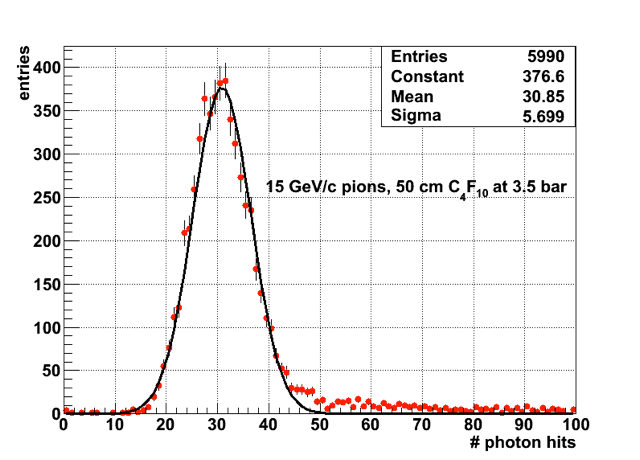}
   \caption{For a charged particle at saturation in 50 cm of  C$_4$F$_{10}$ at 3.5 atm: (left) distribution of the number of photoelectrons per event, and (right) distribution of the number of reconstructed photon cluster per event. Due to photons geometrical overlap each cluster can be originated by more than one photon.}
      \label{hits_clusterC4F10}
\end{figure}

Fig.~\ref{convolutionC4F10} shows the optical properties of all media and the CsI quantum efficiency used in the Monte Carlo simulation. As an example single pion events for normal incidence and for $15^{\circ}$, respectively, are presented in Fig.~\ref{16gevpion-15degrees} in absence of background. In Fig.~\ref{hits_clusterC4F10} the number of detector hits per charged particle at saturation (maximum Cherenkov angle) is shown. This results in $\sim 30$ photoelectrons and the  number of reconstructed photon clusters, $N_{rp}  \sim 9.5$. Note, the cluster can include two or more photons due to
geometrical overlap.
%
%%%%%%%%%%%%%%%%%%%%%%%%%%%%%%%%%%%%%
%%%%%%%%%%%%%%%%%%%%%%%%%%%%%%%%%%%%%
%
%%%%%%%%%%%%%%%%%%%%%%%%%%%%%%%%%%%%%
%%%%%%%%%%%%%%%%%%%%%%%%%%%%%%%%%%%%%
%
%%%%%%%%%%%%%%%%%%%%%%%%%%%%%%%%%%%%%
%%%%%%%%%%%%%%%%%%%%%%%%%%%%%%%%%%%%%

\begin{figure}[!h]
    \centering
   \includegraphics[width=0.8\textwidth]{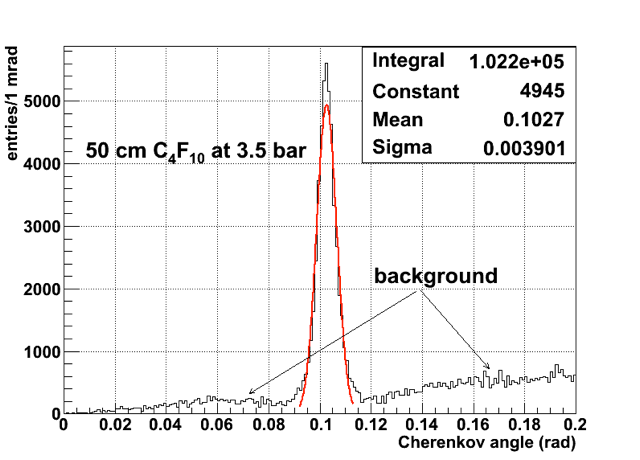}
    \caption{Distribution of reconstructed single photon Cherenkov angle, from
pions at saturation in 50 cm C$_4$F$_{10}$ at 3.5 atm in presence of Pb-Pb background.}
    \label{singlepb}
\end{figure}

\begin{figure}[!h]
  \centering
   \includegraphics[width=0.52\textwidth]{./ThreePart_16gev_3_5bar_50cm.png}
   \includegraphics[width=0.52\textwidth]{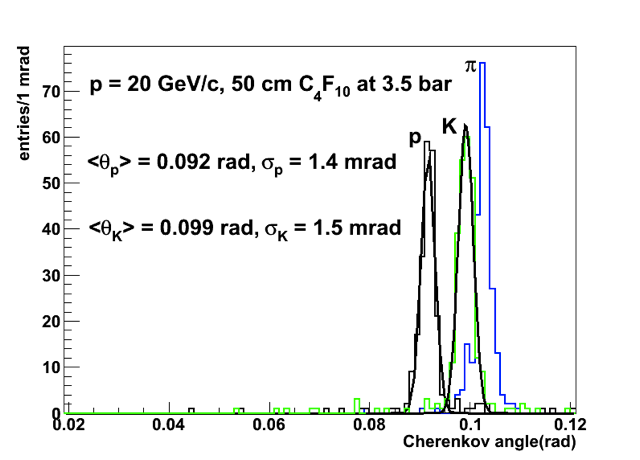}
   \includegraphics[width=0.52\textwidth]{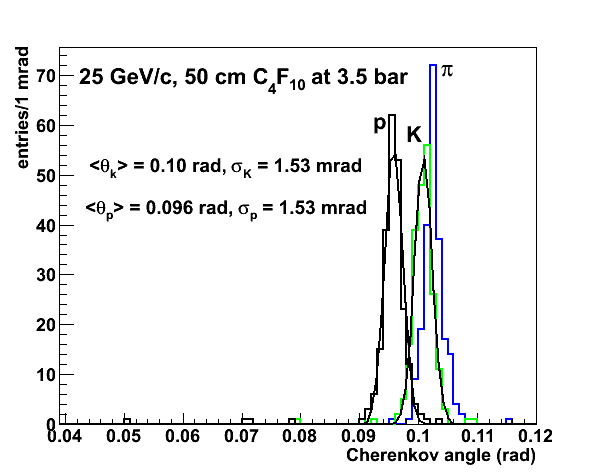}
  \caption{Distribution of ring-averaged Cherenkov angles for pions, kaons, and
protons after the Hough Transform method at $16, 20$ and $25$ GeV/c at 3.5 atm gas radiator pressure.}
      \label{thetaCh}
\end{figure}

Events obtained by embedding Cherenkov rings from pions, kaons and protons, at different momenta, with background Pb-Pb HIJING events at LHC energies, have been analyzed using the same pattern recognition procedure developed for the
HMPID~\cite{hough1,hough2}. Starting from the impact point of charged particles and photons on the chamber, the Cherenkov angle is determined by means of a back-tracing algorithm. Then a Hough Transform (i.e. looking for local maxima in a feature parameter space like the Cherenkov angle of each photon for the analyzed track in this case) is applied to filter out the background and improve the signal of identified particles. Fig.~\ref{singlepb} shows the single photon Cherenkov angle distribution from pions in presence of background. In Fig.~\ref{thetaCh} are presented the ring-averaged Cherenkov angle distributions obtained for pions, kaons and protons at $16, 20$,  and $25$~GeV/c, after the Hough Transform processing.

%%%%%%%%%%%%%%%%%%%%%%%%%%%%%%%%%%%%%
%%%%%%%%%%%%%%%%%%%%%%%%%%%%%%%%%%%%%
%
%%%%%%%%%%%%%%%%%%%%%%%%%%%%%%%%%%%%%
%%%%%%%%%%%%%%%%%%%%%%%%%%%%%%%%%%%%%

The Cherenkov angle resolution at saturation of about $1.5$ mrad, obtained in these simulations, is consistent with the theoretical value calculated for $8-9$ detected photons, including also background.

The summary of the PID performance obtained with the VHMPID using C$_4$F$_{10}$ Cherenkov radiator at 3.5 atm is given in Table \ref{tab:pid35atm} where positive identification lower limits are determined by Cherenkov emission thresholds and minimum $N_{rp}$ for effective identification, while upper limits correspond to $3\sigma$ separation.

\begin{table}[h!]
\centering
  \begin{tabular}{p{1.0in}cc}
\hline
\multicolumn{3}{c}{{\em Momentum range of identification, 3.5 atm}}\\ \hline
Particle type & Absence of signal  & Signal \\
              & (GeV/c) & (GeV/c) \\ \hline \hline
Pions, $\pi$ &  ---  & $2 - 16$\\ 
Kaons, $K$ & --- & $6 - 16$ \\ 
Protons, $p$ &$ 6 - 10$ & $11 - 26$ \\ \hline \hline
\end{tabular}
\caption{Identification capabilities for the VHMPID with C$_4$F$_{10}$ radiator at 3.5 atm.
The lower and upper limits correspond to $N_{rp}>3$ and to $3\sigma$ separation, respectively.}
\label{tab:pid35atm}
\end{table}

\begin{figure}[!h]
    \centering
   \includegraphics[width=0.52\textwidth]{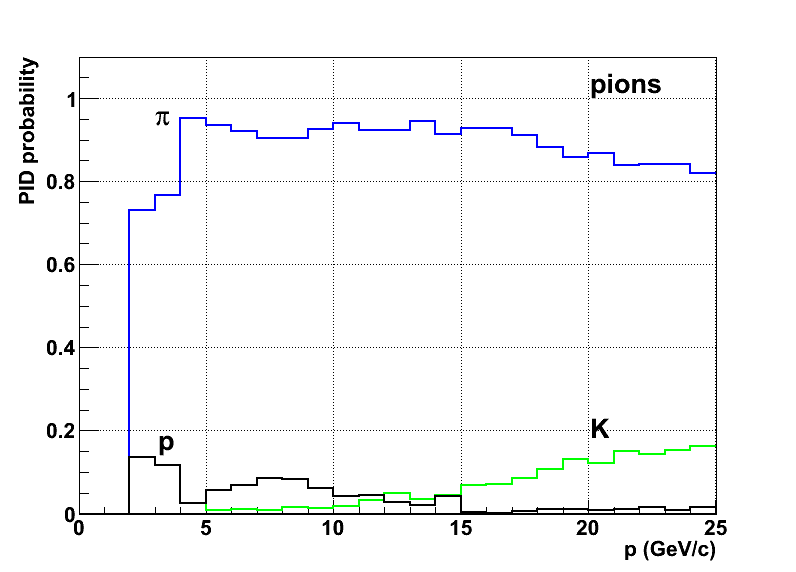}
   \includegraphics[width=0.52\textwidth]{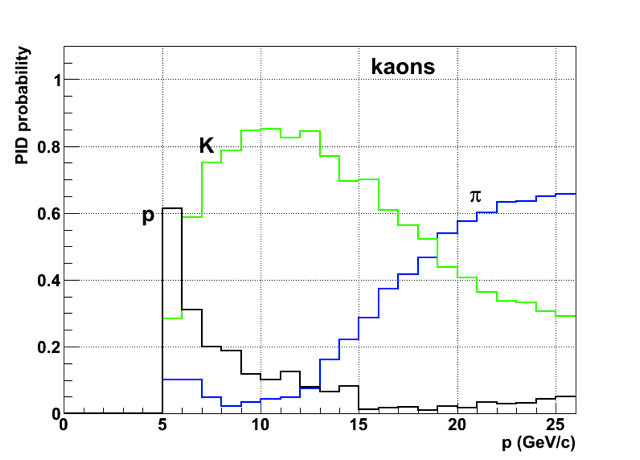}
   \includegraphics[width=0.52\textwidth]{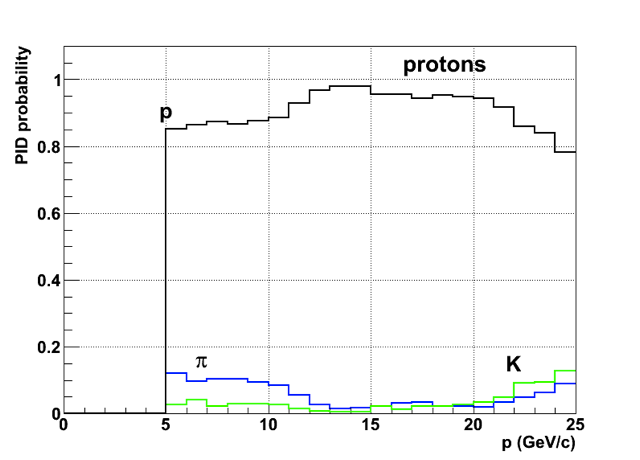}
    \caption{PID efficiency and contamination for $\pi$, $K$ and $p$
      achieved in single particle events embedded in HIJING
      background, at radiator gas pressures of 3.5 atm. The
      probability to be identified as different particle species is
      shown for pions (upper panel), kaons (middle panel), and protons
      (lower panel). }
    \label{contamination}
\end{figure}

Fig.~\ref{contamination} reports the probability of correct identification for pions, kaons and protons and wrong identification as each the other two particle species, at 3.5 atm, achieved in this Monte-Carlo study. The identification of proton below threshold can be achieved simply by exclusion when the probability for being a pion or a kaon deduced from the pattern recognition algorithm is 0 and the particle momentum is in the range below Cherenkov emission threshold of protons. 
%
%%%%%%%%%%%%%%%%%%%%%%%%%%%%%%%%%%%%%
%%%%%%%%%%%%%%%%%%%%%%%%%%%%%%%%%%%%%

\begin{figure}[!h]
    \centering
    \includegraphics[width=0.9\textwidth]{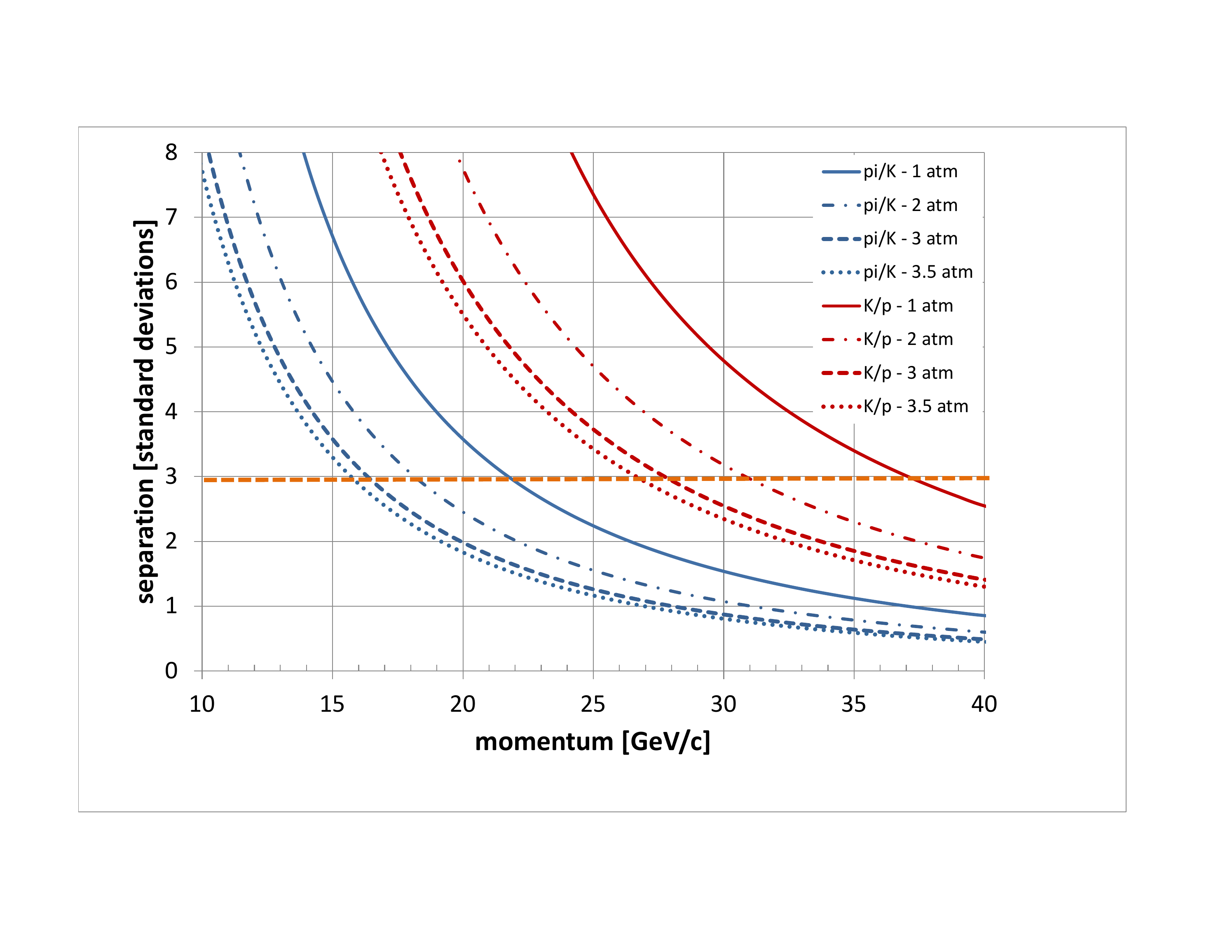}
    \caption{The theoretical separation in unit of standard
      deviation for $\pi/K$ and $K/p$ at 1, 2, 3 and 3.5 atm gas radiator pressure and for 1.5 mrad Cherenkov angle resolution.}
    \label{nsigsep}%
\end{figure}

The theoretical estimation of the particle separation n$_\sigma$ in
unit of standard deviations for C$_4$F$_{10}$ at 1, 2, 3, and 3.5 atm is shown in Fig.~\ref{nsigsep} for a 'conservative' angular resolution of 1.5 mrad, in good agreement with the presented simulation results. 

%%%%%%%%%%%%%%%%%%%%%%%%%%%%%%%%%%%%%
%%%%%%%%%%%%%%%%%%%%%%%%%%%%%%%%%%%%%

%%%%%%%%%%%%%%%%%%%%%%%%%%%%%%%%%%%%%%%%%%%%%%%%%%%%%%
\newpage
\subsection{Test beam results}
\label{Subsection:Detector_performance_testbeam}

\begin{figure}[!h]
    \centering
    \includegraphics[width=0.75\textwidth]{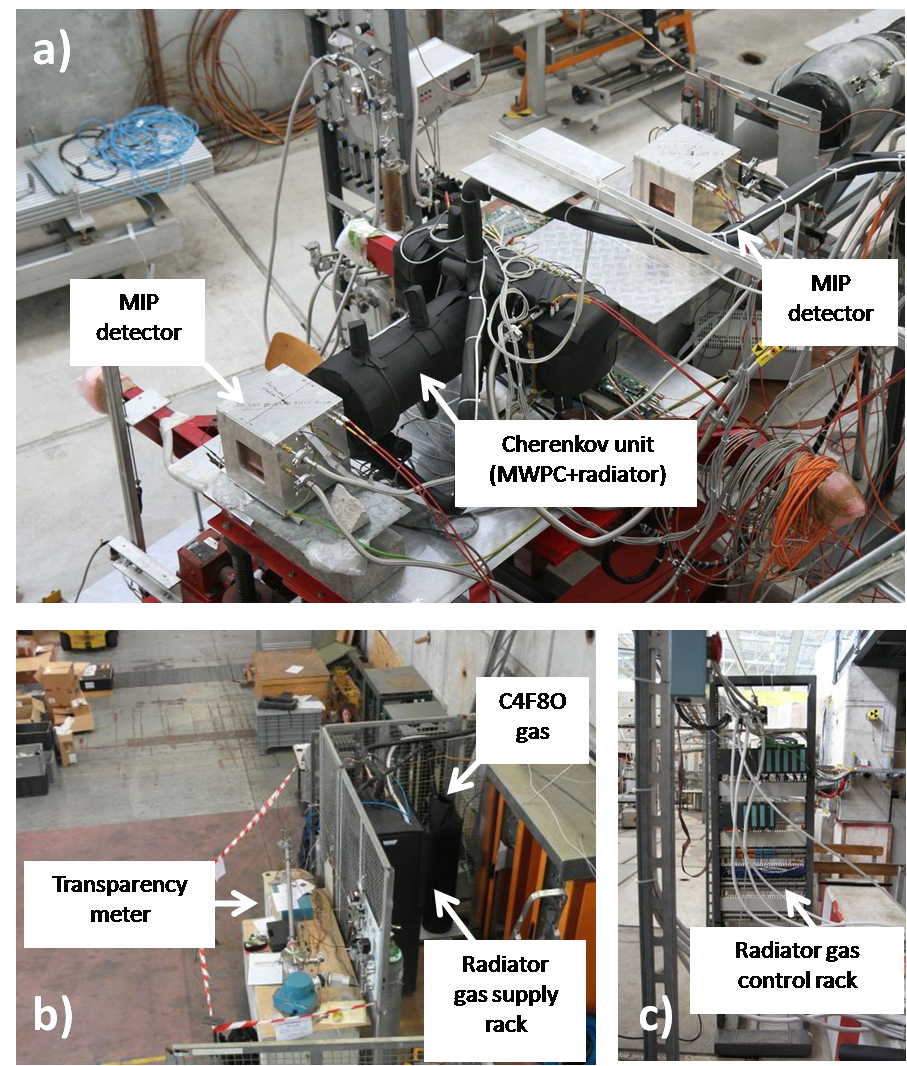}
        \caption{Test beam setup at the CERN/PS: a) the Cherenkov unit (MWPC, radiator volume) and the MIP detectors, b) the radiator gas supply and transparency meter, and c) the radiator gas control rack.}
    \label{setupOct2012}%
\end{figure}

During the past years regular test beam measurements at the CERN/PS have been performed to study the VHMPID layout and its components. The most recent and most complete test setup is shown on Fig.~\ref{setupOct2012}. This prototype is containing all the future VHMPID detector parts: A radiator volume designed to hold a pressure up to 5 bar, a MWPC with a $4 \times 8$ mm$^2$ pad cathode coated with CsI photosensitive film, separated from the radiator volume by a Sapphire window (160 mm in diameter and 5 mm thickness). The setup contains also two Close Cathod Chamber-type (CCC) MIP detectors --- a precise and durable gaseous tracking chamber --- one in front and one behind the Cherenkov unit. The pressure, temperature, and the flow of the C$_4$F$_8$O radiator gas are controlled by a fully automatic system, based on PLC hardware and PVSS software. The transparency of the returning radiator gas was measured using a UV-monochromator based transparency meter, integrated in the gas control system. 
%
%%%%%%%%%%%%%%%%%%%%%%%%%%%%%%%%%%%%
\begin{figure}[!h]
    \centering
    \includegraphics[width=0.49\textwidth]{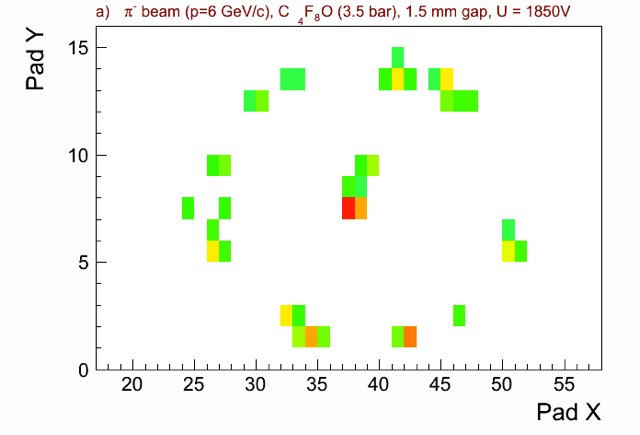}
    \includegraphics[width=0.49\textwidth]{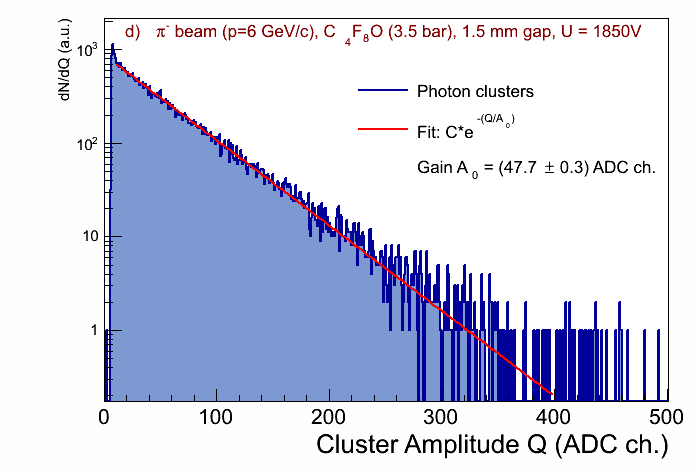}
    \includegraphics[width=0.49\textwidth]{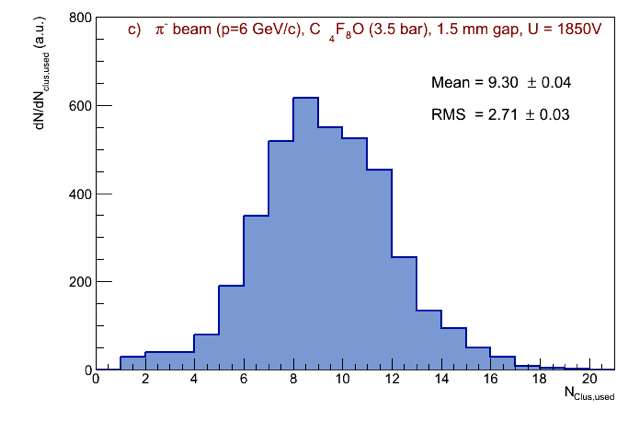}
    \includegraphics[width=0.49\textwidth]{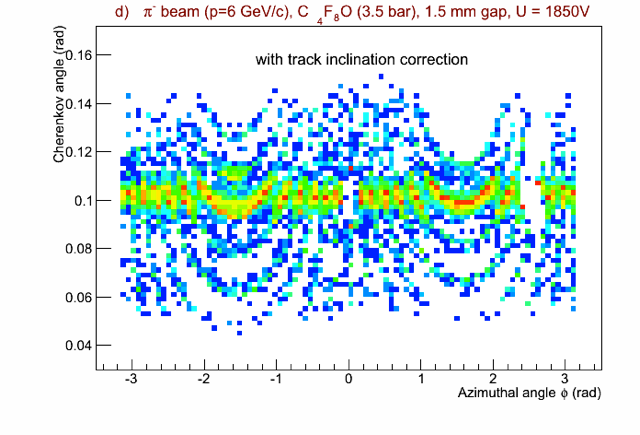}
    \includegraphics[width=0.49\textwidth]{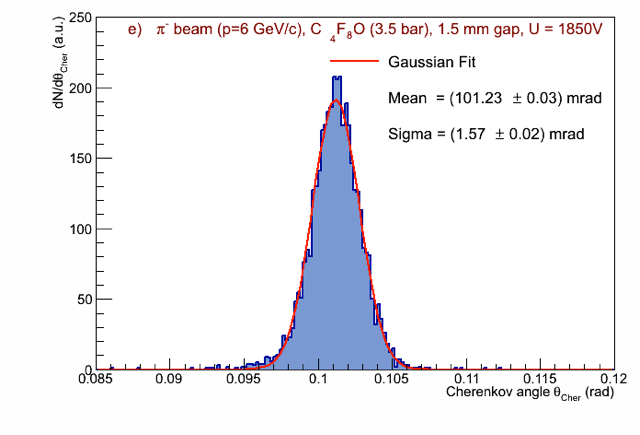}
    \includegraphics[width=0.49\textwidth]{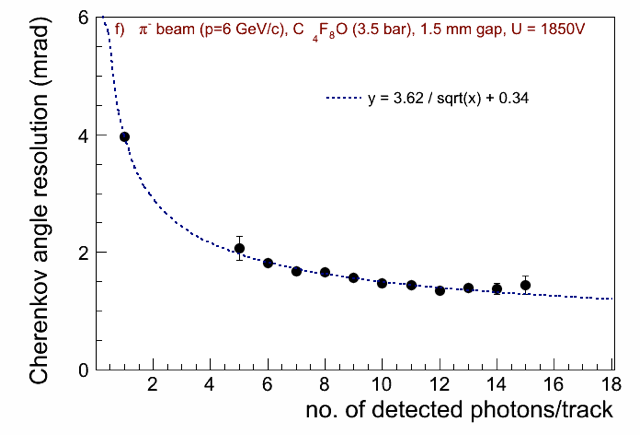}
        \caption{Main testbeam results for 6 GeV/c pions and C$_4$F$_8$O at 3.5 bar: (a) single event, (b) single photon cluster pulse height at 1850 V and a distance between anode and padplane of 1.5 mm, (c) distribution of number of detected photons in the Cherenkov ring fiducial, (d) Cherenkov angle vs azimuthal angle (corrected for the detector--beam alignment), Cherenkov angle distributions per ring (e) and per number of detected photons (f).}
    \label{testbeam}%
\end{figure}

The main results from this test are summarized in Fig.~\ref{testbeam}.  In a single event, Fig.~\ref{testbeam} a, both the MIP and the Cherenkov photon ring are clearly visible. The pulse height (PH) distribution for single photons is shown in Fig.~\ref{testbeam} b and fitted with an exponential function to retrieve the chamber gain parameter $A_{0}$ in ADC channels corresponding to a gaseous gain of $\sim5\times10^{4}$ for a single electron.
 The observed number of photon clusters (Fig.~\ref{testbeam}c) and the ring-averaged Cherenkov angle resolution (Fig.~\ref{testbeam} e) are consistent with those obtained in simulation (see Figs.~\ref{hits_clusterC4F10} and~\ref{thetaCh}) using nominal gas transparency. The most relevant result is that the observed single photon angular resolution of 4.0 mrad after correcting for the detector--beam alignment (Fig.~\ref{testbeam} d) is quite close to the design value, obtained by simulation in Fig.~\ref{singlepb}. Fig.~\ref{testbeam} f shows the angular resolution as a function of the number of photon clusters in the ring from which a ring angular resolution around 1 mrad can be achieved for particles at saturation. 

\begin{figure}[!h]
    \centering
    \includegraphics[width=0.54\textwidth]{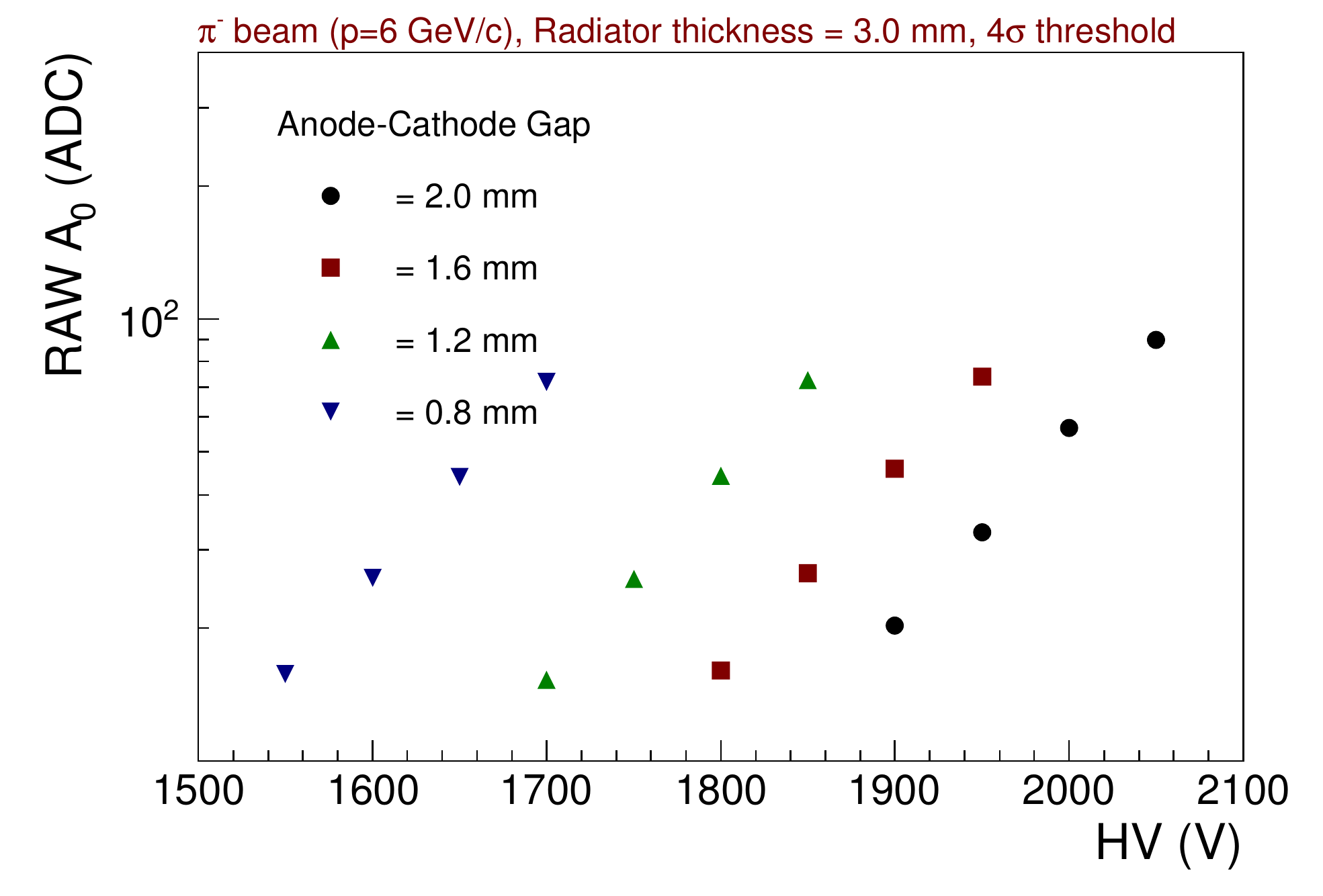}
        \caption{Gain parameter ($A_{0}$) variation with applied HV for four different anode-cathode gaps.}
    \label{vargap-gain}%
\end{figure}

\begin{figure}[!h]
    \centering
    \includegraphics[width=0.49\textwidth]{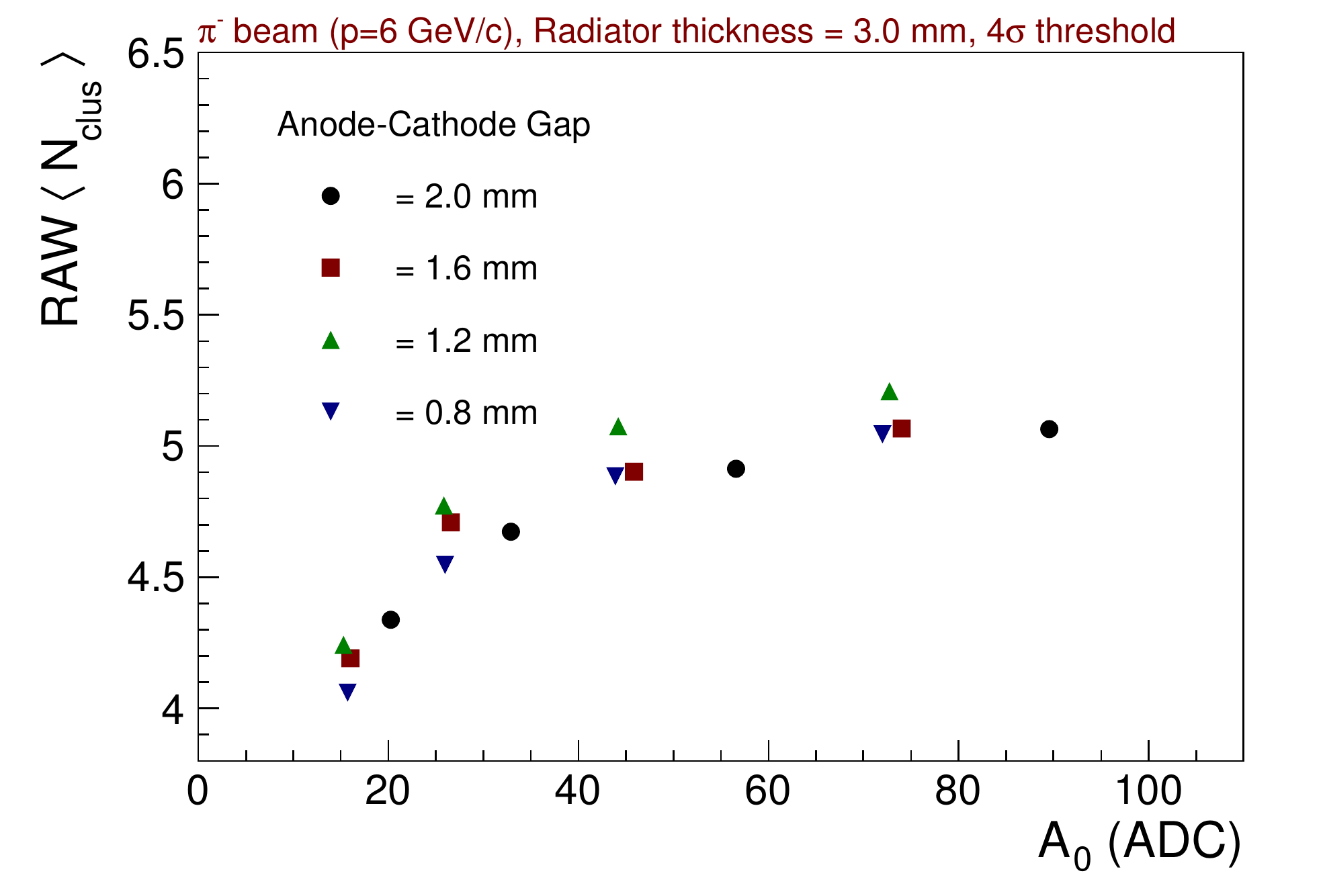}
    \includegraphics[width=0.49\textwidth]{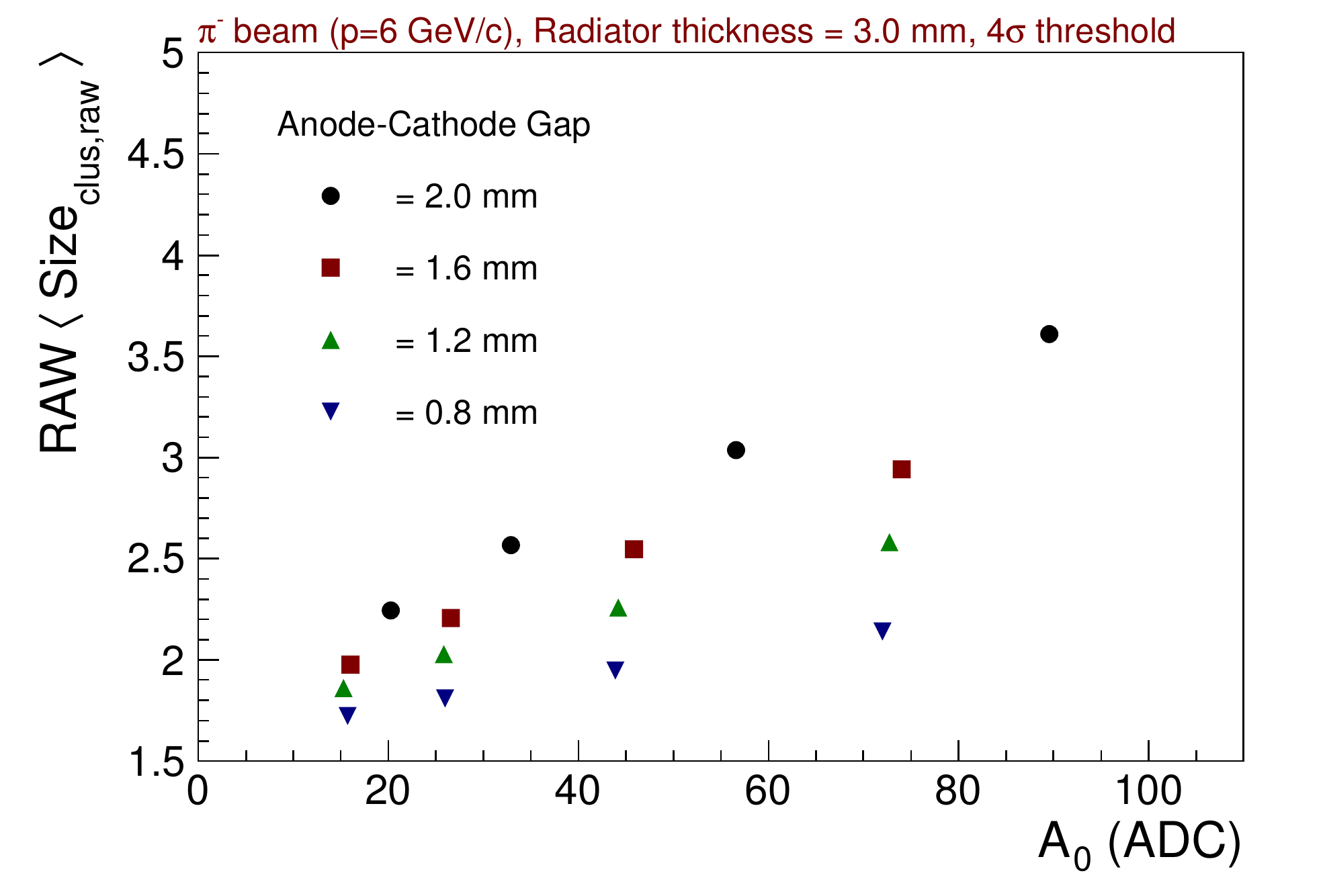}
        \caption{Cluster multiplicities and cluster sizes as a function of the gain parameter $A_{0}$ for four different anode-cathode gaps.}
    \label{vargap-nclu}%
\end{figure}

Different distances between the anode and the padplane of the the MWPC (in the range $0.8-2$ mm) have been studied in a dedicated test with the aim to check the detector performance and the stability of operation with a very small gap (down to 0.6 mm) to achieve an operation at trigger rates in the MHz regime. 
Fig.~\ref{vargap-gain} shows the measured gain in ADC as a function of the HV for different anode--cathode gap values, while in Fig.~\ref{vargap-nclu} the raw cluster multiplicities and sizes are reported,  which are consistent with expectations.

\begin{figure}[!h]
    \centering
  	\includegraphics[width=0.6\textwidth]{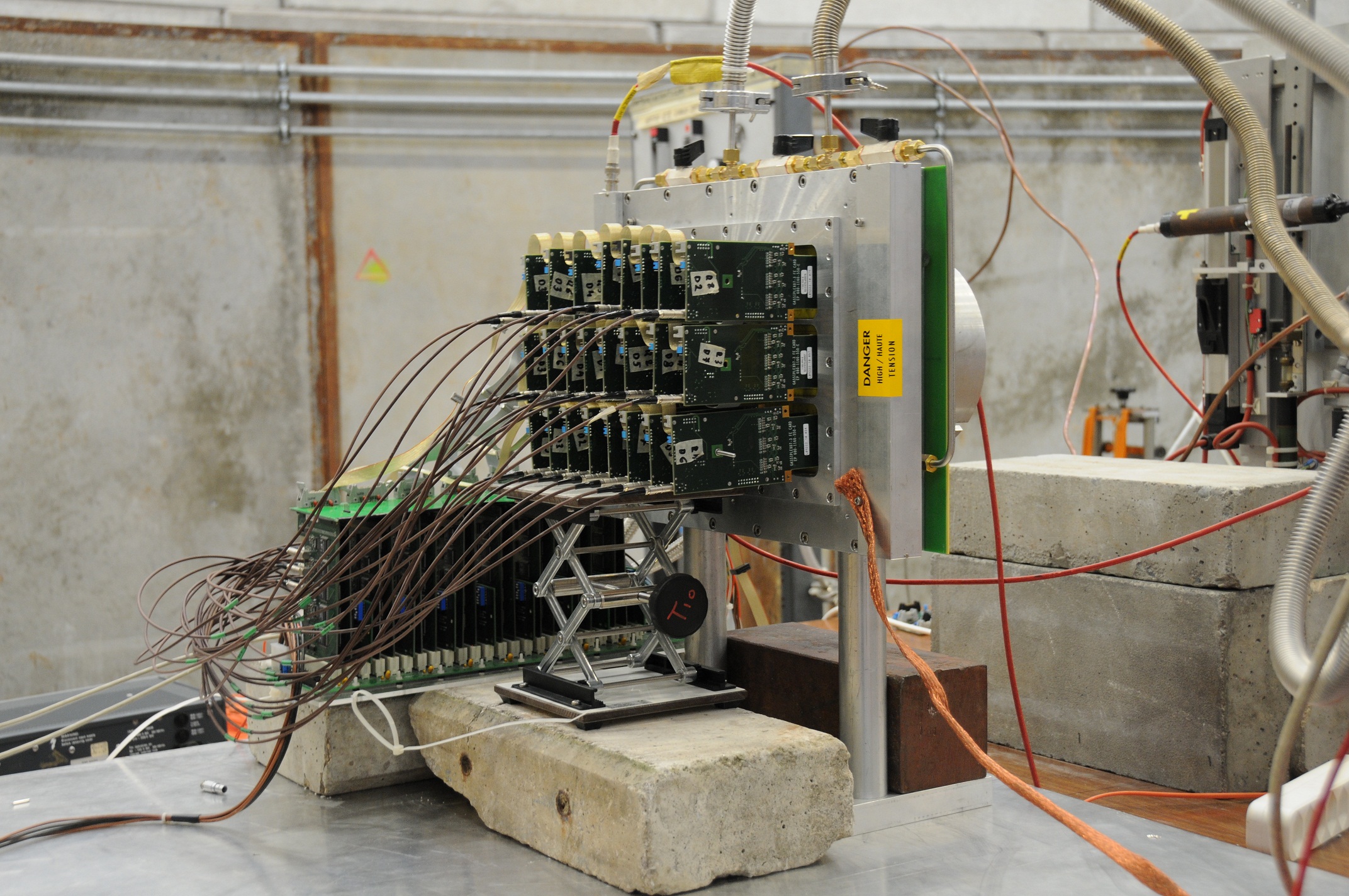}
        \caption{MWPC prototype with the final VHMPID size ($180 \times 240$ mm$^2$).}
    \label{KolkataChamber}%
\end{figure}

Very recently we built a MWPC prototype with the final VHMPID size ($180 \times 240$ mm$^2$) and an anode--padplane gap of 0.8 mm (see Fig.~\ref{KolkataChamber}). It was showing a stable operation in the test beam, which indicates that the desired distance between the anode and the padplane of 0.6 mm could be achievable.

\begin{figure}[htb]
    \centering
    \includegraphics[width=0.7\textwidth]{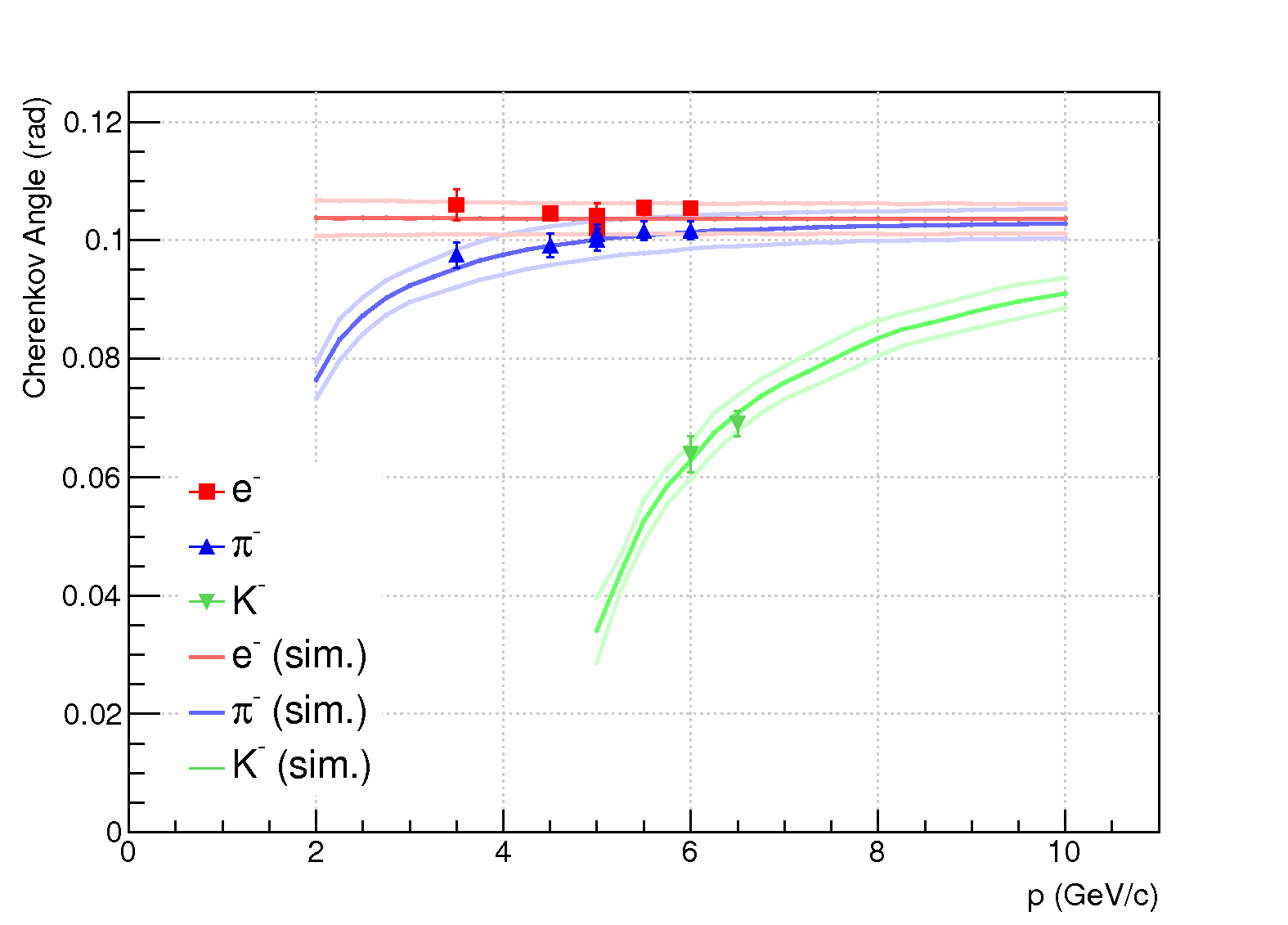}
        \caption{Test beam results
for the Cherenkov angle according to particle momentum drawn together with
simulated distributions for each particle species. Upper and lower
simulated curves denote the width observed in simulations.}%
    \label{testBeamSim}%
\end{figure}

To reproduce and extrapolate the results observed in the test beam setup, Monte Carlo simulations with the full AliRoot + GEANT3 framework have been used. These generally reproduce the qualitative and quantitative features of the test beam results well and can be used to compute expected Cherenkov ring properties not only at the tested particle momenta but also to extend to several configurations that haven't been measured. This can be seen in Fig.~\ref{testBeamSim}, where the simulation curves are compared to test beam ring position and widths for pions, kaons and electrons. Pion and electron separation is achievable below a momentum of 4~GeV/c and kaon rings can be observed starting at 5~GeV/c and above. Further adjustments to the simulations will allow an improved quantitative agreement between Monte Carlo and measurements.
In conclusion, the modified layout of the CsI-MWPC photon detector as well as the radiator gas pressurization and heating concept have been fully validated by the various beam tests, demonstrating that the expected PID performance has been achieved.
%
%

%%%%%%%%%%%%%%%%%%%%%%%%%%%%%%%%%%%%%%%%%%%%%%%%%%%%%%%%%%%%%%%%%%%%%%%%%%%%%%%%%%%%%%%%%%%%%%%%%%
\subsection{HMPID performance}
\label{Appendix:A}

\subsubsection{Detector stability and hardware performance}

\begin{figure}[!h]
\centering
\includegraphics[width=0.49\textwidth]{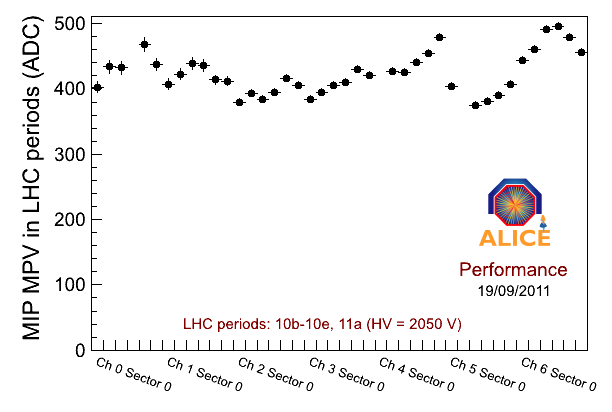}
\includegraphics[width=0.49\textwidth]{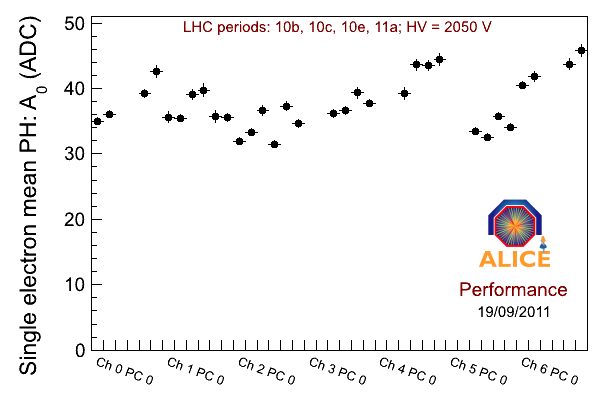}
\caption{Left panel: mean and RMS of the most probable value extracted from the track matched charge particle cluster for each HV sector in the indicated LHC periods. Right panel: mean and RMS of the electronics threshold corrected single electron mean pulse height for each photo-cathode in the indicated LHC periods.}
\label{hmp:fig1}
\end{figure}

\begin{figure}[!h]
\centering
\includegraphics[width=0.49\textwidth]{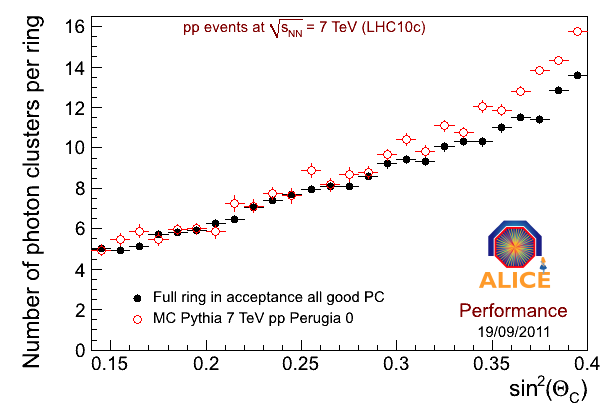}
\includegraphics[width=0.49\textwidth]{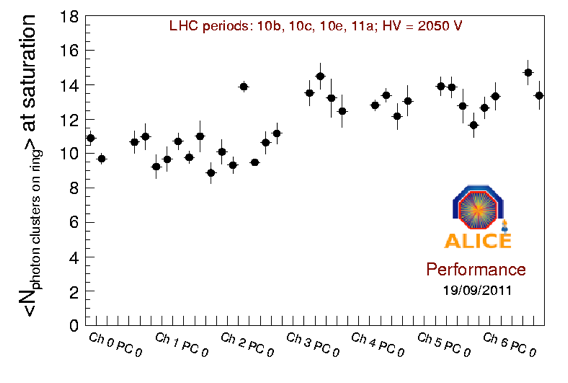}
\caption{Left panel: number of photon clusters in data and in simulation as a function of sin$^{2}$ of the Cherenkov angle for rings fully contained in the photo-cathodes. Right panel: average number of photon clusters extracted for rings at saturation for each photo-cathode in the indicated LHC periods. }
\label{hmp:fig2}
\end{figure}

In the HMPID we can factorize the photon production/detection to three elements: the applied chamber gain, the radiator properties and the Quantum Efficiency (QE) of the CsI photo-cathodes. The performance was studied in beam operations, where detector parameters can change simultaneously over time.
The applied gain on the individual high voltage sectors, monitored by the charge deposition of the MIPs and the single electron pulse height distribution from photon clusters on fully contained rings, is  stable over time as shown in Fig.~\ref{hmp:fig1}.
%%%%%%%%%%%%%%%%%%%%%%%%%%%%%%%%%%%%%%%%%%%%%%%%%%%%%%%%%%%%%%%
%
%%%%%%%%%%%%%%%%%%%%%%%%%%%%%%%%%%%%%%%%%%%%%%%%%%%%%%%%%%%%%%%%
The trend and the absolute number of measured photon clusters on ring is in good agreement with the simulations as shown in Fig.~\ref{hmp:fig2}. In the simulation, the nominal gain (35 ADC) nominal QE and nominal C$_{6}$F$_{14}$ transparency are used~\cite{HMPID-TDR}. While the different photo-cathodes show variation among each other, the average number of photon clusters per ring at saturation is stable for the individual photo-cathodes. The average number of photon clusters per ring does not show decreasing trend  with time~\cite{HMPMolnar}.
The photon production and detection is working efficiently, the QE of the CsI photo-cathodes did not decrease even 6 years after their production.
%%%%%%%%%%%%%%%%%%%%%%%%%%%%%%%%%%%%%%%%%%%%%%%%%%%%%%%%%%%%%%%
%

%%%%%%%%%%%%%%%%%%%%%%%%%%%%%%%%%%%%%%%%%%%%%%%%%%%%%%%%%%%%%%%%
\subsubsection{Challenges in tracking and reconstruction}
\label{sec:ChallengesTrackingAndReconstruction}

Particle identification in the HMPID requires the particle's track to be extrapolated from the central tracking devices of ALICE (ITS, TPC and TRD) and associated with the corresponding cluster of the
minimum ionizing particle in the HMPID cathode plane. Starting from the photon cluster coordinates, a back-tracing algorithm calculates the corresponding Cherenkov angle. Background discrimination is performed exploiting the Hough Transform Method (HTM). To each track is associated a Cherenkov angle $\langle \theta_c \rangle $, obtained as the average of the angles in the same ring. In this way HMPID can identify, on track-by-track basis, pions and kaons between 1 GeV/c and 3 GeV/c and protons from 1.5 GeV/c up to 5~GeV/c.

The HMPID is located $\sim 5$ m from the primary vertex, hence tracks have to be propagated through significant material budget after the TPC, with respect to other RICH detectors. Precise knowledge of the track parameters is essential, since the Cherenkov ring reconstruction resolution depends on them. Reconstructed tracks are propagated up to the HMPID chambers by means of a dedicated algorithm. The first algorithm used was tuned already in the STAR experiment. It propagates the track from the last point in the TPC or TRD detectors to the HMPID chamber planes by means of a simple helix. In the real ALICE environment with its 0.5 T magnetic field and a significant material budget in front the HMPID, this procedure does not allow to obtain a good track angular resolution at the chambers entrance to perform the Cherenkov angle measurement, especially in the low momentum region (below 2 GeV/c). 
\begin{figure}[!h]
    \centering
    \includegraphics[width=0.8\textwidth]{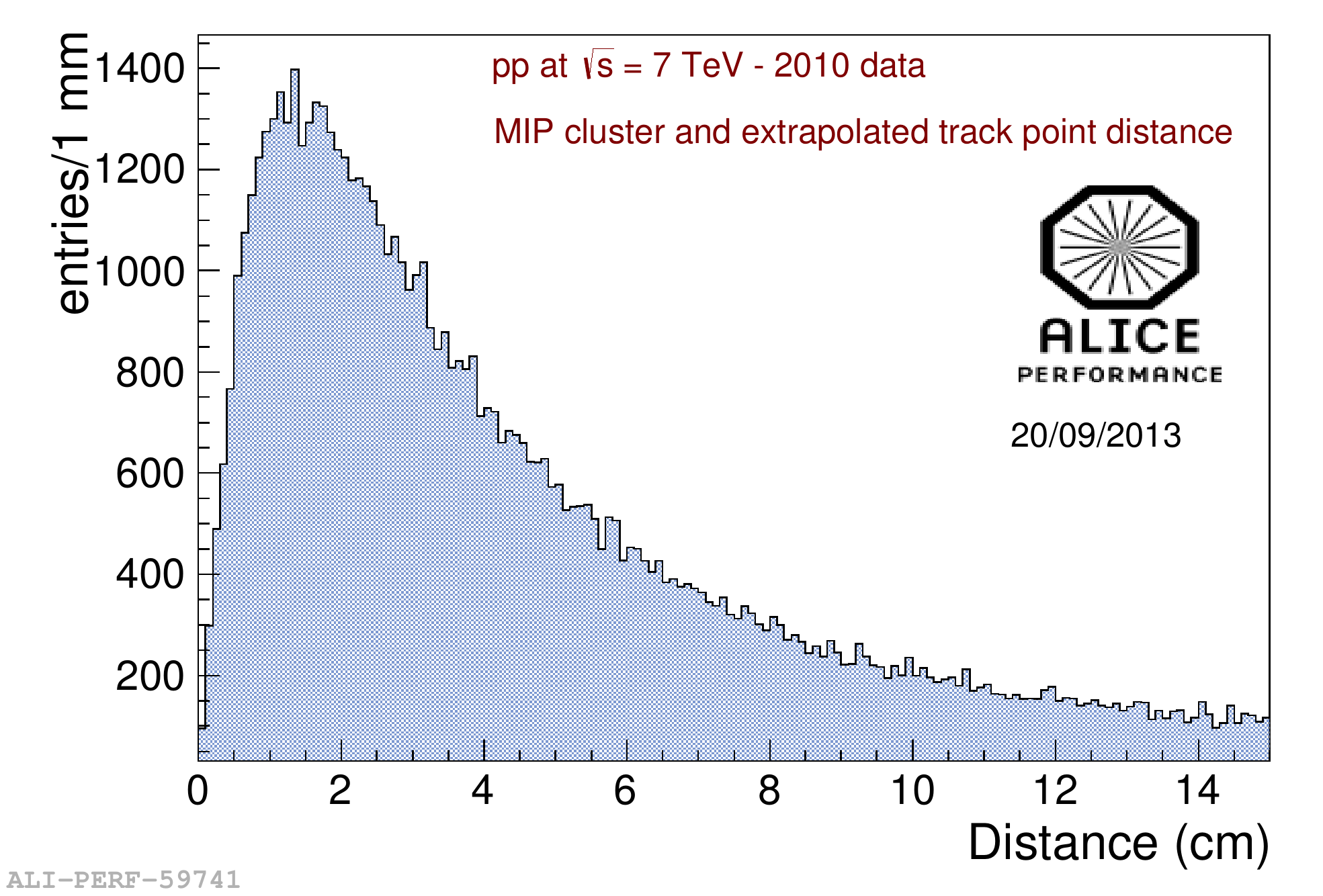}
    \caption{Distribution of the distance between primary tracks intersection point at HMPID plane and the corresponding MIP point for HMPID chamber 2.}
\label{distance}
\end{figure}

In Fig.~\ref{distance} we show the distribution of the distance between primary tracks intersection point at HMPID plane and the corresponding MIP point for $\sqrt{s} = 7.0$ TeV pp data; most of the tracks have a distance above 2 cm. To obtain satisfactory Cherenkov angular resolution it is necessary to apply a strong cut ($\textless$ 1 cm) on the distance between the extrapolated track point and the corresponding MIP point, loosing about the 60\% of the available track statistics. Significant effort has been devoted to improve the tracking. For optimal PID performance one needs the $p/p_T$ directly at the PID detector.  Since the Kalman filter can provide the best estimated $p/p_T$ at the interaction point (first track point) and the PID detector front face, we simply had to modify the tracking software output. In the improved tracking the running track is picked up at the last TPC point and propagated up to the HMPID through the TRD and TOF. The improved extrapolation algorithm takes better into account the energy loss and the dependence of the magnetic field value on the distance from the interaction point. In particular with the new procedure it is possible to exploit the precise knowledge (1 mm precision) of the HMPID MIP information in the track fitting. %inserted: 
The improved tracking information brings the resolution of the Cherenkov angle close to the design values. 
%
%not approved
%\begin{figure}[!h]
%    \centering
%    \includegraphics[width=0.6\textwidth]{./TrackAngularResolution.png}
%    \caption{Track angular resolution at HMPID plane for unconstrained (black) and constrained (blue) tracks as a function of the inverse of the transverse momentum multiplied for the sign of the track charge.}
%\label{TrackAngleRes}
%\end{figure}

\begin{figure}[!h]
\centering
\includegraphics[width=0.48\textwidth]{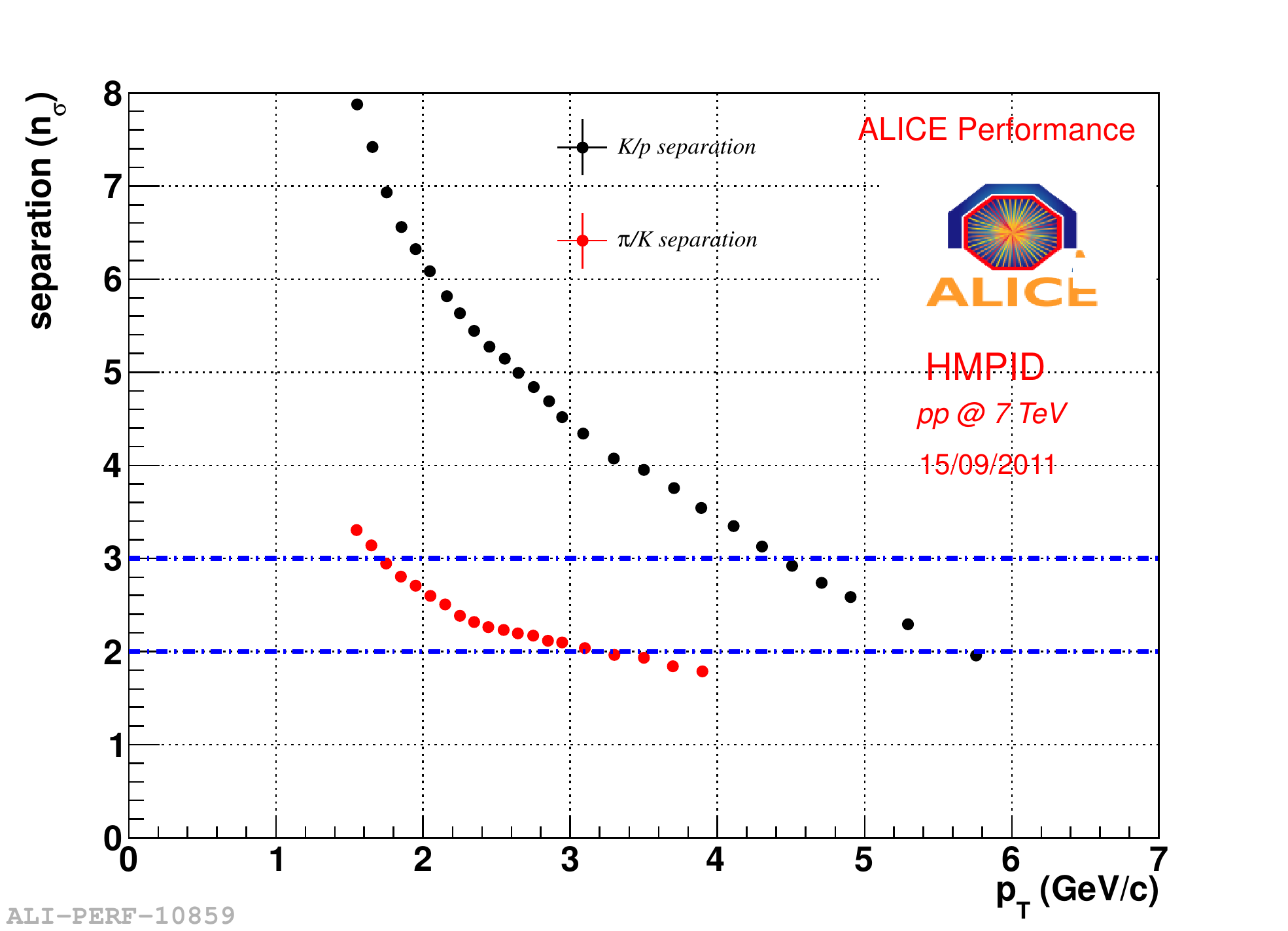}
\includegraphics[width=0.48\textwidth]{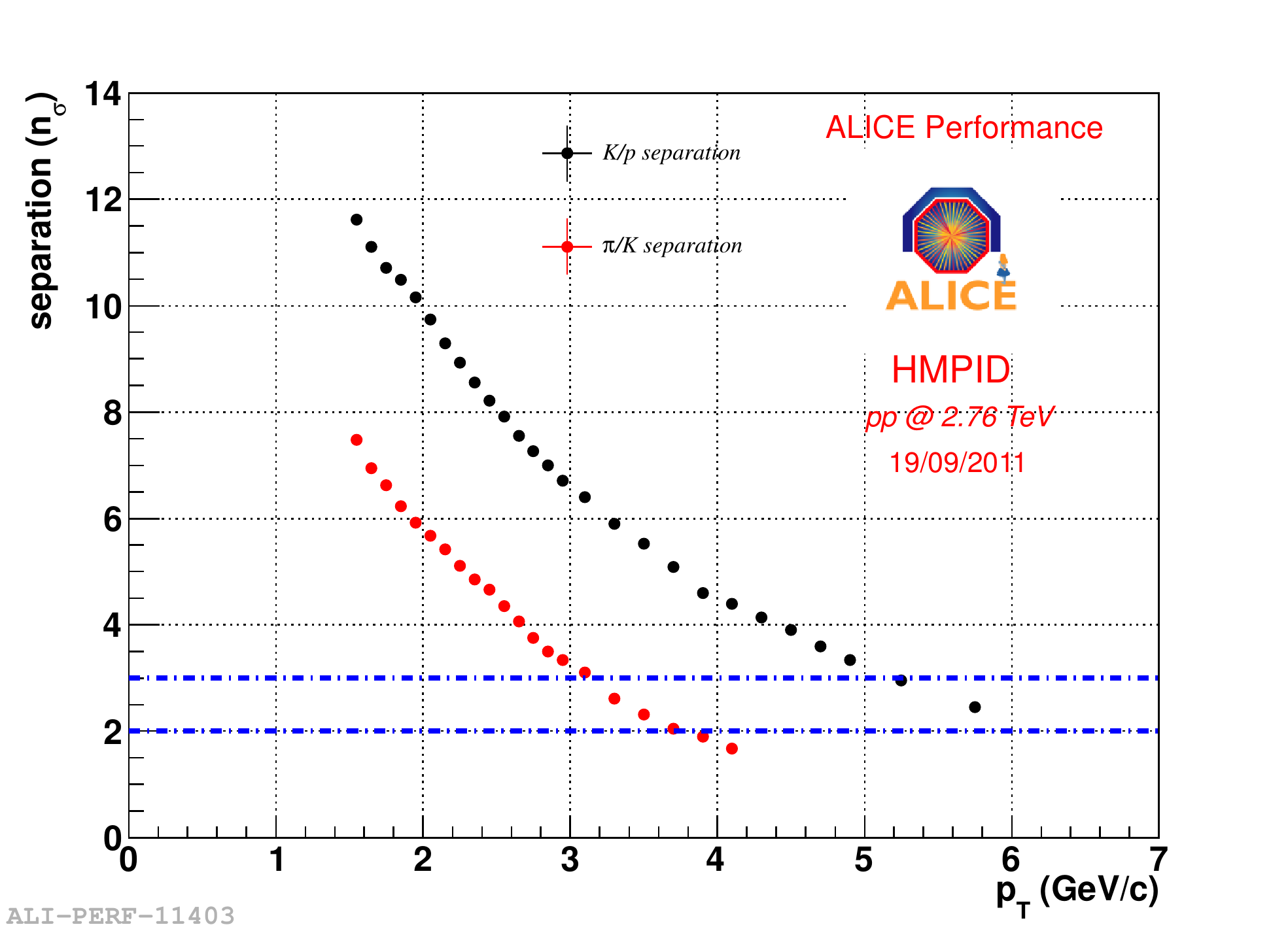}
\caption{PID separation achieved in HMPID as a function of the charged particle momentum for
early tracking (left panel) and improved tracking (right panel). Blue lines represent the target 2 and 3 $\sigma$ separation~\cite{BarilePhD}.}
\label{hmp:fig3}
\end{figure}

%removed because figure removed:
%In Fig.~\ref{TrackAngleRes} we show the track angular resolution for constrained and unconstrained tracks, the significant improvement obtained is evident. The constrained track is refitted using MIP information, while the other one is the results of a simple track propagation up to the HMPID planes. Improved tracking information brings the resolution of the Cherenkov angle close to the design values.
%Fig.~\ref{TrackAngleRes} also shows that the VHMPID will operate in the momentum range, where the angular resolution is maximum ($1/p< 0.2$ c/GeV). 
Fig.~\ref{hmp:fig3} shows the improvement in PID separation between the standard and the improved tracking algorithm. The $\pi/K$ and $K/p$ separation are slightly exceeding the expected target values.
Note that the design values are obtained from test beam results with perfect knowledge of the track (perpendicular incidence) while in the current running conditions at 0.5 T solenoid field, the average track incident angle is $\sim  20^{\circ}$.
%%%%%%%%%%%%%%%%%%%%%%%%%%%%%%%%%%%%%%%%%%%%%%%%%%%%%%%%%%%%%%%%
%
%%%%%%%%%%%%%%%%%%%%%%%%%%%%%%%%%%%%%%%%%%%%%%%%%%%%%%%%%%%%%%%%
%
%%%%%%%%%%%%%%%%%%%%%%%%%%%%%%%%%%%%%%%%%%%%%%%%%%%%%%%%%%%%%%%%
\begin{figure}
\centering
\includegraphics[width=0.47\textwidth]{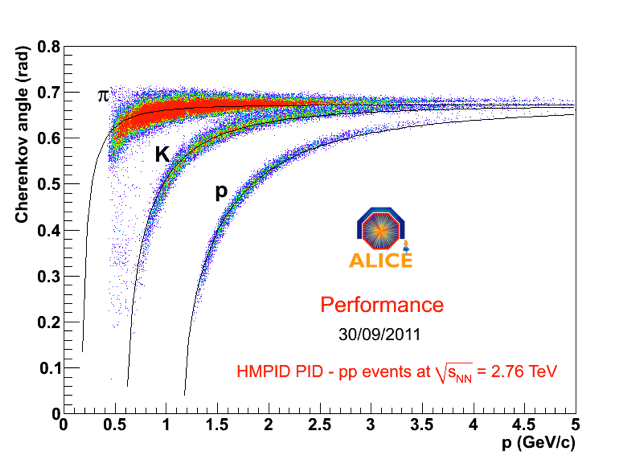}
\includegraphics[width=0.47\textwidth]{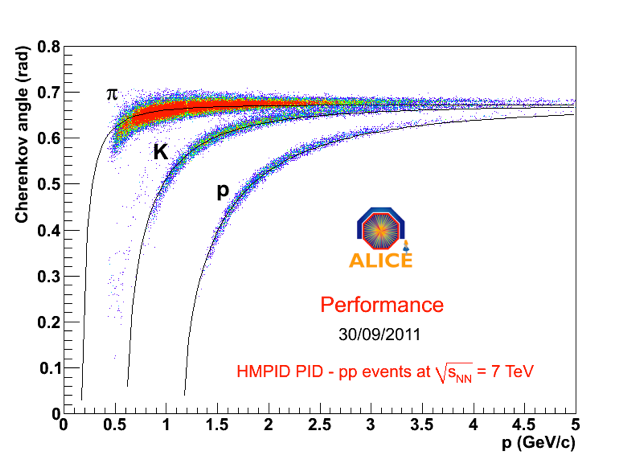}
\caption{The $\theta_{C}$ as a function of the charged particle momentum with improved tracking algorithm. Curves represent the theoretical expectation for the nominal $C_{6}F_{14}$ index of refraction~\cite{HMPID-TDR}.}\label{hmp:fig4}
\end{figure}
%%%%%%%%%%%%%%%%%%%%%%%%%%%%%%%%%%%%%%%%%%%%%%%%%%%%%%%%%%%%%%%%
%
%%%%%%%%%%%%%%%%%%%%%%%%%%%%%%%%%%%%%%%%%%%%%%%%%%%%%%%%%%%%%%%%
\begin{figure}[!h]
\centering
\includegraphics[width=0.6\textwidth]{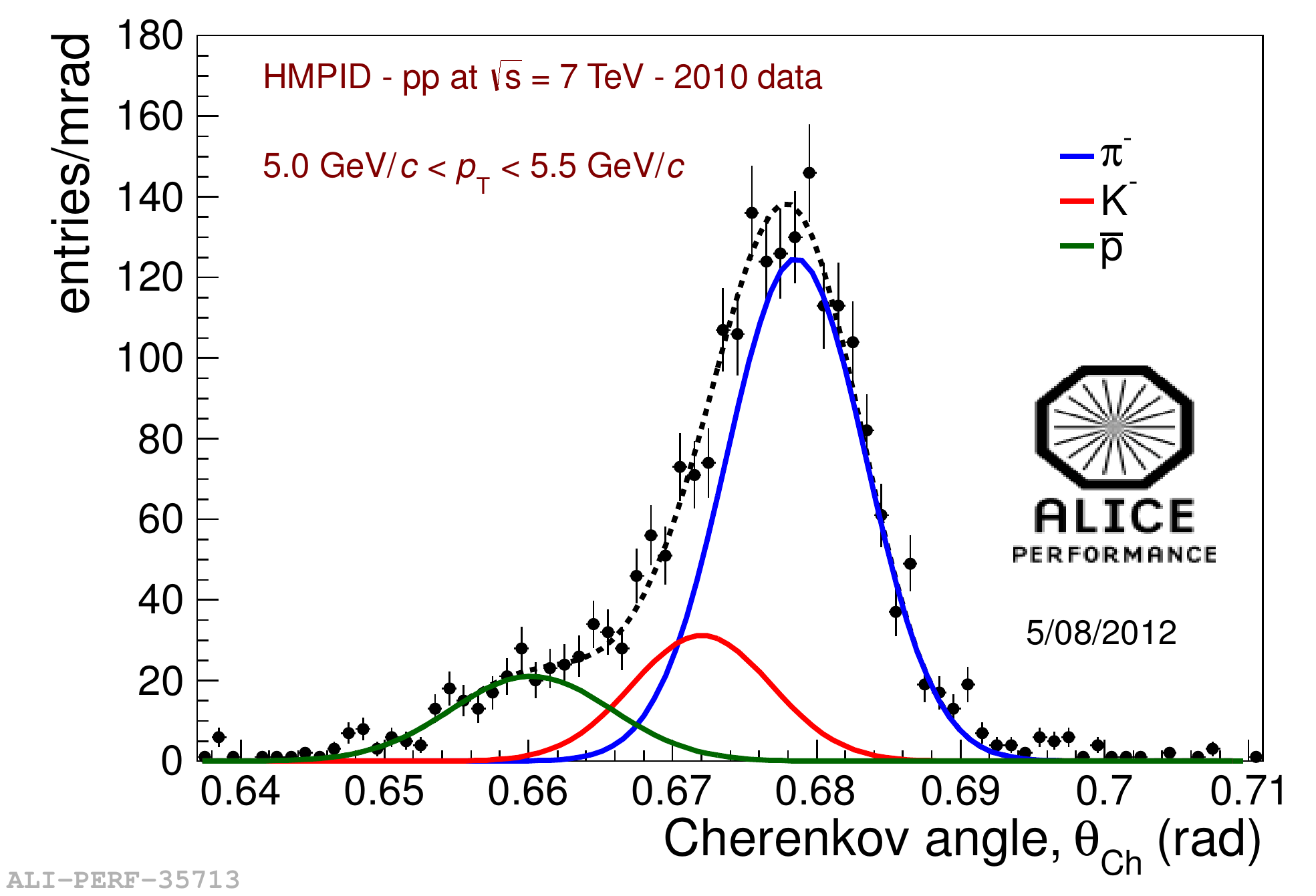}
%not approved
%\includegraphics[width=0.47\textwidth]{./spectraPositiveINEL-eps-converted-to.pdf}
%\includegraphics[width=0.47\textwidth]{./spectraNegativeINEL-eps-converted-to.pdf}
\caption{Example of the three Gaussian fit to negative particles and identified raw particle spectra measured in the HMPID with the improved tracking for positively and negatively charged particles~\cite{BarilePhD}.}
\label{hmp:fig5}
\end{figure}
\begin{figure}[!h]
\centering
\includegraphics[width=0.47\textwidth]{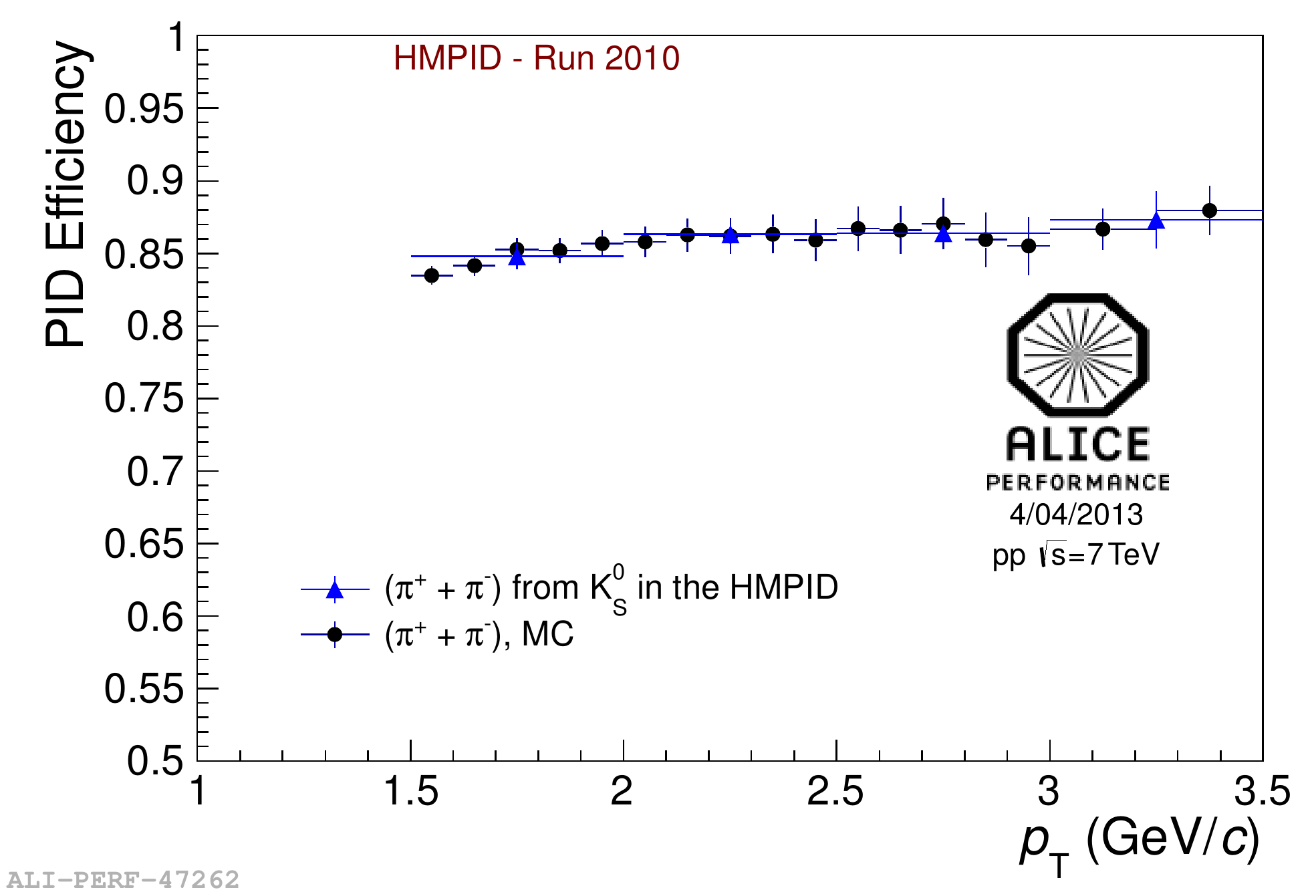}
\includegraphics[width=0.47\textwidth]{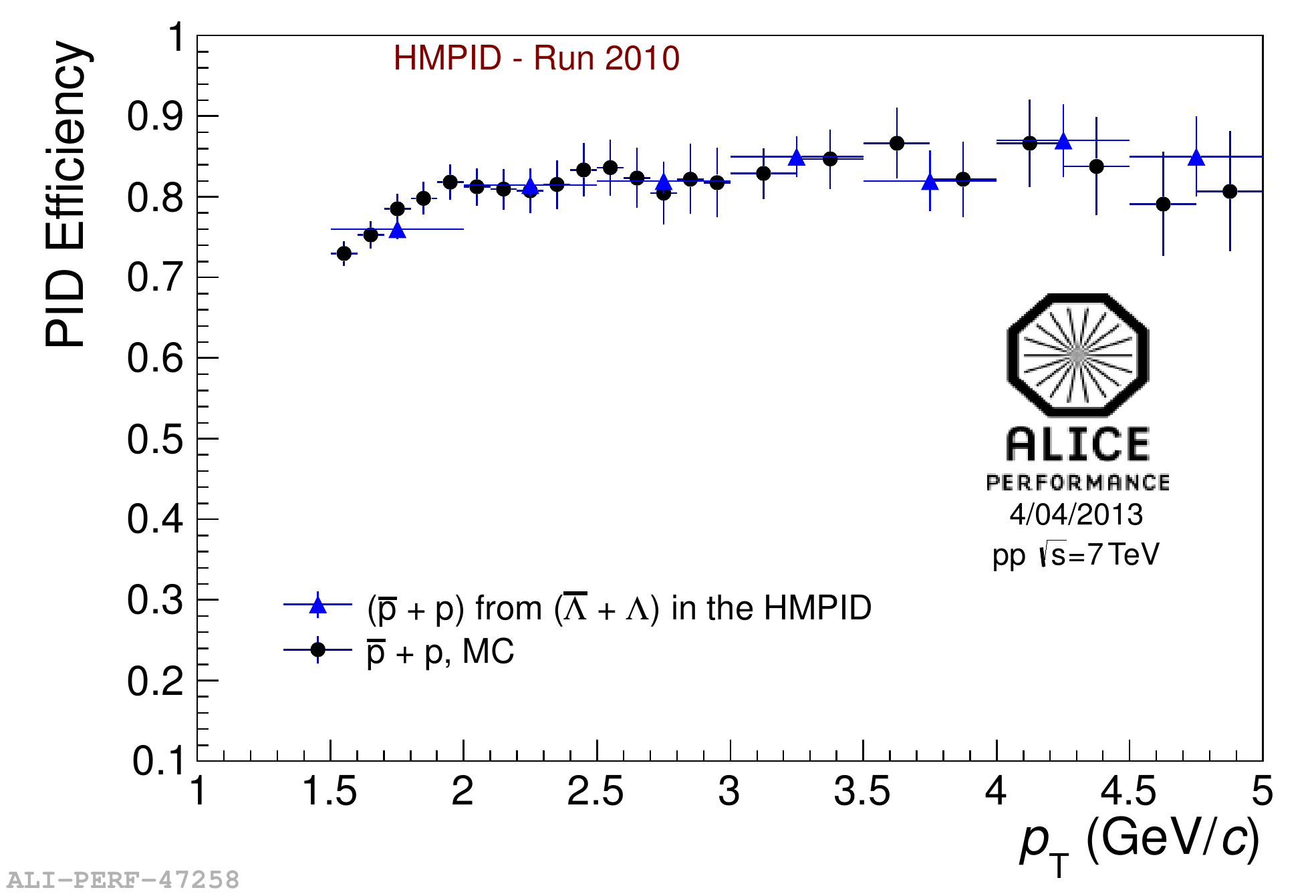}
\caption{Pion and proton identification efficiency obtained from V0 decays and MC simulations~\cite{BarileQM11,BarilePhD}.}
\label{hmp:fig7}
\end{figure}
%%%%%%%%%%%%%%%%%%%%%%%%%%%%%%%%%%%%%%%%%%%%%%%%%%%%%%%%%%%%%%%%

\subsubsection{PID results}

The HMPID detector allows to perform particle identification either on a track-by-track basis or on a statistical basis. For track-by-track identification it is required to have at least 3$\sigma$ separation between the particle species. PID in HMPID relies on the correlation between the reconstructed Cherenkov angle and the track momentum, as shown in Fig.~\ref{hmp:fig4}. The bands are well separated and show good agreement with the expected theoretical Cherenkov angle.

Particle yields can also be extracted by means of statistical unfolding from a three Gaussian fit of the Cherenkov angle distribution in a narrow momentum range. Fig.~\ref{hmp:fig5} shows an example of this fit in pp collisions at $\sqrt{s} =$ 7.0 TeV with the improved tracking procedure. 
% MW: removed this sentence, since the figure is not approved
% and the fully corrected spectra of p$/\bar{{\textrm p}}$, $K^{+}/K^{-}$, and $\pi^{+}/\pi^{-}$ in the momentum region where a PID separation with more than 3$\sigma$ is achievable. 
The VHMPID upgrade (with a radiator gas pressure of 3.5 atm) would allow to identify particles on a track-by-track basis starting from 2 GeV/c for pions and 5 GeV/c for protons. 

\begin{figure}[!h]
\centering
%\includegraphics[width=0.5\textwidth,height=10cm]{./HMPID_rTPC_090213.pdf}
%\caption{The p$\bar{{\textrm p}}$ spectra from the ITS,TPC,TOF, and HMPID ($3$~GeV/c $< p_T < 6$~GeV/c) combined and from the relativistic dE/dx method of the TPC.}\label{hmp:fig8}
\includegraphics[width=0.5\textwidth,height=10cm]{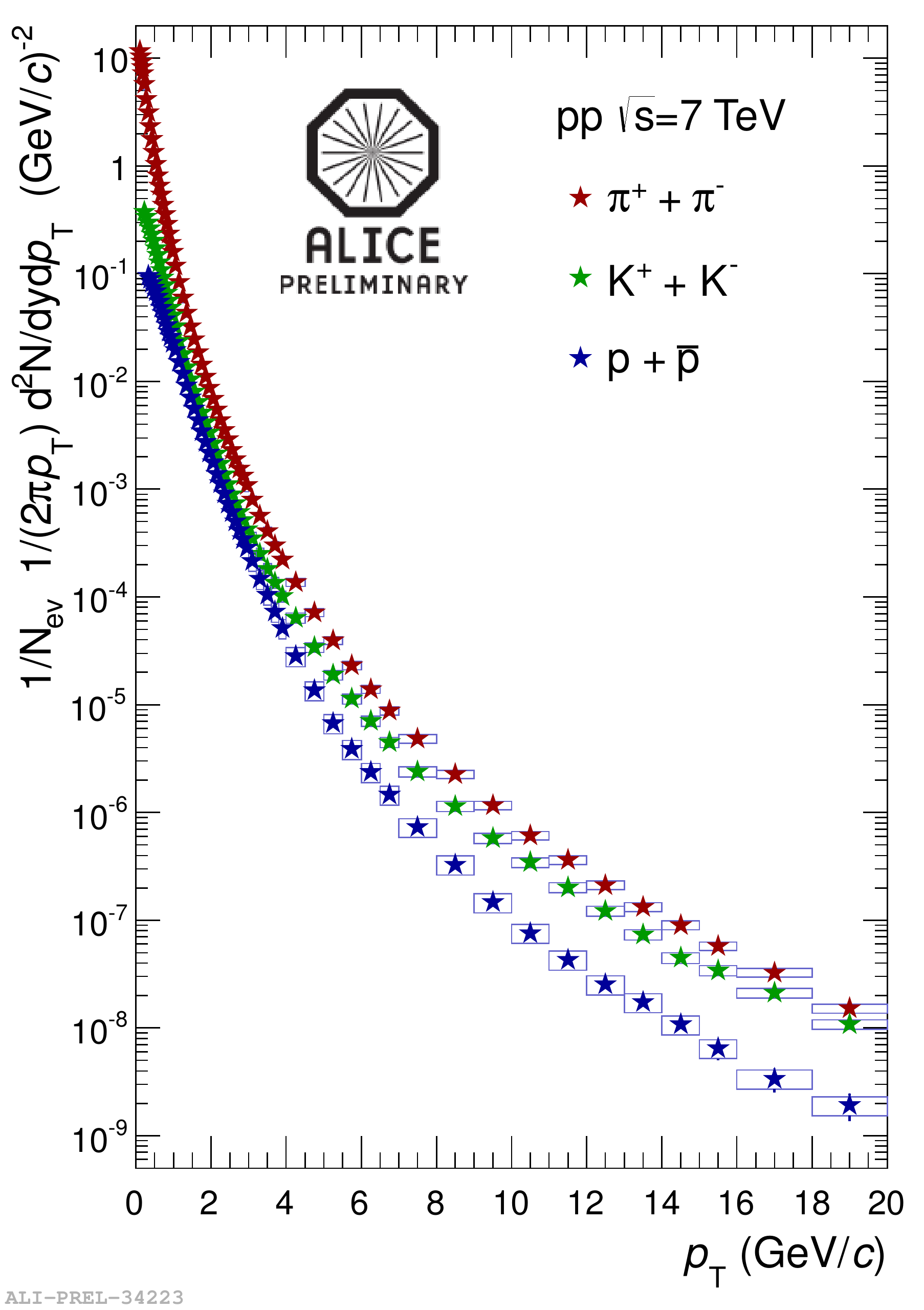}
\caption{Transverse momentum spectra for (anti-)protons, charged pions and kaons in pp collisions at $\sqrt{s} = 7.0$ TeV. Low momenta are from the combined spectra: TPC-ITS-TOF and HMPID ($3$~GeV/c $< p_\mathrm{T} < 6$~GeV/c). High momenta ($ p_\mathrm{T} > 3$~GeV/c) from the relativistic dE/dx method of the TPC. Statistical and systematic uncertainties are shown as error bars and grey areas, respectively.}\label{hmp:fig8}
\end{figure}

Proton and pion identification efficiencies have been evaluated from real data, exploiting V0 decays that allow to select a clean sample of protons and pions, and from MC simulations. As seen in Fig.~\ref{hmp:fig7}, both methods are in good agreement in the whole momentum region. The kaon efficiency has been extracted from MC simulations only, as well as the acceptance correction, tracking efficiency and anti-proton absorption.
Charged hadron spectra and ratios ($\overline{p}$/$p$, $K$/$\pi$, $p$/$\pi$) and their corresponding
systematic errors have been obtained and combined with those from the other ALICE PID detectors, using $\sqrt{s} = 7.0$ TeV pp data
reconstructed with the improved tracking procedure. As an example, Fig.~\ref{hmp:fig8} shows transverse momentum spectra for (anti-)protons, charged pions and kaons using the combination of the ITS, TPC, TOF, and HMPID information as well as the relativistic dE/dx method of the TPC.

%%%%%%%%%%%%%%%%%%%%%%%%%%%%%%%%%%%%%%%%%%%%%%%%%%%%%%%%%%%%%%%%%%%%%%%%%%%%%%%
\newpage
\section{Physics Performance Studies}
\label{PhysicsPerformance}

The combined VHMPID/Calorimeter coverage will, for the first time in relativistic heavy ion collisions, allow full jet reconstruction with particle identification back to back with the existing EMCal. The main measurement which will benefit from such an integrated detector system is the determination of particle identified fragmentation functions in heavy ion and proton-proton collisions, since the jet energy can be determined unambiguously and thus leads to a strict analysis of the fractional momentum. Furthermore the proposed detector combination enables us to do actual jet-jet correlation measurements with particle identification. Since the VHMPID physics goals depend strongly on the proposed detector setup and {\sl vice versa}, we present Monte Carlo generated simulations and theoretical calculations in order to estimate yields to be measured by the VHMPID. Our analysis was focused on two main directions:
\begin{itemize}
\item
The integrated physics properties based on identified single particle hadron and resonance yields  in pp and Pb-Pb collisions (see sections~\ref{pp-yields}, \ref{pbpb-yields}  and \ref{Ch-ratios}.)

\item
Hadon-hadron, and jet-hadron capability studies including intra-jet and inter-jet studies. We present results on hadron-hadron and jet-hadron correlations with the combined VHMPID/DCal detector in conjunction with EMCal and TPC measurements. (see section~\ref{sec:corr}.)
\end{itemize}

For our analysis we used the VHMPID described in section~\ref{sec:det-layout} and \ref{Section:Detector_performance}. The proposed pressurized Cherenkov radiator allows us to identify particles within a wide range of momenta, as plotted on Table~\ref{tab:pid35atm}.With these capabilities the VHMPID directly extends the track-by-track identification capabilities of the existing HMPID detector, which is sensitive up to $p_T$ $<$ 6 GeV/c. All simulations assume the full final VHMPID coverage, i.e. stages 1-3 of the staged implementation approach. The addition of stage 3 does not affect the simulations much except to allow for a fully symmetric coverage, which helps slightly to reduce the systematic uncertainty in the jet-jet and jet-hadron correlations. The relevance of the addition of stage 2 is highlighted in section~\ref{sec:corr}, where we show that certain correlation measurements, in particular the ones that require jet reconstruction in the VHMPID acceptance, are not possible without stage 2. It is also worthwhile pointing out that all simulations assume fully integrated DCal modules backing the VHMPID modules across the full VHMPID coverage, which is of relevance again mostly to the jet based results in section~\ref{sec:corr} such as the determination of the medium modification of particle identified fragmentation functions.

%%%%%%%%%%%%%%%%%%%%%%%%%%%%%%%%%%%%%%%%%%%%%%%%%%%%%%%%%%%%%%%%%%%%%%%%%%%%%%%%%%
\subsection{High-$p_T$ physics in proton-proton collisions}
\label{pp-yields}

The relevance of high momentum identified particle spectra measured in elementary hadron collisions has been documented in detail in several publications related to NLO and fragmentation function (FF) calculations~\cite{Hirai:2010,Albino:2010}. Recent claims by CDF that the charged hadron spectrum at very high-$p_T$ might show a violation of factorization~\cite{Aaltonen:2009} is only the latest evidence that high momentum particle fragmentation is not well constrained, neither experimentally, nor theoretically~\cite{Albino:2010_2}.

Hadron identification is very important for global fits of FFs. Any hadron-hadron data for $p_T > 2$ GeV/c is suitable for fitting. Although $e^+e^-$ data are very precise, they do not constrain the gluon FF very well, and do not constrain the differences between positively- and negatively-charged hadron production (valence FFs) nor some other differences between quark FFs (non-singlets). On the other hand, hadron- (and charge-sign) identified hadron-hadron data do constrain all these FFs components, and thus complement any hadron-identified $e^+e^-$ data. Inclusion of identified spectra in fits will also provide competitive constraints on the strong coupling constant as in Ref.~\cite{Albino:2010}. Fits to unidentified hadron data are also performed but these data are often contaminated by particles other than those of interest, such as electrons.

RHIC pp reaction data for all particles and Tevatron pp reaction data from the CDF collaboration for $K$s and $\Lambda/\bar{\Lambda}$ have recently been included in FF parameterizations to improve the constraints on the gluon fragmentation, the quark flavour separation and also, in the case of the RHIC data, to determine the charge-sign asymmetry FFs. Hadron mass effects were included in the calculation of these data as well~\cite{Albino:2008a}.
Disagreements between FNAL and RHIC measurements with theoretical predictions at high momentum were attributed to the fact that at higher collision energies (FNAL, LHC) perturbation theory eventually fails, if the fraction of available momentum $z = 2p_T \cosh (y /\sqrt{s})$ taken away by the produced hadron becomes too low.
Still, the uncertainty band for the NLO calculations is very large in the comparison to experimental data, which is due to the fact that even the most sophisticated fragmentation function sets, be it DSS~\cite{Florian:2007}, AKK~\cite{Albino:2008b}, or HKNS~\cite{Hirai:2007} suffer from a lack of particle identified data, in particular, baryonic high-$p_T$ spectra. Although the pion spectra are well constrained, neither the kaon nor the proton or lambda spectra are measured to high accuracy. Within the model the pion, kaon and proton FF are added to yield the charged particle spectrum, but the proton and kaon spectra are constrained only through measurements from V0's (Lambdas and $K$s out to $p_T \approx 10$ GeV/c). Recent measurements using relativistic rise (relativistic dE/dx) in STAR have not been sufficiently reliable to extract more precise values~\cite{Xu:2009}.

\begin{figure}[!h]
\centering
    \includegraphics[width=\textwidth]{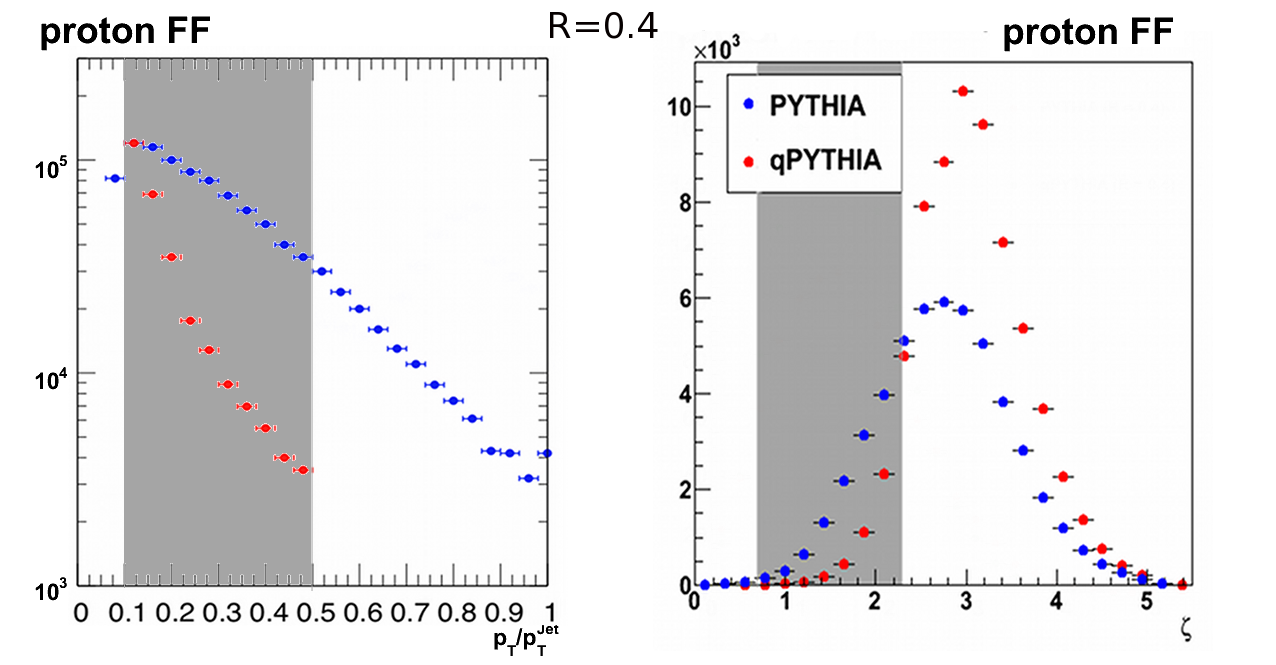}
    %\vspace{-1.0cm}
    \caption{Proton fragmentation function measurements in the VHMPID based on 50-60 GeV PYTHIA and qPYTHIA jets. Comparison as function of a.) the fractional momentum z, b.) the inverse quantity $\xi$ = ln (1/z). The shaded areas show the z and $\xi$ ranges where the VHMPID measures the spectrum.}
    \label{pp-p-ff}%
\end{figure}

The capabilities of the VHMPID for this type of measurement have been simulated by reconstructing the proton fragmentation function based on the expected yearly rate of inclusive jets with a jet energy of 50-60 GeV. Fig.~\ref{pp-p-ff}(left) shows a comparison of the fractional momentum spectrum (z=p$_{T}$/p$_{T}^{jet}$) between unmodified PYTHIA jets and medium modified jets based on qPYTHIA. Fig.~\ref{pp-p-ff}(right) shows the same comparison as a function of $\xi$= ln(1/z). The areas in which the VHMPID measurement determines the spectrum are shaded. It is apparent that the high efficiency in the particle identification in the VHMPID not only enables us to make a high precision fragmentation function measurement in proton-proton collisions, it also allows for the determination of medium modification effects on the order of a factor 1.2 or higher. This point is emphasized in section~\ref{sec:corr} where the ratio of the measured proton fragmentation functions in pp and PbPb collisions is shown with realistic statistical and systematic error bars applied (Fig.~\ref{fig:PID-FF}).
%
%%%%%%%%%%%%%%%%%%%%%%%%%%%%%%%%%%%%%
%%%%%%%%%%%%%%%%%%%%%%%%%%%%%%%%%%%%%
%

In summary, at present, elementary proton reaction data are crucial for the extraction of the poorly constrained gluon and valence quark FFs, and therefore future accurate measurements from the LHC will be most welcome. The already achieved high accuracy in inclusive cross section measurements at the LHC show that the constraints on as (and FFs) from a single experiment measuring identified charged particles will be highly significant.

%%%%%%%%%%%%%%%%%%%%%%%%%%%%%%%%%%%%%%%%%%%%%%%%%%%%
\subsection{High momentum particle rates in the VHMPID in PbPb collisions}
\label{pbpb-yields}

\subsubsection{High momentum hadron yields in heavy ion collisions}

%\vspace{-2.cm}
\begin{figure}[htb]
 \centering
\includegraphics[width=0.47\textwidth,height=9cm]{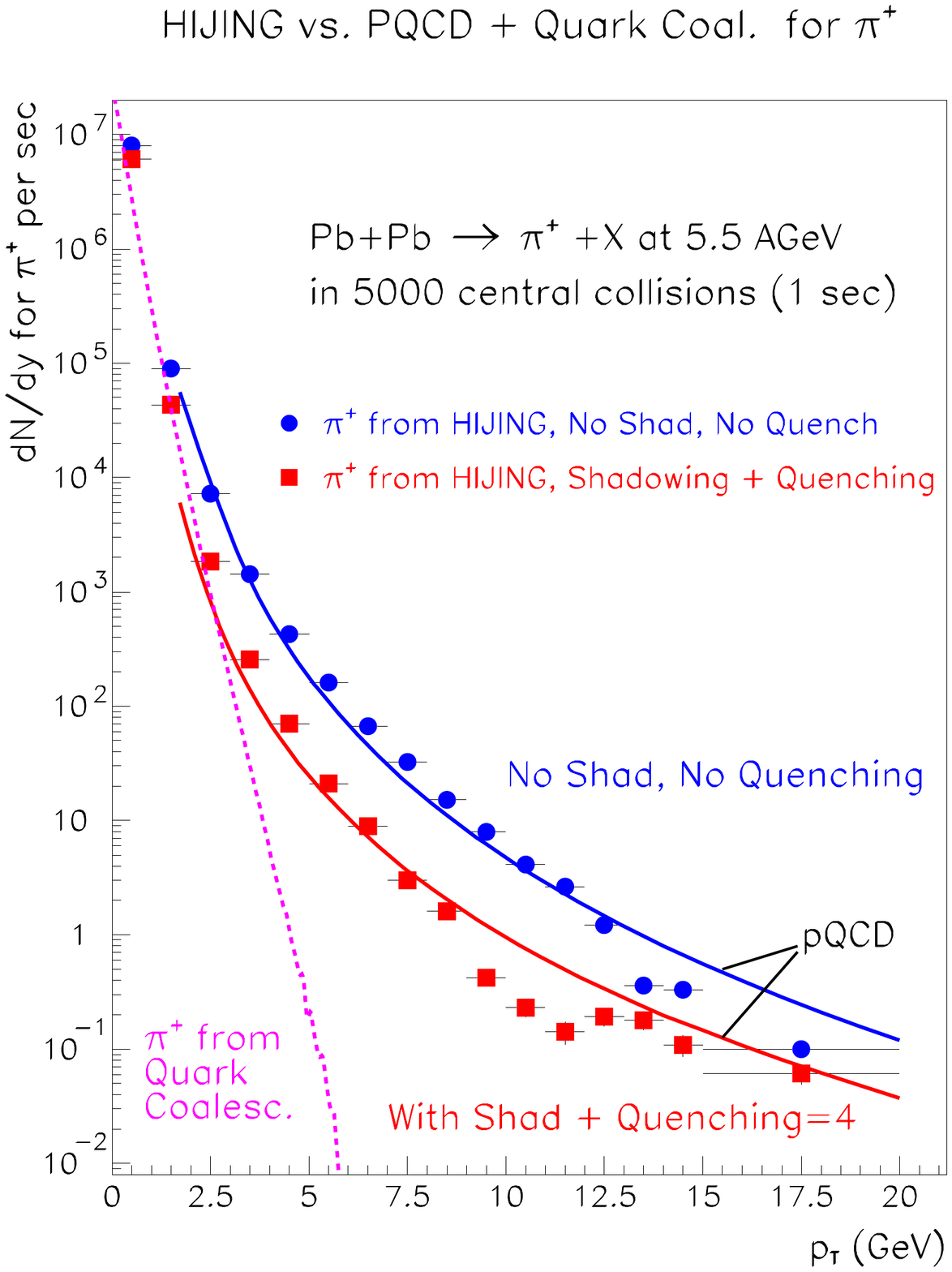}
\includegraphics[width=0.47\textwidth,height=10cm]{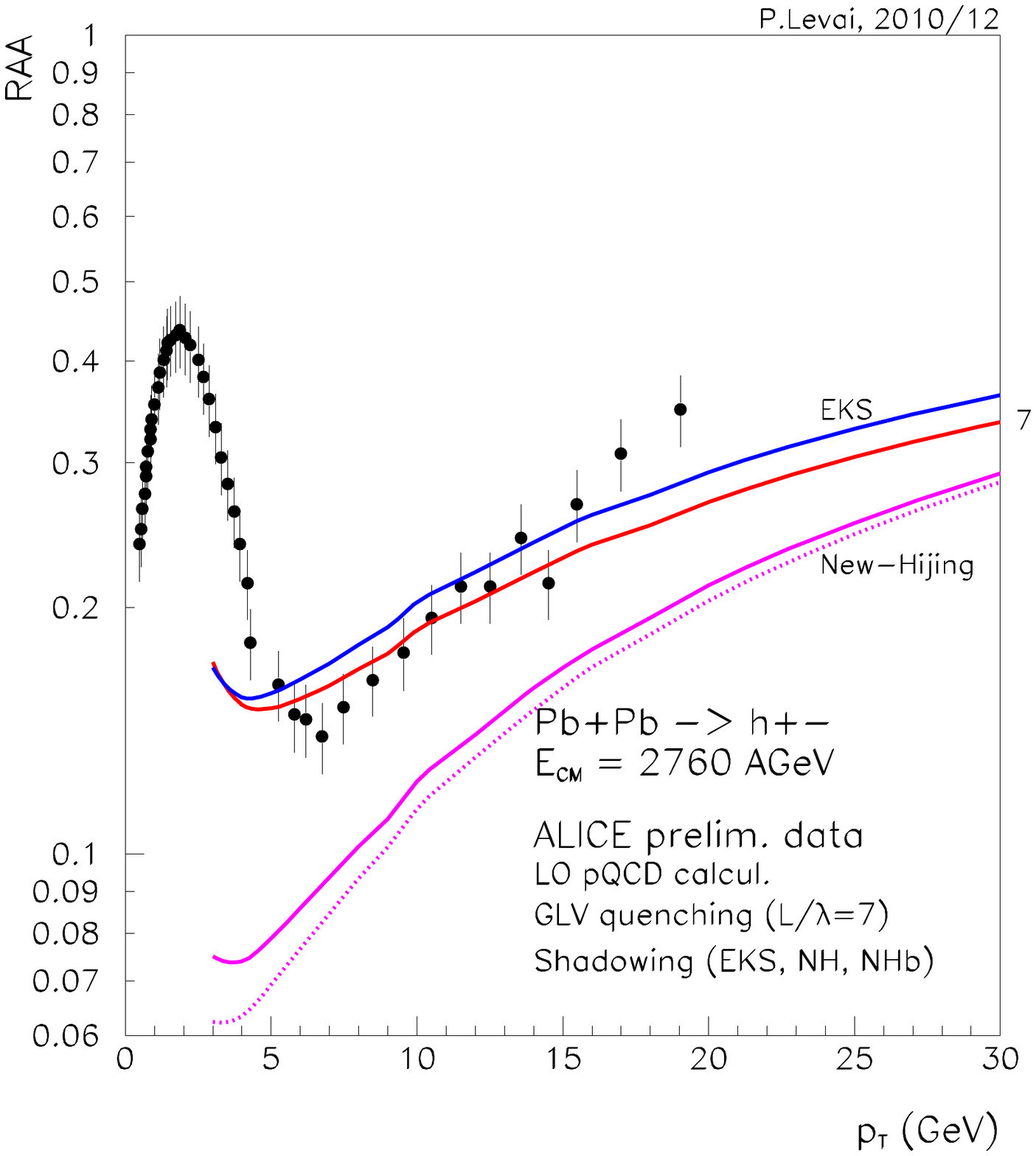}
\caption{Left panel: Comparison of pQCD results and HIJING simulation for $\pi^+$ production in central Pb-Pb collisions at $\sqrt{s} = 5.5$ $A$TeV for a 5 Khz readout rate for central events ~\cite{Levai:2008zz}. Right panel: The measured nuclear modification factor at $\sqrt{s}= 2.76 \, A$TeV, $R^{h^{\pm}}_{PbPb}(p_T)$ based on Ref.~\cite{LevaiIndia} using data from Ref~\cite{ALICE-data2011}.}
\label{hijingpb10000}
\end{figure}
%%%%%%%%%%%%%%%%%%%%%%%%%%%%%%%%%%%%%

The integrated properties of hadron production in nucleus-nucleus collisions have been widely studied in the past, especially through nuclear suppression factor measurements, jet identification and jet-medium correlation spectra. At RHIC and LHC energies these investigations are connected to hard probes of the quark-gluon plasma (QGP). RHIC measurements reached the level of performance to obtain high-statistics data on unidentified charged hadrons (using TPCs) and neutral pions (using calorimeters), but identified charged hadron measurements are still missing for the intermediate and high transverse momentum regimes. As it was pointed out by the latest RHIC, ALICE and CMS measurements, Refs.~\cite{RHIC_QM2011,ALICE_QM2011,CMS-QM2011} hadron species exhibit specific differences in their high momentum properties. To understand the origin of these deviations requires event-by-event  PID measurements, which the VHMPID can perform.

In this section we estimate the expected yields, and number of hadron events measured in the VHMPID modules. We use two methods: a Monte Carlo-based hadron generator (HIJING~\cite{hijing}) and a perturbative QCD-based parton model calculation~\cite{YiZhang:2001}. These models are generally in good agreement with earlier experimental data however, theoretical estimates for high-$p_T$ hadron yields in Pb-Pb collisions contain uncertainties especially in the lower-intermediate momentum region of the VHMPID  ($6$ GeV/c $< p_T <  25$ GeV/c). The agreement between the pQCD model and the HIJING generated spectra can be seen on the {\sl left panel} of Fig.~\ref{hijingpb10000} for central Pb-Pb collisions at 5.5 $A$TeV center of mass energy. Points are for $\pi^+$ spectra calculated from HIJING, and lines are for the obtained pQCD results. Another {\sl dashed line} also plotted based on the coalescence model~\cite{LevaiIndia} for the comparison below the pQCD regime ($p_T\leq 1-2$ GeV/c). The pQCD and HIJING calculations had the same parameter settings for the geometry and we used the HIJING shadowing~\cite{hijing}, and GLV jet quenching  at the final state~\cite{GLV}. We normalized the data for the yields assuming 5 kHz = 5000 event/s central collisions to tape rate. This is the nominal number for high luminosity running in case the high level trigger will not provide sufficient information to reduce the ALICE per event data volume. Should the HLT tracking become operational the rate figures shown here can simply be increased by an order of magnitude to reflect the full central collisions interaction rate of 50 kHz.

In order to scale the plot to the anticipated yearly rate in the VHMPID one has to multiply the dN/dy with the duration of a Pb-Pb run at $5.5 \, A$TeV which is equivalent to $1.2 \times 10^6$ seconds, and the proper VHMPID coverage (stages 1 and 2 of the VHMPID integration = 24\% of the TPC acceptance) . 

\begin{table}[h!]
\centering
  \begin{tabular}{lcccc}
\hline
$p_T$ range & Quenching & $N_{\pi+}$ & $N_{K^+}$ & $N_{p^+}$ \\  \hline \hline
 & \sc{no} & 240/s  &  90/s  & 180/s \\
$[5 \ \rm{GeV/c} : 9 \ \rm{GeV/c}]$ & & & &  \\
 & \sc{yes} & 60/s & 24/s & 66/s \\
\hline
 & \sc{no} &  30/s  &   12/s  & 18/s \\
$[9 \ \rm{GeV/c} : 25 \ \rm{GeV/c}]$ & & & &  \\
 & \sc{yes} & 6/s & 3/s & 4/s \\
\hline \hline
\end{tabular}
\caption{Event yields for identified charged particles: $ \pi^+ $, $ K^+ $, and $ p^+ $  within the VHMPID geometry in the given transverse momentum ($p_T$) ranges for $0-10\%$ central Pb-Pb collisions and at $\sqrt{s}= 5.5 A$TeV.}
\label{tab:PID-PbPb-yields}
\end{table}

Table~\ref{tab:PID-PbPb-yields} shows a breakdown of all anticipated particle-identified rates at 5 kHz central collisions. The calculated results were based on the HIJING generator: without jet quenching and with jet quenching denoted by "{\sc no}" and "{\sc yes}" respectively. The quenching leads to a high momentum yield suppression of about $\sim 4-5$, similarly as predicted by the GLV method in Refs.~\cite{GLV, BGG}. To put these numbers in perspective, we can expect around $20,000$ protons above 20 GeV/c in a yearly Pb-Pb run from central collisions at a 5 kHz readout rate.

Although the VHMPID coverage is only about $1/4$ that of the TPC, the yearly proton yield in the $5-25$ GeV/c range is roughly a factor three higher than in the TPC due to the stringent cuts required in the TPC analysis in order to isolate the protons in the relativistic
dE/dx spectrum. In summary, the VHMPID is capable to measure the identified hadron yields at excellent level, even in case of the strong suppression of the yields.

%%%%%%%%%%%%%%%%%%%%%%%%%%%%%%%%%%%%%%%%%%%%%%%%%%%%%%%%%%%%%%%%%%%%%%%%%%%%%%%
\subsubsection{High-momentum hadronic resonance production in heavy ion collisions}

In addition to the identification of ground state hadrons, the VHMPID also enables the reconstruction of high momentum hadronic resonances. Since many of these excited states decay in the medium they have been suggested as a signature for chiral restoration as well as a measure of medium properties such as the medium lifetime and the modification of the fragmentation process in the medium.

\begin{figure}[!h]
\centering
\includegraphics[width=0.6\textwidth]{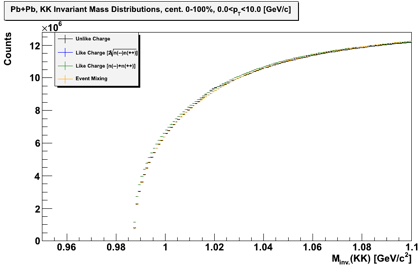}
\includegraphics[width=0.38\textwidth,height=6.2cm]{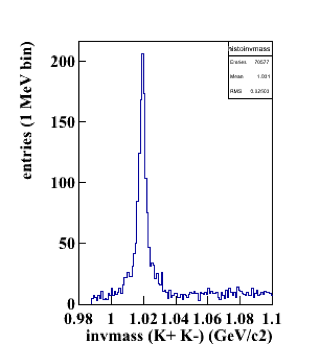}
\caption{Left panel: Invariant mass distribution for $K^+K^-$ pairs in min. bias Pb-Pb collisons at 5.5 $A$TeV. Right: reconstructed $\phi$-meson peak in VHMPID
acceptance based on 5 Million central HIJING events (unquenched).}
\label{fig:resonance-tpc}
\end{figure}

The VHMPID adds a unique aspect to these studies since it allows the reconstruction of very high momentum resonances through simultaneous detection of both decay daughters. As an example we have studied the decay of the $\phi$-meson into $K^+K^-$ pairs. {\sl Left panel} of Fig.~\ref{fig:resonance-tpc} shows the present status of the $\phi$-meson reconstruction in the Pb-Pb year-1 data based on 3.6 Million analyzed minimum bias events recorded with the TPC. The resulting signal-to-background ratio is 0.02 and although a reliable signal can be extracted, the association of any actual $\phi$-meson to a jet or even a particular particle correlation function will fail. Thus, medium modifications to the fragmentation or the level of chirality can only be studied on a statistical basis with large statistical and systematic uncertainties. Resonance extraction on a jet by jet basis will only be possible
with the VHMPID.

The {\sl right panel} of Fig.~\ref{fig:resonance-tpc} shows a simulation of the $\phi$-meson peak reconstructed in the VHMPID from identified kaon pairs with both kaons above the $p_T > 5$ Gev/c threshold. This result is based on 5 Million unquenched central Pb-Pb HIJING events. The signal to background ratio is $12:1$, i.e. a 500-fold improvement to the low momentum measurement in the TPC. A good portion of this improvement stems from the tight $p_T$-cut on the decay daughters and the relative background free environment at high momentum. In order to better assess the improvement factor we ran a simulation of the aforementioned HIJING high $P_T$ $\phi$-meson sample without PID cuts. In this case the S/B ratio is $1.6:1$, so we still anticipate an order of magnitude improvement when using the VHMPID. In addition the unambiguous identification of the decay daughters not only improves the S/B ratio it also allows the correlation of these resonances with jets or other hadrons (see later in section~\ref{sec:corr}).
In terms of anticipated rate, the number of high-$p_T$ $\phi$'s (both kaons above 5 GeV/c) reconstructed in the VHMPID is about 4\% of the number in the TPC. Based on unquenched HIJING this corresponds to about 0.0002 reconstructed high-$p_T$ $\phi$-mesons per central Pb-Pb event or about 1/s assuming an 5 kHz central event rate a level-0. Thus the measurement of these resonance in the VHMPID is not statistics limited.

%%%%%%%%%%%%%%%%%%%%%%%%%%%%%%%%%%%%%%%%%%%%%%%%%%%%%%%%%%%%%%%%%%%%%%%%%%%%%%%%%%%%%%%%%%%%%%%%%%
\subsection{Ratios of identified particles at high-$p_T$}
\label{Ch-ratios}

\begin{figure}[!h]
\centering
\includegraphics[width=0.75\textwidth]{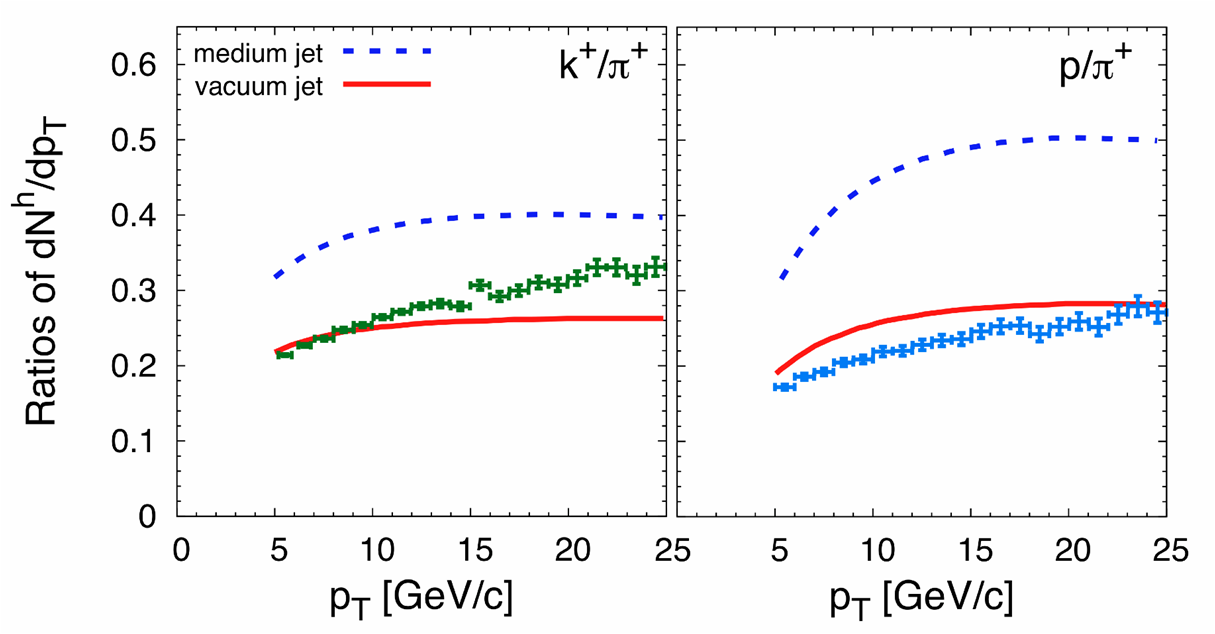}
\caption{Calculated $K^+/\pi^+$ ({\sl left panel}) and $p^+/\pi^+$ ({\sl right panel}) ratios based on PYTHIA calculations
using the VHMPID efficiencies and systematic uncertainties (based on the annual yield of triggered 50 GeV jets) compared to theoretical pQCD predictions for vaccum and medium jets ~\cite{Sapeta:2007ad}.}
\label{fig:PID-ratios}
\end{figure}

Measuring identified particle ratios based on the spectra shown in the previous section is a unique way to test medium modifications of the fragmentation process, as was suggested by e.g. Ref.~\cite{Sapeta:2007ad}. Differences between the production of mesons/baryons, particles/anti-particles, and/or strange and heavy-flavor content hadrons are predicted for higher energies and low-$x$ values. Predictions differ strongly, though, and the need for precise track-by-track measurements is obvious. In Fig.~\ref{fig:PID-ratios}  we show the prediction by ~\cite{Sapeta:2007ad} for a specific identified particle ratio (in this case p/$\pi$) in reconstructed jets with 100 GeV jet energy. This measurement is a perfect example for the strength of the VHMPID/Cal combination in the ALICE central barrel. The jet first is fully reconstructed in the calorimeter, which leads a reliable determination of the jet energy. In step 2 the VHMPID determines the particle species of all high momentum hadrons in the same jet cone track-by-track. The anticipated statistical and systematic error bars for the jet energy and the particle ratio are shown in the data points in the figure. The particle yields were based on a non-quenched PYTHIA simulation. As shown the VHMPID can resolve hadron ratios down to a $15-20\%$ error. The specific Sapeta\,--\,Wiedemann quenching model predicts flavor, baryon number, and hadron mass differences for particle ratios in in-medium jets due the specific nature of the gluon splitting function used to describe the partonic energy loss. Thus, these measurements are relevant to determine the energy loss mechanism in medium and in particular specific state properties such as the path-length dependence and the transport coefficient. This study will require track-by-track particle identification and a dedicated jet trigger since the ratio modification is also dependent on the initial parton momentum (i.e. the measured jet energy). 

%%%%%%%%%%%%%%%%%%%%%%%%%%%%%%%%%%%%%%%%%%%%%%%%%%%%%
\subsection{High momentum particle correlations}
\label{sec:corr}

The following sections detail our studies of jet-jet, jet-hadron and di-hadron correlations. For all cases a statistics estimate is given.  Di-hadron correlation statistics were determined for intra-jet (both particles in the VHMPID) and inter-jet correlations (one particle in the VHMPID, one particle in the TPC). For the case of correlation measurements the staged implementation of the VHMPID plays a significant role. Di-hadron and jet-hadron measurements can be performed with stage 1 (VHMPID/DCal) alone. Jet-jet correlations require at least stage 2 (doubling of the VHMPID/Cal acceptance) and ideally stage 3 (symmetrizing of the acceptance through
inclusion of PHOS region).

%%%%%%%%%%%%%%%%%%%%%%%%%%%%%%%%%%%%%%%%%%%%%%%%%%
\subsubsection{Jet-hadron and jet-jet correlations}

We have studied the probability of measuring jet-jet and jet-hadron correlation spectra by triggering on reconstructed jets in the EMCal in central PbPb collisions. The EMCal is located  back to back (i.e. 180 degree in $\phi$ and similar $\eta$-coverage) with the proposed VHMPID/Cal device. We use the di-jet rates published in the EMCal Physics Performance Report (PPR) \cite{EMCal-PPR} and the DCal TDR and scaled by the anticipated high luminosity interaction rate to estimate the statistics for jet triggered measurements with the VHMPID. 

For the jet-jet studies we have assumed the full coverage of stages 1 and 2 of the VHMPID implementation which leads to around 25\% coverage of the TPC volume. About half of all EMCal triggered di-jets have their partner reconstructed in the combined acceptance of stage 1 and stage 2. Jets with cone size up to $R=0.7$ are contained in the VHMPID/Cal although the gap in the middle of the cone due to the PHOS leads to about a $30\%$ systematic error in the jet energy determination. This error would be significantly reduced in stage 3 of the VHMPID implementation (down to about $5\%$ ). 

For jet-jet correlations we use as an example the number of EMCal triggered di-jets with 100 GeV jet energy each. Based on our simulations we expect around $2.5 \times 10^{6}$ di-jets in the VHMPID coverage assuming a 100\% efficient jet trigger.

In order to perform jet-hadron correlations one has to determine the track identification probability. Around $70\%$ of the jets identified in the calorimeter behind the VHMPID have a reconstructed leading particle within the acceptance of the VHMPID including detector gaps. On average the leading particle carries about $ 20\% $ of the jet energy. Thus, most jets in the $40-140$ GeV energy range have a leading particle reconstructed in the VHMPID.

The jet-hadron correlations enable an unambiguous measurement of particle identified fragmentation functions
as long as the jet energy can be determined with good accuracy on the EMCal side, either through a combination of neutral (EMCal) and charged (TPC) energy or through the measurement of a photon on the EMCal side.

\begin{figure}[!h]
\centering
\includegraphics[width=0.5\textwidth]{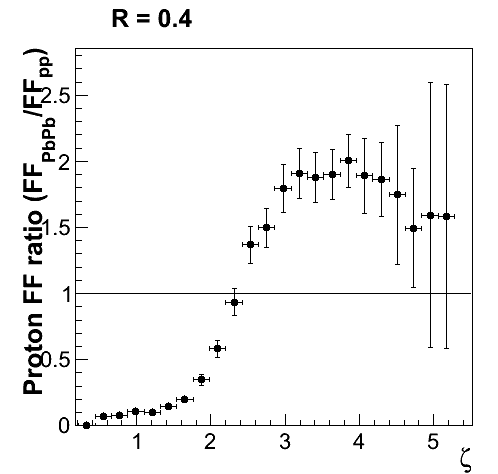}
\caption{Ratio of proton fragmentation functions simulated for pp and PbPb collisions for 50 GeV jets using the PID uncertainties for TPC, TOF and VHMPID and the jet energy uncertainties for the EMCal. Jets are reconstructed with a cone algorithm using $R=0.4$ for the cone size. The statistics are based on one PbPb run at present luminosities. The simulation uses PYTHIA for pp and qPYTHIA embedded in HIJING for Pb-Pb. The VHMPID contributes in the region of $\xi = 0.7-2.3$, i.e. across the full suppression region.}
\label{fig:PID-FF}
\end{figure}

As an example we have simulated the accuracy to which a medium modification measurement of the proton fragmentation function can be performed. The anticipated medium effect is based on a comparison between  PYTHIA and qPYTHIA  predictions for 50 GeV jets [6, 7] as shown in Fig.~\ref{fig:PID-FF}. The VHMPID makes a significant contribution to this plot by enabling the proton identification in the range from $\xi = 0.7$ to 2.3 ($\xi = \ln(E_{jet}/p_{had}))$. In this range the measurement is clearly not statistics limited. The systematic error bars on the proton yield ($y$-axis) and the jet energy ($x$-axis) are included. 

It is worthwhile pointing out though that the limiting factor in the fragmentation function determination will always be the required correction to the jet energy (unfolding), which can only be achieved on a statistical basis. The track-by-track identification of the VHMPID still improves considerably over a statistical leading particle measurement, since the track information can be used to constrain the event class which is selected to perform the unfolding procedure. Furthermore, since the VHMPID can identify particles track-by-track, the unfolding of the identified fragmentation functions becomes similar to unfolding the charged hadron fragmentation function. This is not the case with statistical PID alone, where one has to unfold the jet energy and the  particle ID, i.e. both numerator and denominator in the fractional momentum variable.

%%%%%%%%%%%%%%%%%%%%%%%%%%%%%%%%%%%%%%%%%%
\subsubsection{Hadron-hadron correlations}
\label{sec:h-h-corr}

Charged hadron-hadron correlations have been extensively studied since the ISR and S$p\bar{p}$S experiments in the mid 70's. These measurements have provided a good tool to investigate the fragmentation process in addition to full jet reconstruction. More recently these correlation measurements have been applied in relativistic heavy-ion collisions, where full jet reconstruction is increasingly difficult, in order to determine medium effects such as back-to-back jet suppression and modifications to the fragmentation process. In addition, the discovery of long range correlation structures in pp, p-Pb and Pb-Pb collisions at the LHC is based on two particle correlations~\cite{CorrCMS:2012,CorrALICE:2012}. The various hadronization and hydrodynamical models of the space-time evolution of the hot and dense color medium which have been employed to describe these phenomena rely heavily on assumed parton and particle distributions. The hadron spectra and extracted fragmentation functions are integrated distributions which mix up all possible contributions of parton-hadron channels via the convolution of initial state, QCD, and final state properties of a high energy collision. The only way to decouple specific hadronization contributions is to test the conservation of quantum numbers by following track-by-track the baryon number and flavor specific emission patterns of hadrons from partonic medium which can be performed by the VHMPID detector. 

A VHMPID based di-hadron correlation study involves either the reconstruction of one of the charged particles in the TPC or the reconstruction of both charged particles in the VHMPID. Depending on the configuration we can study inter- as well as intra-jet correlations. The more general inter-jet correlations require a charged hadron somewhere in the TPC and a high-$p_T$ particle identified in the VHMPID. Since we can assume that both of these high-$p_T$ particles are due to jet fragmentation the expected statistics should be close to the ones we deduced for the jet-hadron correlations. We used HIJING events to estimate the rate, which, as expected, exceeds the rate for jet-hadron correlations by the acceptance difference between EMCal and TPC. We estimate around 15 $\times$ 10$^{6}$ pairs per Pb-Pb run can be reconstructed in TPC and VHMPID .

The more restrictive intra-jet correlations require two charged particles in the VHMPID acceptance above the $p_T$ threshold for reconstruction. Clearly this kind of correlation is statistics limited, but not necessarily due to the size of the VHMPID. Since the detector contains jets up to $R=0.7$ the lack of statistics is largely due to the lack of two high-$p_T$ particles in the same jet in particular for jets below $50$ GeV jet energy. We estimate that for jets with energy larger than $120$ GeV the leading and the sub-leading particle should exceed the VHMPID $p_T$-threshold and since in this case we do not require di-jets we can scale the number from the high luminosity EMCal simulations for inclusive jets, which is about $5 \times 10^{6}$ in the EMCal coverage for a jet energy of $120$ GeV. We expect around 3 Million identified di-hadron pairs in the VHMPID per year when taking into account the detection efficiency. These numbers will be slightly enhanced through jets in which the leading particle is a hadronically decaying resonant state, which in most cases leads to both decay daughters being reconstructed inside the jet cone, i.e. inside the acceptance of the VHMPID. 

\begin{figure}[!h]
    \centering
    \includegraphics[width=0.47\textwidth]{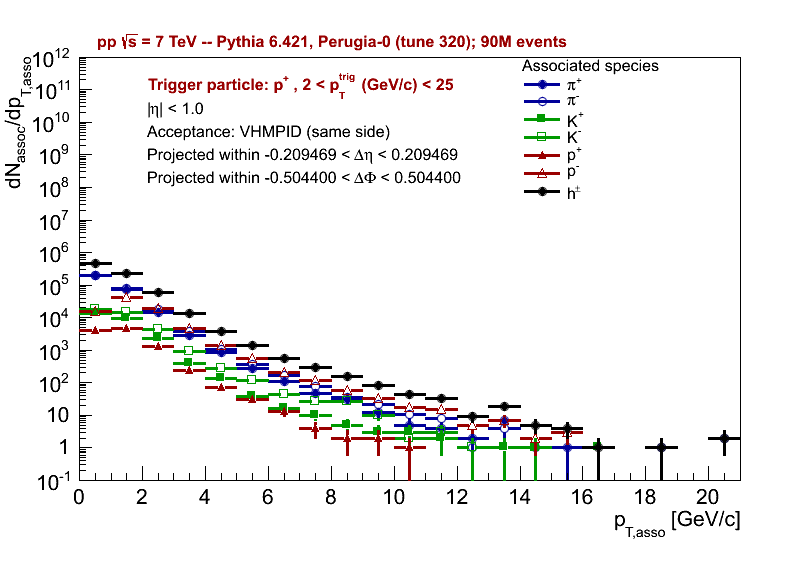}
    \includegraphics[width=0.47\textwidth]{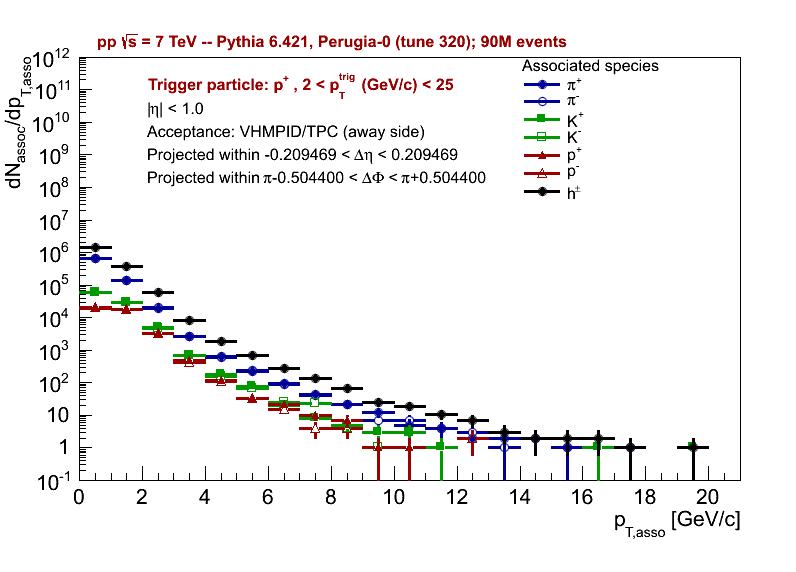}
   \includegraphics[width=0.47\textwidth]{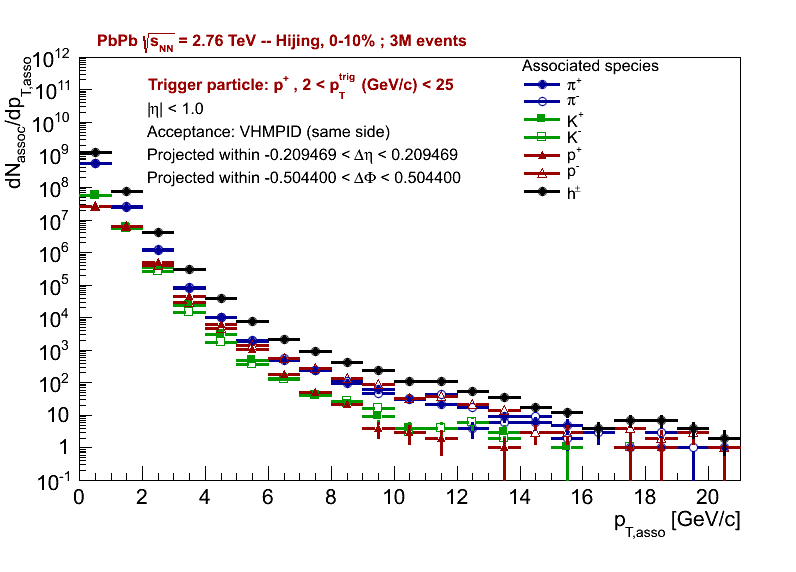}
    \includegraphics[width=0.47\textwidth]{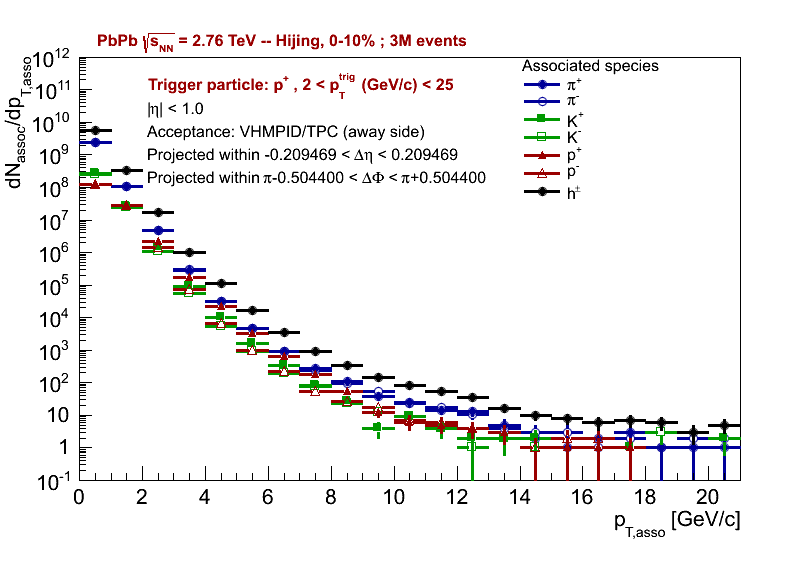}   
    \caption{PID-triggered spectra in the VHMPID acceptance for cases of PYTHIA, pp at 7 TeV ({\sl upper row}) and HIJING, $0-10\%$ central Pb-Pb at 2.76 $A$TeV center of mass energy ({\sl lower row}). {\sl Left panels} are for same side correlations in the VHMPID acceptance and {\sl right panels} show the away-side correlation in a VHMPID-sized TPC patch.}
\label{fig:PID-triggered_spectra}
\end{figure}

In Fig.~\ref{fig:PID-triggered_spectra} we show the number of associated identified particles in the VHMPID for trigger particles identified in the VHMPID ({\sl same side}) or an identical size acceptance in the TPC ({\sl away side}). The simulations show that for any particle species combination the statistics is sufficient to perform a detailed correlation analysis (based on 90 million PYTHIA and 3 million $0-10\%$ most central HIJING events including jet-quenching). The total yield per heavy-ion run ($50\%$ duty cycle) can be obtained by the multiplying the HIJING spectra with a factor $\sim 20,000$ assuming $50$ kHz event rate.  In order to calculate correlation yield in pp running per month the PYTHIA spectra need to be multiplied by $\sim 25,000$ assuming a $2$ MHz event rate.
%%%%%%%%%%%%%%%%%%%%%%%%%%%%%%%%%%%%%
%%%%%%%%%%%%%%%%%%%%%%%%%%%%%%%%%%%%%
The main correlations worthwhile studying with identified tracks are related to the hadronization mechanism, in particular in the $5-10$ GeV/c range where heavy-ion collisions exhibit significant particle dependent differences to elementary proton-proton collisions. For example, it was shown by ALICE~\cite{corr:1} that the baryon-over-meson ratio (for light and strange particles) in the intermediate momentum range is enhanced by more than a factor three in Pb-Pb compared to pp. Since this unique kinematic range separates the low momentum thermal bulk production from the pure parton fragmentation at high momentum, the anticipated production mechanism is of particular interest. Competing models of correlated (based on modified parton fragmentation, e.g. HIJING) and uncorrelated (thermal coalescence, e.g. AMPT~\cite{AMPT} or EPOS~\cite{EPOS1,EPOS2}) origin have been proposed, and only particle identified correlation measurements as a function of momentum and emission angle enable us to distinguish between different mechanisms. Furthermore the more general question of quantum number conservation during the phase transition from the partonic to the hadronic world can be addressed only through correlations beyond the simple baryon/meson ratios. Baryon--antibaryon correlations will tell us about antimatter production, strange flavor correlations will potentially exhibit flavor differences during the hadron emission from the deconfined state. To test the performance of the VHMPID on PID-triggered hadron-hadron correlations we used a sample of 90 million proton-proton collisions at 7 TeV center of mass energy, generated by AliRoot using the PYTHIA Perugia0 (tune 320). For this analysis we applied the VHMPID acceptance and geometry described in section~\ref{sec:det-layout}. The strongly correlated particle production in these elementary collisions serves as a baseline for upcoming heavy ion studies.
%%%%%%%%%%%%%%%%%%%%%%%%%%%%%%%%%%%%%
\begin{figure}[!h]
    \centering
    \includegraphics[width=0.47\textwidth]{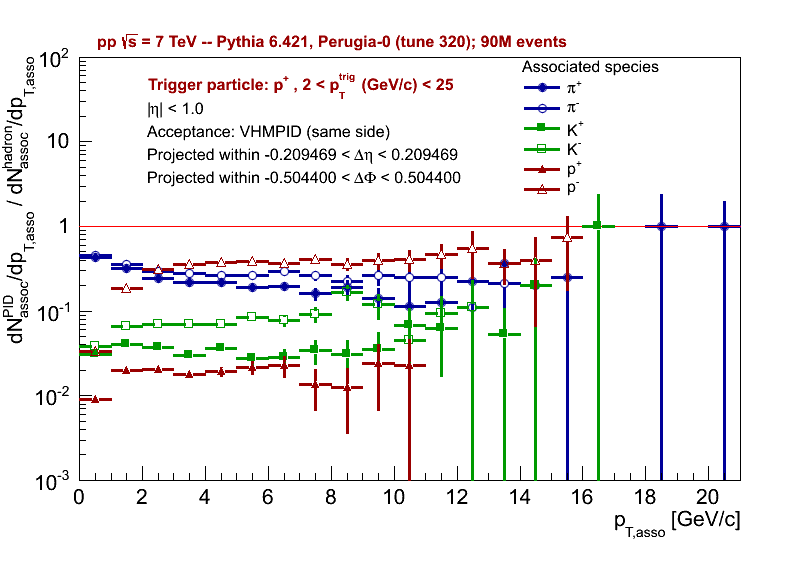}
    \includegraphics[width=0.47\textwidth]{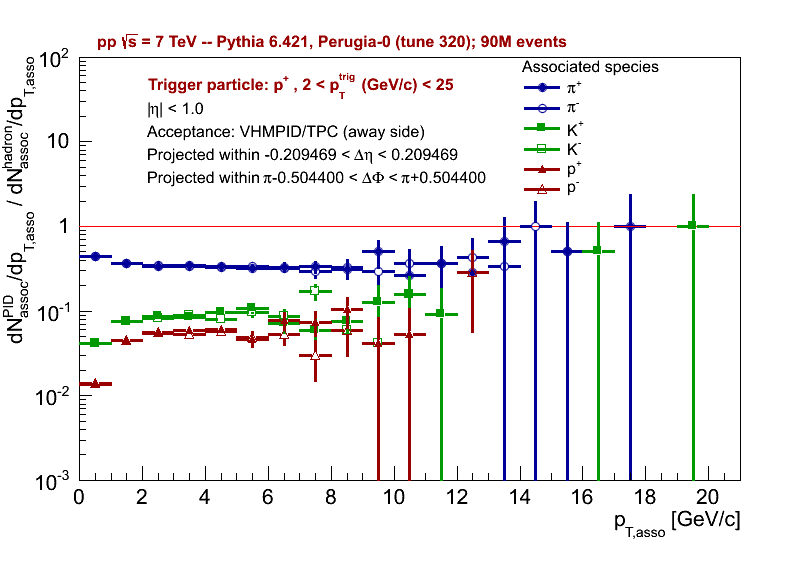}
    \includegraphics[width=0.47\textwidth]{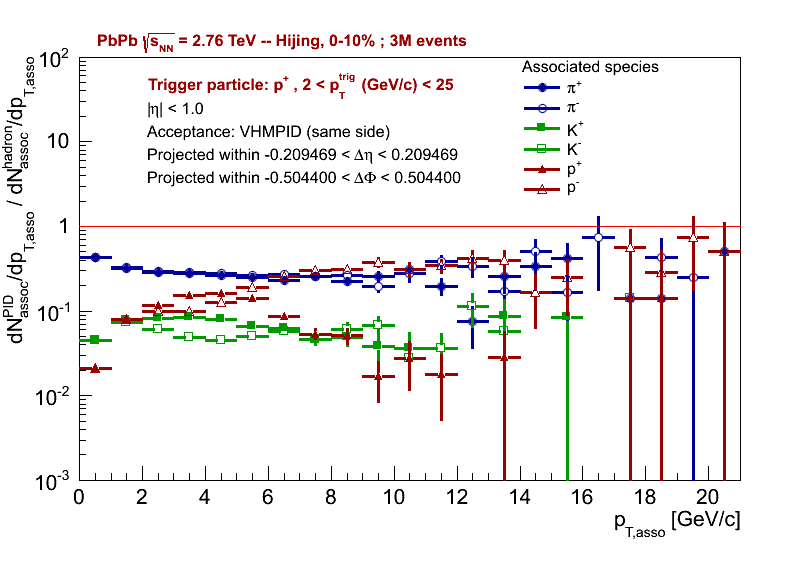}
    \includegraphics[width=0.47\textwidth]{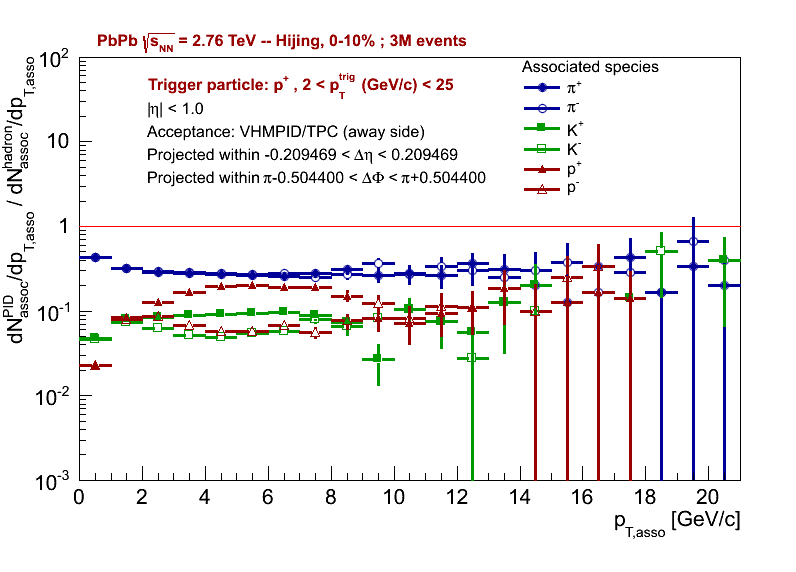}
    \caption{PID-triggered yield relative to the untriggered yield.}
\label{fig:PID-triggered_ratio}
\end{figure}
%%%%%%%%%%%%%%%%%%%%%%%%%%%%%%%%%%%%%

The first study is based on a strong correlation between protons and antiprotons in the unmodified fragmentation process. The {\sl the upper left panel} of Fig.~\ref{fig:PID-triggered_ratio} shows the $\bar{p}-p$ correlation in a single jet cone (same side) which is about an order of magnitude larger than the $p-p$ correlation in the same cone and independent of particle momentum. In contrast the away-side correlations ({\sl upper right hand panel} of Fig.~\ref{fig:PID-triggered_ratio}) exhibit an equal probability for $p-p$ and $\bar{p}-p$. This shows that in an unmodified fragmentation process the baryon number and charge is conserved and leads to highly correlated distributions in the same phase space. It will be relevant to measure this correlation strength in heavy ion collisions in order to determine the quantum number conservation pattern in the hadronization process from an extended equilibrated medium. As an example, quenched HIJING which is shown in the {\sl lower panels} of Fig.~\ref{fig:PID-triggered_ratio} exhibits a peculiar pattern on the same and away side in the $2$ GeV/c $<p_T<8$ GeV/c range in order to account for the baryon-meson anomaly measured at RHIC and LHC~\cite{ALICEbaryon-meson}. 

%%%%%%%%%%%%%%%%%%%%%%%%%%%%%%%%%%%%%
\begin{figure}[!t]
    \centering
\includegraphics[width=0.47\textwidth]{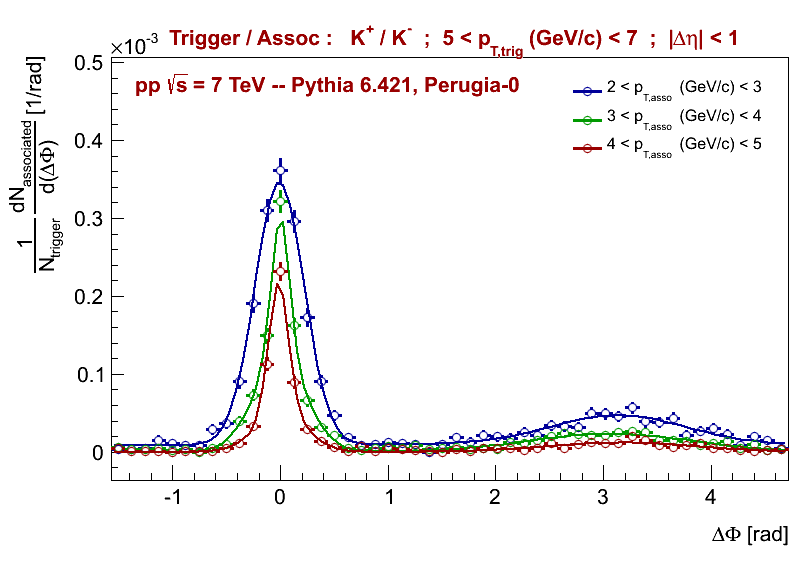}
\includegraphics[width=0.47\textwidth]{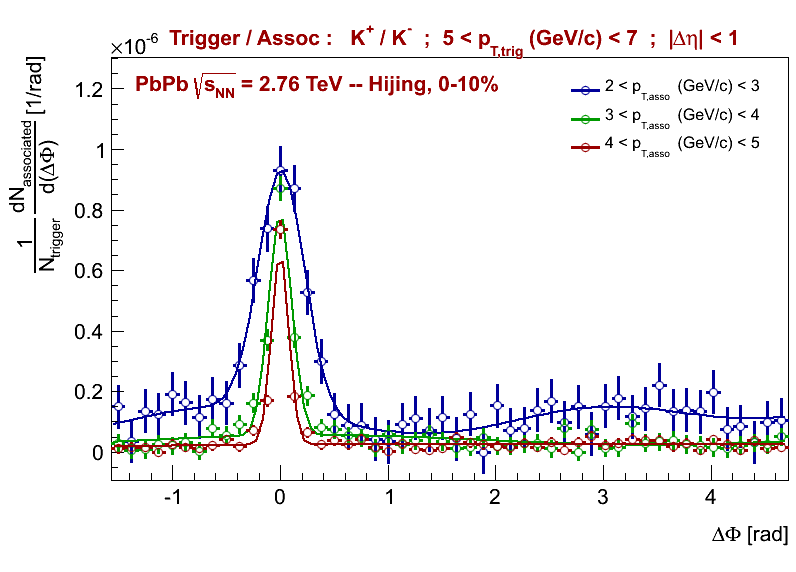}

\includegraphics[width=0.47\textwidth]{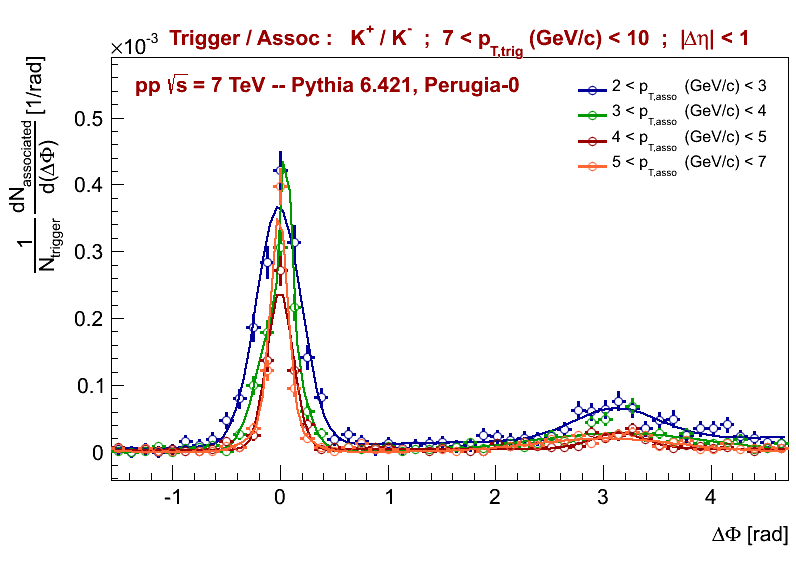}
\includegraphics[width=0.47\textwidth]{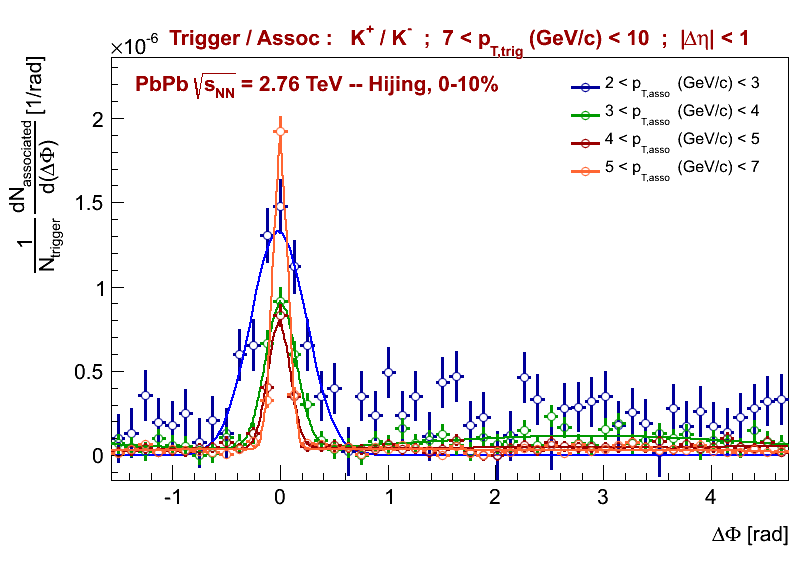}

\includegraphics[width=0.47\textwidth]{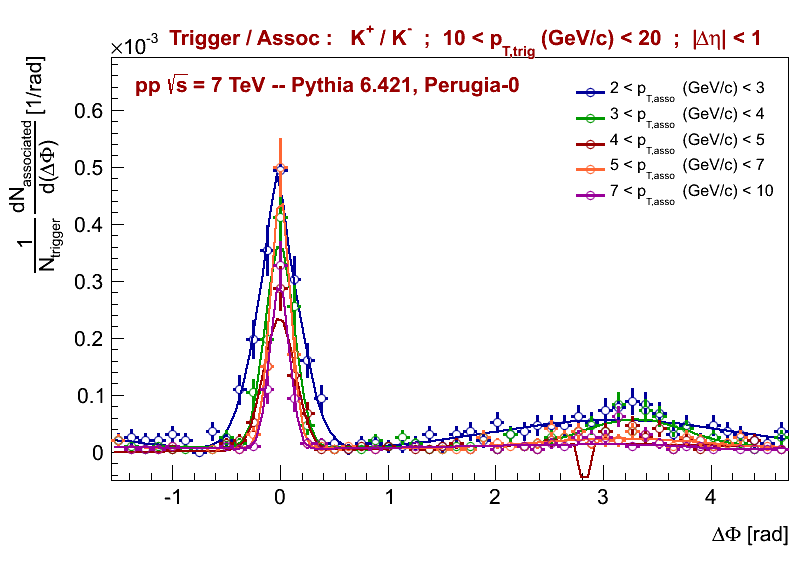}
\includegraphics[width=0.47\textwidth]{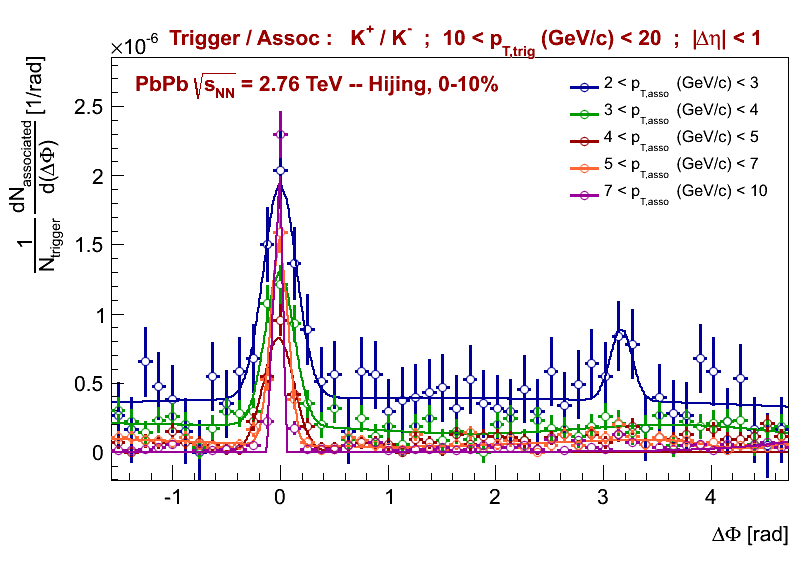}
    \caption{PID-triggered angular correlation for the $K^+$-triggered $K^-$ spectra integrated in the $\left| \Delta \eta \right| < 1$ range. Note, to avoid overlap, some of the curves are shifted.}
\label{fig:PID-triggered_corr}
\end{figure}
%%%%%%%%%%%%%%%%%%%%%%%%%%%%%%%%%%%%%

For such a measurement the same side correlations can be measured with both particles identified and reconstructed in the VHMPID, whereas the away side correlations will rely on a trigger particle identification in the VHMPID and an associated particle assignment based on a statistical sample in the TPC. 

%We have also investigated a sample of 300k events of $0-10\%$ central Pb-Pb collisions at $2.76A$TeV center of mass energy. These events were calculated in AliRoot as well, but using HIJING Monte Carlo Generator, including HIJING shadowing and jet quenching. We also used the VHMPID geometry and acceptance for this calculation. It is interesting to see that the effect observed in pp PYTHIA sample has been modified, so PID-triggered ratios present a medium dependent probe (Fig.\ref{corrfig:3}). However, we note near side correlations originating from the conservation of charges is seen on Fig.~\ref{corrfig:4}. 

An example for flavor dependent measurements, using VHMPID and TOF, is shown in Figs.~\ref{fig:PID-triggered_corr}. It shows that the same side flavor and charge correlations decrease as a function of the transverse momentum of the associated, $p_{T,assoc}$ and trigger particle, $p_{T, trig}$ (the width narrows) compared to the away side correlations which stay roughly constant when the trigger particle momentum is raised. A comparable HIJING calculation shown in the {\sl right hand side of} Fig.~\ref{fig:PID-triggered_corr}, reveals a factor of three increase in the same side correlation amplitude and a constant amplitude on the away side. The question is whether this effect is unique to the strange flavor particles and whether this kinematic pattern persists when one hadronizes out of an extended thermal medium. 

Many more examples of identified particle correlation patterns have been predicted in pp and AA collisions~\cite{corr:2,corr:3,corr:4,corr:5,corr:6,corr:7,corr:8} and can be investigated with a track-by-track PID detector, such as the VHMPID. These include specific patterns in heavy quark fragmentation which could shed light on the $J/\Psi$ production mechanism~\cite{corr:2}, predictions on PID differences between quark and gluon jets~\cite{corr:3,corr:4}, proton--antiproton differences~\cite{corr:5}, and unique effects due to diquarks in the partonic medium~\cite{corr:6,corr:7,corr:8}.

\subsection{Event-by-event PID-spectral shape analysis}
\label{ebyePID}

A large variety of theoretical descriptions of high heavy-ion collisions and high energy nuclear effects originate from the non-perturbative nature of QCD at low momentum (small-$x$). Meanwhile perturbative QCD seems to work well at high-$p_T$ for the strongly interacting processes. The interconnections between perturbative/non-perturbative on one side and statistical/hydrodynamical calculations on the other side have not been resolved. The observed Boltzmann distribution-like behavior at low momenta supports the idea of a thermalized set of degrees of freedom, while the high-$p_T$ tail of the spectra follow a power law given by the field equations without thermodynamics.

\subsubsection{The non-extensive statistical approach}

To deal with this ambivalent behavior a new theoretical approach has been proposed recently~\cite{Tsallis,Biro}. Within the framework of non-extensive statistical theory one can go beyond the first order approximation in the thermodynamical description, which leads to different entropy formulae and non-additive composition rules~\cite{Biro:2012,Barnafoldi:2012ts}. These result in a Tsallis distribution for the energy spectra without explicitly specifying the underlying microscopical processes. The Tsallis distribution features both properties of the momentum or energy distribution functions that were pointed out above. In addition, it fits the experimental data~\cite{cmstsallis,alicetsallis} using a temperature-like ($T$) and a power-like ($1-q$) parameter.
\begin{figure}[!h]
    \centering
    \includegraphics[width=10.5cm]{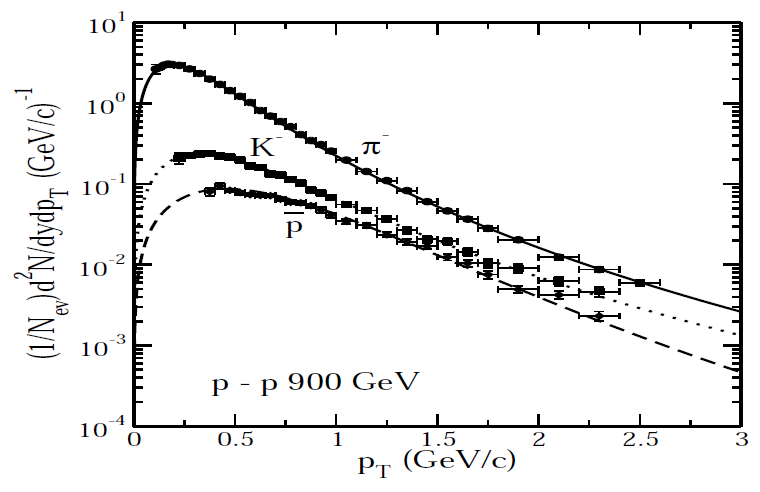}
    \includegraphics[width=0.47\textwidth]{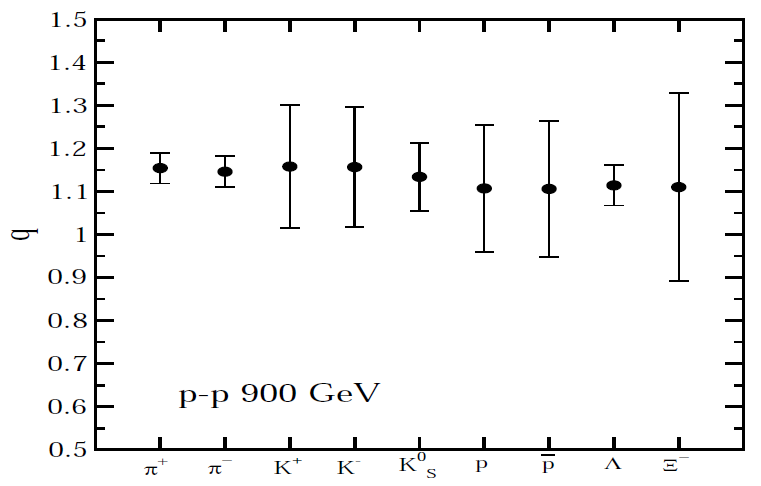}
    \includegraphics[width=0.47\textwidth]{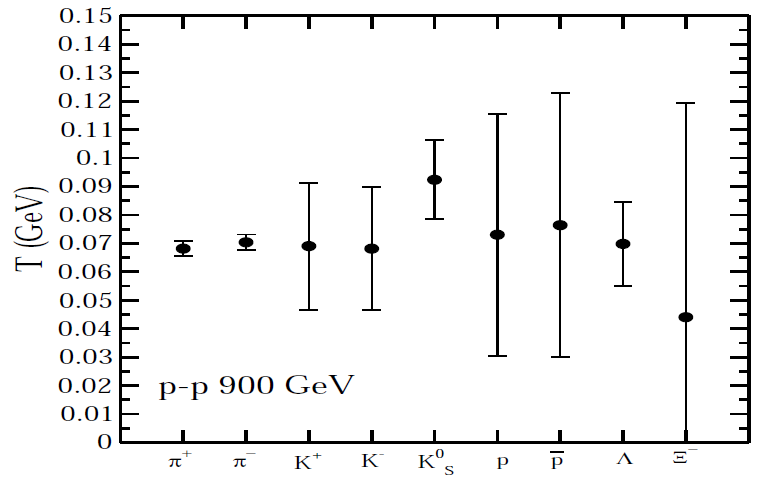}
    \caption{Tsallis fits on ALICE~\cite{ALICE:pid} and CMS~\cite{CMS:pid} data, using a modified-Hagedorn model described in Ref~\cite{Worku}}
\label{fig:Worku-Tsallis}
\end{figure}

In the {\sl upper panel} of Fig.~\ref{fig:Worku-Tsallis} identified spectra measured by the ALICE~\cite{ALICE:pid} and by CMS~\cite{CMS:pid} collaborations are fitted in the low-$p_T$ regime using a modified Hagedorn model. The resulting Tsallis parameters with errors are shown on the lower panels. For a better fit, precise measurements of the the high momentum tail of the PID spectra are necessary. From a theoretical point of view %, the meaning of the Tsallis fit parameters has not been clarified yet. 
no detailed microscopical model has been developed for the partonic/hadronic evolution due to the lack of precise track-by-track data. However, there are various interpretations of the physical meaning of $q$-parameter~\cite{Worku,Becattini,Wilk,Begun,Biro,Urmossy}.

\begin{figure}[!h]
    \centering
    \includegraphics[width=0.47\textwidth]{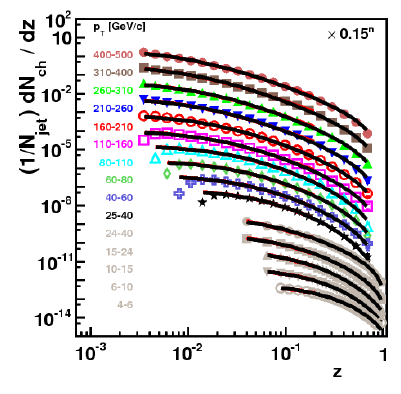}
    \includegraphics[width=0.45\textwidth]{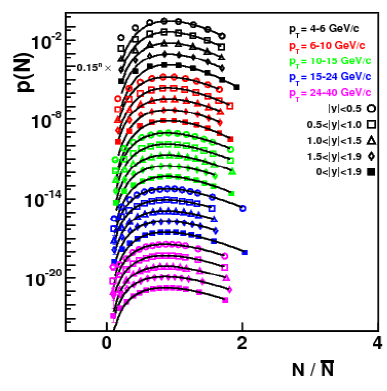}
    \caption{Tsallis fits on jet spectra and jet-multiplicity in Ref~\cite{Urmossy}}
\label{fig:Urmossy-Tsallis}
\end{figure}

The general agreement in this fast evolving, new, and rich field is that track-by-track measurements of identified particle spectra, in parallel with the measurement of event multiplicity with PID, are necessary to understand the microscopic processes in more detail. Track-by-track measurements of particle identified spectra will help us to determine the canonical or microcanonical nature of the system as well as the strength and origin of any multiplicity fluctuations. As an example, Fig.~\ref{fig:Urmossy-Tsallis} presents a simultaneous fit of various spectra and multiplicity distributions in proton-proton collision. 
The {\sl left hand side} shows the momentum spectra, the {\sl right hand side} shows the multiplicity distributions. The fits are based on a $q$-microcanonical model described in Ref.~\cite{Urmossy}.

A similar fit has been done for identified particle spectra data at $0.9$, $2.76$, and $7$ TeV LHC energies measured in proton-proton collisions~\cite{Urmossy:2012pb}. Since transverse momentum spectra is available for identified hadrons at low $p_T$, thus identified hadron ($\pi^{+}$, $K$, and $p$) data can be fitted. 
Moreover, assuming Euler gamma distribution for the multiplicity distribution --- as superstatistics --- Tsallis distribution can be derived a  the transverse energy, $E_T$ and multiplicity, $N$ event-by-event fluctuations assume the Tsallis being of the statistical manifold.  

%%%%%%%%%%%%%%%%%%%%%%%Eddig OK

\subsubsection{Measureable effects}

The PID and track-by-track capabilities in the high-momentum tail of the particle distributions are the main features of the VHMPID detector. Measurements with this would provide necessary input beyond the usual integrated/event-averaged measured properties, in order to resolve the ambiguities that result a more complete understanding of the microscopic processes contributing to the particle production at relativistic energies. Here are some key questions from theoretical point of view:

\begin{itemize}

\item {\sl What is responsible for the power low tail measured at high-$p_T$?} 

Power law tails can be easily derived from self-similar statistical processes. On the other hand (p)QCD dynamics only generates power low distributions by itself.

\item {\sl Can we assume thermodynamical equilibrium for high-$p_T$ particles?}

High momentum hadrons are taught to be non-equilibrium degrees of freedom, however the source of the low- and high-$p_T$ particles are the same. In the usual thermodynamical picture there is a need for thermodynamical equilibrium (e.g. to define $T$), while in higher order thermodynamics (generally with Tsallis distribution) has no such an issue. 

\item {\sl What is the origin of the 'collectivity'? Is it coming from 'quark level' or 'hadron level'?} 

In order to determine the degrees of freedom of the theory, first we need to find the proper scaling parameter (e.g. energy variable: $p_T$, $m_T$, $\beta \gamma (m_T-v \cdot p_T) $, $\beta X$) of the theory --- even measuring it in a cutted phase space (see Fig.~\ref{fig:biroscaling}. The proper scaling can give us information whether dimensionality or finite-volume effects are responsible for the Tsallis distribution. (Latter can cause larger fluctuation.) 
\begin{figure}[!h]
    \centering
    \includegraphics[width=0.8\textwidth]{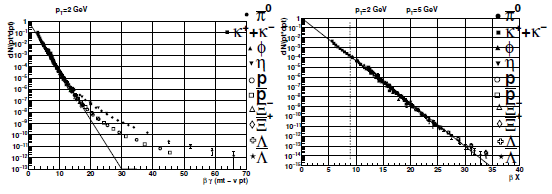}
    \caption{Scaling on hadron spectra at RHIC energies in a blast-wave model Ref~\cite{BiroScaling:2012}}
\label{fig:biroscaling}
\end{figure}
\item {\sl Is there difference between baryon and meson formation? What is the statistical origin of this (e.g coalescence, fragmentation, etc.)?}

To understand this would be important to map differences between microscopical processes in baryon and in meson formations~\cite{Biro:2008km,Biro:2008zzb}, what is e.g. predicted as follows,
%\begin{equation}
$$ \frac{q_{meson}-1}{q_{baryon}-1} = \frac{3}{q_{quark}-1} \ \frac{q_{quark}-1}{2} = \frac{3}{2} \ \ . $$
%\label{biro32rule}
%\end{equation}
Here $q_{meson}$ and $q_{baryon}$ are for the $q$ parameters for a given identified meson and baryon spectra respectively. We note, Tsallis parameter, $T$ might be also differ for different species due to their masses or formation -- which is a natural consequence of the Tsallis formalism~\cite{BiroIDGas,VanKazan}.

\end{itemize}

%%%%%%%%%%%%%%%%%%%%%%%%%%%%%%%%%%%%%%%%%%55
\subsubsection{Proposed measurements}

Recent PID techniques are eligible for low-momentum measurements especially where high statistics available without overlapping in the dE/dx --- up to $2$ GeV/c~\cite{cmstsallis,alicetsallis,Adare:2010fe}. Here, precise determination of the Tsallis parameter $T$ can be obtained by usual experimental setups of the available existing detectors. The precise description of the low-momentum part refers for the 'body' of the Tsallis distribution or for the temperature of the thermal Boltzmann spectra in the thermalized limit. 

To determine the distribution's tail (given by the parameter $q$) we propose track-by-track PID measurements at high momentum in both proton-proton and heavy-ion collisions with the following aims: 
\begin{itemize}

\item Test of the scaling parameter (energy-variable) in even-by-event PID measurements. 

\item Test of the 'rule of 3/2' given above, via measuring and fitting high-precision PID spectra up to $10-20$ GeV/c. 

\item Tsalls-like behavior is more relevant in fragmentation dominated hadronization e.g. in $e^+e^-$ and pp where, the test of the interchange between thermal (Boltzmann) and non-thermal (pQCD) distribution is highly motivated. Especially within the 3 GeV/c $ < p_T < 10$ GeV/c  transverse momentum region in central and mid-central heavy ion collisions. There Tsallis distribution fits better and smoother than other models~\cite{Biro:2008km}. Moreover test of the energy or multiplicity fluctuation required track-by-track PID measurements at high-$p_T$. 

These measurements would provide a stronger constraints for existing theoretical descriptions, especially understanding hadronization at microscopical level.  

\item By the formation of hot dense media in high-energy central nucleus-nucleus collisions hadronization is expected to be modified at the final state (e.g. jet-quenching). This would result changes in the measured PID spectra and in multiplicity distributions especially in the 2 GeV/c $ < p_T < 20$ GeV/c momentum region. 

A detailed species-dependent mapping of this regime would determine the modification of the hadronization in parallel mapping the media via the measured signatures of the quark-gluon plasma.

\end{itemize}

%%%%%%%%%%%%%%%%%%%%%%%%%%%%%%%%%%%%%%%%%%%%%%%%%%%%%%%%%%%%%%%%%%%%%%%%%%%%%%%%%%%%%%%%%%%%%%%%%
\section{Conclusions}

The RHIC results have hinted at the relevance of PID at high transverse momenta on a track-by-track basis. In addition, theory predictions for the higher energies clearly point at the necessity to identify hadrons out to 25 GeV/c. We propose to build a new detector, which will enhance significantly the physics capabilities of ALICE exploring this new regime of hard processes in proton-proton and heavy-ion collisions.
The proposed baseline VHMPID detector, using a pressurized gaseous Cherenkov radiator coupled
to a MWPC-based CsI photo-detector, will allow charged hadron identification in a tunable momentum range starting at $5$ GeV/c up to $25$ GeV/c. These measurements will be unique at the LHC. No other experiment will match these capabilities, and no other experiment will be able to correlate these unique results to low momentum PID and other measurements obtained with the existing ALICE detector.

%%%%%%%%%%%%%%%%%%%%%%%%%%%%%%%%%%%%%%%%%%%%%%%%%%%%%%%%%%%%%%%%%%%%%%%%%%%%%%%%%%%%%%%%%%%%%%%%%
\section*{Acknowledgement}

The authors wish to thank A. Bream, T. Schneider, M. Van Stenis and C. David from CERN/PH-DT for their precious help and contributions in many phases of the project. A. Catinaccio (CERN/PH-DT) and D. Perini (CERN/EN-MME) are also acknowledged for their useful and constructive support.
The collaboration acknowledges also J. Harris and N. Smirnov from Yale University for their contribution to the project.
This work was supported in part by Mexico project PAAPIT IN115808 and Conacyt P79764-F, National Research
Foundation of Korea (NRF), National Science Foundation (USA) under Grants No.  NSF-PHY-0968903 and NSF-PHY-1305280, Hungarian OTKA grants PD73596, NK62044, NK77816, CK77719, CK77815, NK106119, and H07-C 74164, NIH TET 10-1 2011-0061, ZA-15/2009, and E\"otv\"os University. Authors (GGB,DV) also
thank the J\'anos Bolyai Research Scholarship of the Hungarian Academy of Sciences.

%%%%%%%%%%%%%%%%%%%%%%%%%%%%%%%%%%%%%%%%%%%%

%%%%%%%%%%%%%%%%%%%%%%%%%%%%%%%%%%%%%%%%%%%%%%%%%%%%%%%%%%%%%%%%%%%%%%%%%%%%%%%%%%%%%%%%%%%%%%%%%%%%%%%%%

\newpage

\input{vhmpid_bib_v22.tex}